\documentclass[a4paper,11pt]{article}
\pdfoutput=1
\usepackage{jcappub}
\usepackage[utf8]{inputenc}
\usepackage[T1]{fontenc}
\usepackage{microtype}
\usepackage{bm}
\usepackage[dvipsnames,svgnames,x11names]{xcolor}
\usepackage{aas_macros}
\usepackage{multirow}
\usepackage{physics}

\let\L\relax
\DeclareMathOperator{\L}{\mathcal{L}}
\DeclareMathOperator{\deltaD}{\delta^\textsc{d}}

\DeclareMathOperator{\erfc}{erfc}
\renewcommand{\d}{\mathrm{d}}
\newcommand{\vr}{{\bm r}}
\newcommand{\vx}{{\bm x}}
\newcommand{\vq}{{\bm q}}
\newcommand{\vPsi}{{\bm \Psi}}
\newcommand{\vk}{{\bm k}}
\newcommand{\vp}{{\bm p}}
\newcommand{\vP}{{\bm P}}
\newcommand{\vDelta}{{\bm\Delta}}
\newcommand{\rhobarm}{{\bar\rho_\mathrm{m}}}
\newcommand{\rhocrit}{{\rho_\mathrm{c}}}
\newcommand{\Om}{{\Omega_\mathrm{m}}}
\newcommand{\Tree}{\mathrm{T}}
\newcommand{\PM}{\mathrm{PM}}
\newcommand{\xs}{x_\mathrm{s}}
\newcommand{\Eul}{\mathrm{E}}
\newcommand{\Lag}{\mathrm{L}}

\title{
Cosmological simulation in tides:
power spectra,
halo shape responses,
and shape assembly bias
}

\author[a]{Kazuyuki Akitsu,}
\author[b]{Yin Li}
\author[c,a]{and Teppei Okumura}

\affiliation[a]{Kavli Institute for the Physics and Mathematics of the Universe (WPI), 
UTIAS, The University of Tokyo,Kashiwa, Chiba 277-8583, Japan}
\affiliation[b]{Center for Computational Astrophysics
\& Center for Computational Mathematics, Flatiron Institute,
162 5th Avenue, New York, NY 10010, USA}
\affiliation[c]{Institute of Astronomy and Astrophysics, Academia Sinica,
Roosevelt Road, Taipei 10617, Taiwan, ROC}

\emailAdd{kazuyuki.akitsu@ipmu.jp}
\emailAdd{yinli@flatironinstitute.org}
\emailAdd{tokumura@asiaa.sinica.edu.tw}

\abstract{
The well-developed separate universe technique enables accurate
calibration of the response of any observable to an isotropic
long-wavelength density fluctuation.
The large-scale environment also hosts tidal modes that perturb
all observables anisotropically.
As in the separate universe, both the long tidal and density modes
can be absorbed by an effective anisotropic background,
on which the interaction and evolution of the short modes
change accordingly.
We further develop the tidal simulation method, including proper
corrections to the second order Lagrangian perturbation theory (2LPT)
to generate initial conditions of the simulations.
We measure the linear tidal responses of the matter power spectrum,
at high redshift from our modified 2LPT,
and at low redshift from the tidal simulations.
Our results agree qualitatively with previous works,
but exhibit quantitative differences in both cases.
We also measure the linear tidal response of the halo shapes,
or the shape bias, and find its universal relation with the linear
halo bias, for which we provide a fitting formula.
Furthermore, analogous to the assembly bias, we study
the secondary dependence of the shape bias, and discover
for the first time the dependence on the halo concentration and axis ratio.
Our results provide useful insights for studies of the intrinsic alignment
as a source of either contamination or information.
These effects need to be correctly taken into account when one uses intrinsic alignments of galaxy shapes as a precision cosmological tool.
}

\begin{document}
\maketitle
\flushbottom

\section{Introduction}
\label{sec:intro}

In large-scale structure (LSS) surveys, what we expect to observe is the long-range correlation of biased tracers (e.g. galaxy number density field) mediated by long-wavelength perturbations.
Long-wavelength perturbations have an impact on the formation and evolution of the small-scale structure via nonlinear mode-couplings induced by gravity.
Because of the equivalence principle, the leading-order effects of the long-wavelength gravitational potential on the local physics arise from the second derivative of the gravitational potential,
which can be decomposed into the large-scale overdensity and tidal fields.
Therefore, it has been a fundamental task of LSS cosmology to investigate how the large-scale overdensity and tidal field generate the long-range correlation of biased tracers.

The separate universe simulations provide us with a powerful means to accurately measure or calibrate the local response of various statistics, such as the power spectrum and halo mass function, to the large-scale overdensity \cite{Li_etal:2014a, Wagner_etal:2014,Baldauf_etal:2015,Li_etal:2015,Lazeyras_etal:2015}.
The homogeneity and isotropy of the large-scale overdensity leads to the simple prescription 
for the modification, i.e., just changing the cosmological parameters according to the amplitude of the local overdensity.
On the other hand, large-scale tidal field breaks isotropy and thus
the local background embedded into the large-scale tides is no longer Friedmann–Lema{\rm \^\i }tre–Robertson–Walker (FLRW) universe but becomes rather  anisotropic~\cite{BondMyers96I, Akitsu_etal:2016}.
Recently there have appeared some works that incorporate the large-scale tidal field in $N$-body simulations
by introducing anisotropic scale factors~\cite{Schmidt_etal:2018, Stucker_etal:2020, Masaki:2020a, Masaki:2020b}.
These simulations enable us to isolate effects of the large-scale tidal field 
since we can impose homogeneous tides in the entire simulation box.
In other words, utilizing such simulations we can robustly measure tidal responses separately from other effects.

In this paper, we implement the large-scale tidal field in $N$-body simulations
together with the appropriate initial condition generator where we solve the second order Lagrangian perturbation theory (2LPT) in an anisotropic background. 
Using our simulations, we measure two kinds of tidal responses: 
the matter power spectrum response and the halo shape response.
The large-scale tidal field makes the local clustering pattern anisotropic,
which potentially mimics other anisotropic signatures such as the redshift-space distortion and Alcock-Paczy\'nski effect~\cite{Akitsu_Takada:2017,LiSchmittfullSeljak17,Akitsu_etal:2019}.
The tidal response of the matter power spectrum measured from our simulations can give a rough estimate
of such contaminants on small scales. The response also allows us to compute the covariance of weak lensing power spectra~\cite{Barreira_etal:2017}.
Since no conclusive results have yet been reached about the amplitude or scale-dependence of the tidal response of the matter power spectrum at high redshifts due to possible numerical artifacts~\cite{Stucker_etal:2020,Masaki:2020b},
we show the tidal response of the matter power spectrum from our modified 2LPT.

The density field is not the only one affected by the large-scale tidal field; 
shapes of galaxies and halos are naturally affected as well.
Indeed, these shapes are aligned with each other
even at large separation due to the large-scale tidal field, known as ``intrinsic alignment''~\cite{Catelan_etal:2000,Hirata_Seljak:2004, Mandelbaum_etal:2005}.
The intrinsic alignment is usually considered as a source of the error 
in measuring weak lensing signals from galaxy imaging surveys.
However, given that its origin is the underlying large-scale tidal field, 
it should have their own cosmological information.
For example, the intrinsic alignment potentially probes the energy budget of the universe~\cite{Taruya_Okumura:2020,Kurita_etal:2020}, the stochastic gravitational waves background~\cite{Schmidt_Jeong:2012, Schmidt_etal:2013}, and the angular-dependent primordial non-Gaussianity~\cite{Schmidt_etal:2015,Akitsu_etal:2020,Kogai_etal:2020}.
In order to extract the cosmological information from measurements of intrinsic alignments, it is of importance to test theoretical models of the intrinsic alignment.
Our simulation allows us to directly examine the simplest model of the intrinsic alignment, the so-called linear alignment or tidal alignment model~\cite{Catelan_etal:2000,Hirata_Seljak:2004}, which predicts shapes of galaxies or halos linearly aligned with large-scale tides.
Using the well-controlled tidal simulations, 
we quantify the strength of alignments over a wide redshift and mass range and explore the secondary dependence of the alignment strength on halo properties other than halo mass.

This paper is laid out as follows.
In Section \ref{sec:method}, we present how to absorb the large-scale tidal and density fields into the simulation background and modifications of the 2LPT initial condition in anisotropic background.
Section \ref{sec:sim} summarizes our simulation specifications.
In Section \ref{sec:results},
we show results on both the tidal response of the matter power spectrum and the halo shapes.
We give conclusion and discussion in Section \ref{sec:discuss}.
Appendices \ref{app:2spt}-\ref{app:force} provide the details of computation and our modifications.
We show the convergence test of our simulations by comparing ours with the usual separate universe simulations in Appendix~\ref{app:SU_comp}.
Results of the different shape definitions used in the main text are summarized in Appendix~\ref{app:nonreduced}.

\section{Methodology}
\label{sec:method}

A uniform (DC) tidal or density field preserves the translational symmetry,
and can be modeled effectively as a time-dependent coordinate transformation.
By doing this, we separate out the DC modes, and absorb
these including their evolution into an effective background, which we can
include in $N$-body simulations~\cite{Gnedin:2011kj}.
In this section we present analytical derivations and numerical implementations
of this method.
Many of the results have already been obtained in the recent literature~\cite{Stucker_etal:2020}.
Here we simplify the derivations, and present for the first time the modulation
of the tidal modes on the second order Lagrangian perturbation theory (2LPT).

\subsection{Model uniform tidal field by coordinate transformation}
\label{sub:background}

In Newtonian cosmology, the large-scale effect of infinitely long-wavelength (i.e. DC)
density and tidal modes on a dark matter particle can be absorbed by a coordinate
transformation
\begin{equation}
    r_i = a_{ij} x_j,
    \label{eq:pos}
\end{equation}
where $\vr$ is the physical coordinate of the dark matter particle, and
$a_{ij}$ is a symmetric matrix that absorbs the DC modes so that the
large-scale displacement is isotropic in the $\vx$ coordinate.
We normalize $a_{ij}$ to the scale factor of global expansion $a\,\delta_{ij}$
in the absence of any DC-mode, in which case $\vx$ is reduced to the usual comoving
coordinates.

From Eq.~\eqref{eq:pos} we can immediately separate the physical velocity $u_i$ into
the expansion of a local background and a peculiar component.
\begin{equation}
    u_i \equiv \dot r_i = H_{ij} r_j + v_i,
    \label{eq:vel}
\end{equation}
where the overdot denotes a time derivative, $H_{ij} \equiv \dot a_{ik}
[a^{-1}]_{kj}$ describes a local anisotropic Hubble expansion, and $v_i\equiv
a_{ij}\dot x_j$ is the peculiar velocity.

The dark matter particles follow the Newtonian equation of motion
\begin{equation}
    \dot u_i = - \frac{\partial}{\partial r_i} (\Phi + \phi),
    \label{eq:acc}
\end{equation}
where we split the gravitational potential into an effective background potential $\Phi$ and a peculiar potential $\phi$.
By plugging Eq.~\eqref{eq:vel} into Eq.~\eqref{eq:acc}, we see that the acceleration also splits into
a local background expansion and a peculiar piece,
which are respectively driven by $\Phi$ and $\phi$,
\begin{align}
    \ddot a_{ik} [a^{-1}]_{kj} r_j &= - \partial\Phi / \partial r_i,
    \label{eq:Friedmann} \\
    \dot v_i + H_{ij} v_j &= - \partial\phi / \partial r_i.
    \label{eq:eom}
\end{align}

One can regard Eq.~\eqref{eq:Friedmann} as a modified Friedmann equation.
The large-scale stress due to the DC modes is absorbed into $\Phi$, leaving $\phi$ sourced only by local structures,
\begin{equation}
    \nabla_\vr^2 \phi = 4\pi G \rhobarm (1 + \Delta_0) \delta,
    \label{eq:poisson_local}
\end{equation}
where $\rhobarm$ is the mean density of matter, $\Delta_0$ is the
large-scale overdensity relative to $\rhobarm$,
and $\delta$ denotes the overdensity with respect to the local background
density $\rhobarm (1 + \Delta_0)$.
The mass conservation between $a\to0$ and a later time requires
$\rhobarm (1 + \Delta_0) \d^3\vr = \rhobarm a^3 \d^3\vx$, or
\begin{equation}
    1 + \Delta_0 \equiv \frac{a^3}{\det a_{ij}}.
    \label{eq:D0=det}
\end{equation}
Without loss of generality, we can simplify the equations and the numerical
implementation by rotating the simulation box to align with the principal axes of the
DC tides, so that $a_{ij} =
a_i \delta_{ij}$ and $H_{ij} = H_i \delta_{ij} = \dot a_i \delta_{ij} / a_i$
with their off-diagonal degrees of freedom eliminated.
Let us define $\Delta_i$ as the relative difference of $a_i$ to $a$:
\begin{equation}
    1 + \Delta_i \equiv \frac{a_i} a, \quad i = 1, 2, 3.
\end{equation}
Combining the above two equations implies the mass conservation, 
\begin{equation}
    (1+\Delta_0) (1+\Delta_1) (1+\Delta_2) (1+\Delta_3) = 1.
    \label{eq:D0123}
\end{equation}
We also have
\begin{equation}
    H_i = H + \frac{\dot{\Delta}_i}{1+\Delta_i}
    \simeq H +\dot{\Delta}_i,
    \label{eq:Hi}
\end{equation}
where $H$ is the global expansion rate.
The approximation holds at high redshifts when $\Delta_i \ll 1$.

This rotation also diagonalizes the DC tidal field,
the traceless part of the Hessian of the large-scale potential,
to $\tau_{ij} = \tau_i \delta_{ij}$.
Now we can write down the effective background potential
that absorbs the DC density and tidal modes
\begin{equation}
    \Phi = \frac{2}{3}\pi G\rhobarm(1+\Delta_0) r^2
        - \frac{\Lambda}{6} r^2
        +2 \pi G\rhobarm \tau_i r_i^2.
    \label{eq:ani_phi_background}
\end{equation}
The DC density modulation $1+\Delta_0$ is determined through Eq.~\eqref{eq:D0=det},
and the last term is sourced by the DC tidal mode.
One can easily verify that substituting the first two terms without $\Delta_0$
into Eq.~\eqref{eq:Friedmann} gives rise to the usual Friedmann equation.

To determine the evolution of $\Delta_i$ in the presence of $\tau_i$,
we plug Eq.~\eqref{eq:ani_phi_background} into Eq.~\eqref{eq:Friedmann}
and subtract the usual Friedmann equation. Then it reads
\begin{equation}
    \ddot \Delta_i + 2H\dot \Delta_i
    = -4\pi G \bar{\rho}_m\left(\tau_i + \frac{1}{3}\Delta_0\right) (1+\Delta_i).
    \label{eq:Deltai_eq}
\end{equation}
Note that the addition of $\Delta_0$ in Eq.~\eqref{eq:ani_phi_background} is valid
even at nonlinear level.
This can be verified by setting $\tau_i=0$ to reproduce the evolution of the
spherical collapse model (e.g.\ \cite{Wagner_etal:2014}).
For anisotropic simulations we compute anisotropic scale factors in the following subsections
by solving Eq.~\eqref{eq:Deltai_eq} and Eq.~\eqref{eq:D0123} numerically,
using the matter dominated initial conditions.

Before proceeding to the next subsection, we can derive some analytic solutions
for better understanding.
Linearizing Eq.~\eqref{eq:Deltai_eq} drops the $1 + \Delta_i$ factor
on the right hand side, and yields an equation with the same form
as that of the linear growth function $D$, if we replace 
$- (\tau_i + \Delta_0/3)$ with $\Delta_i$.
Because both $\tau_i$ and $\Delta_0$ are proportional to $D$,
the linear-order solution is simply given by
\begin{equation}
    \Delta_i^{(1)} = - \tau_i - \frac{\Delta_0^{(1)}}{3}.
    \label{eq:Di_linear}
\end{equation}
Unsurprisingly, the linear anisotropic correction to the scale factor is the sum of
the isotropic one $ - \Delta_0^{(1)} / 3$ and the tidal mode.
Summing over $i$ and using the traceless constraint $\sum_i\tau_i = 0$,
one can verify
\begin{equation}
    \Delta_0^{(1)} + \sum_i \Delta_i^{(1)} = 0,
    \label{eq:D0123_linear}
\end{equation}
which is equivalent to Eq.~\eqref{eq:D0123} in the linear theory limit.

The linear tide $\tau_i$ also induces the second order overdensity that is consistent with the perturbation theory prediction.
We derive the second order solution for $\Delta_0$ in App.~\ref{app:2spt},
and show here the solution for the matter dominated era,
\begin{equation}
\Delta_0^{(2)} =
\frac57 {\Delta_0^{(1)}}^2 + \frac27 \sum_i {\Delta_i^{(1)}}^2
= \frac{17}{21} {\Delta_0^{(1)}}^2 + \frac27 \sum_i \tau_i^2.
\end{equation}
This is consistent with the result from the second-order standard perturbation theory (e.g.\ \cite{sherwin2012}).

\subsection{Lagrangian perturbation theory and initial conditions}
\label{sub:lpt}

Due to gravity, the long-wavelength modes are coupled to the short ones
and affect their growth.
Here we solve the leading-order anisotropic perturbations to the first and second order
Lagrangian displacement, and use the results to generate initial conditions
for $N$-body simulations.
The long modes $\Delta_0$ and $\Delta_i$ are assumed to be first order in the following derivations.

The Lagrangian perturbation theory follows the evolution of the displacement
field $\vPsi(\vq)$, a mapping from a particle's Lagrangian position $\vq$
to its Eulerian position $\vx$:
\begin{equation}
    x_i = q_i + \Psi_i({\vq}).
    \label{eq:disp}
\end{equation}
Before shell crossing, the overdensity is simply related to its Jacobian determinant by
\begin{equation}
    \delta = \Bigl| \frac{\partial{\vx}}{\partial{\vq}} \Bigr|^{-1} - 1
           = \bigl| \delta_{ij} + \Psi_{i,j} \bigr|^{-1} - 1,
    \label{eq:J}
\end{equation}
where $\Psi_{i,j} \equiv \partial\Psi_i / \partial q_j$.

To derive the master equation of the Lagrangian perturbation theory,
we substitute Eq.~\eqref{eq:disp} into Eq.~\eqref{eq:eom},
\begin{equation}
    \ddot \Psi_i^{(1)} + 2 \sum_i H_i \dot \Psi_i^{(1)} = - \sum_i \frac1{a_i}
    \frac{\partial\phi}{\partial r_i}.
\end{equation}
Taking the derivative with respect to $x_i$ and then summing over $i$,
we obtain the master equation
\begin{equation}
\sum_{ij}~
    \Bigl| \frac{\partial{\vx}}{\partial{\vq}} \Bigr|
    \bigl[\delta_{ij} + \Psi_{i,j}  \bigr]^{-1}
    \Bigl[ \ddot{\Psi}_{i,j} + 2H_i\dot{\Psi}_{i,j} \Bigr]
    = 4 \pi G \rhobarm (1+\Delta_0)
    \Bigl( \Bigl| \frac{\partial{\vx}}{\partial{\vq}} \Bigr| - 1 \Bigr),
    \label{eq:master}
\end{equation}
where we have used Eq.~\eqref{eq:poisson_local} and Eq.~\eqref{eq:J}, and the chain rule
$\partial / \partial x_i = \bigl[ \partial q_j / \partial x_i \bigr] \partial / \partial q_j$.

Now let us start with the Zel'dovich approximation (ZA; the linear Lagrangian
perturbation theory~\cite{zeldovich1970}).
Keeping only the leading order displacement terms in Eq.~\eqref{eq:master} leads to
\begin{equation}
    \sum_i \ddot \Psi_{i,i}^{(1)} + 2 \sum_i  H_i \dot \Psi_{i,i}^{(1)}
    = \frac{3}{2}H^2 \Om(a)  (1 + \Delta_0) \sum_i  \Psi_{i,i}^{(1)}.
    \label{eq:disp_1st}
\end{equation}
In deriving the above equation we have used
$\bigl| \partial{\vx} / \partial{\vq} \bigr|^{-1} \simeq 1 + \sum_i\Psi_{i,i}$
and $\Om(a) \equiv \rhobarm / \rhocrit = 8 \pi G \rhobarm / 3 H^2$,
with $\rhocrit$ being the critical density.

At linear order the vorticity in $\Psi_{i,j}^{(1)}$ decays,
so the growing displacement solution is a potential flow
$\Psi_i^{(1)} = - \partial\psi_W^{(1)}/ \partial q_i \equiv -\psi_{W,i}^{(1)}$,
with the potential $\psi_W^{(1)}$ sourced by the overdensity in Lagrangian space:
\begin{equation}
    \nabla_{\vq}^2 \psi_W^{(1)} \equiv \sum_i\psi_{W,ii}^{(1)} = \delta^{(1)}.
\end{equation}
In terms of $\psi_W^{(1)}$, Eq.~\eqref{eq:disp_1st} is
\begin{equation}
 \sum_i \ddot \psi^{(1)}_{W,ii} + 2 \sum_iH_i  \dot\psi^{(1)}_{W,ii}
   - \frac32  \Om(a) (1 + \Delta_0) \sum_i\psi^{(1)}_{W,ii} = 0.
   \label{eq:psiW_1st}
\end{equation}
The subscript $W$ here denotes local quantities inside a \emph{window},
within which the DC density and tidal modes can be nonzero.

This is in contrast with the usual linear growth equation, which describes
the evolution of the short modes in the global background where the long modes vanish.
It can be obtained by setting $H_i \to H$ and $\Delta_0 \to 0$ in the above equation,
\begin{equation}
    \sum_i\ddot{\psi}^{(1)}_{ii} + 2H\sum_i\dot{\psi}^{(1)}_{ii} - \frac{3}{2}H^2\Om(a) \sum_i\psi^{(1)}_{ii} = 0.
    \label{eq:psi_1st}
\end{equation}
The solution to this equation gives the usual time dependence by the linear growth function,
$\psi^{(1)} \propto D(t)$.

In the presence of the long modes, the displacement potential $\psi^{(1)}_W$
receives corrections of order ${\cal O}(\psi^{(1)} \Delta)$.
We denote this correction as $\epsilon^{(1)}$ so that the solution of Eq.~\eqref{eq:psiW_1st} can be written as
\begin{equation}
    \psi_{W}^{(1)} = \psi^{(1)} + \epsilon^{(1)}.
\end{equation}
Then using Eq.~\eqref{eq:Hi}, Eq.~\eqref{eq:psiW_1st}, and Eq.~\eqref{eq:psi_1st}
one can show that $\epsilon^{(1)}$ satisfies
\begin{equation}
    \sum_i\ddot{\epsilon}^{(1)}_{,ii} + 2H \sum_i\dot{\epsilon}^{(1)}_{,ii} - \frac{3}{2}H^2\Om(a)  \sum_i\epsilon^{(1)}_{,ii} = -2 \sum_i\dot{\psi}_{,ii}^{(1)}\dot{\Delta}_i
    + \frac{3}{2}H^2\Om(a)\Delta_0\sum_i \psi_{,ii}^{(1)}.
    \label{eq:eps_1st}
\end{equation}
We can solve this equation by rewriting it in Fourier space
\begin{equation}
    \ddot \epsilon^{(1)} +2H \dot \epsilon^{(1)}
    - \frac32 \Om(a) \epsilon^{(1)}
    = - 2\dot \psi^{(1)} \sum_i\hat p_i^2 \dot \Delta_i
    + \frac32 H^2 \Om(a) \psi^{(1)} \Delta_0.
\end{equation}
Here in this subsection (and App.~\ref{app:2lpt}) we use $\vp$ to denote
the local Lagrangian space wavevector, with $\hat\vp$ being its direction.
The above equation clearly shows that the effect of the long modes manifests in the quadrupolarly
direction-dependent Hubble drag, whose coefficients depend on the growth
history of the long modes $\Delta_i(t)$.

To solve the above equation we can first decompose it as
\begin{equation}
    \epsilon^{(1)}(a, \hat{\vp}) = \sum_i\hat p_i^2 \epsilon^{(1)}_i(a),
    \label{e}
\end{equation}
with each component $\epsilon^{(1)}_i(a)$ satisfying
\begin{equation}
    \ddot \epsilon_i^{(1)} + 2H \dot \epsilon_i^{(1)}
    - \frac32 \Om(a) \epsilon_i^{(1)}
    = - 2 \dot \psi^{(1)} \dot \Delta_i + \frac32 H^2\Om(a) \psi^{(1)} \Delta_0.
    \label{eq:ei}
\end{equation}
Note that $\epsilon_i^{(1)}$ here is different from $\epsilon^{(1)}_{,ii}$ in Eq.~\eqref{eq:eps_1st}.
For the matter dominated era, assuming that $H = 2/3t$, $\Om(a) = 1$, and
the long modes are well sub-horizon ($\Delta_0, \Delta_i \propto D^{(1)}$),
we obtain
\begin{equation}
    \epsilon^{(1)}_i =
    -\frac47 \Delta_i \psi^{(1)} + \frac37 \Delta_0 \psi^{(1)}.
\end{equation}

Now we can write down the Fourier-space correction to ZA due to the long modes
\begin{equation}
    \epsilon^{(1)} = \psi^{(1)} 
    \Bigl( \frac37 \Delta_0 - \frac47 \sum_i\Delta_i \hat p_i^2 \Bigr).
\end{equation}
This can be understood as a direction-dependent modulation on the linear growth function
\begin{equation}
    D_W(t, \hat\vp) = D(t)
    \Bigl(1 + \frac{3}{7}\Delta_0 - \frac{4}{7} \sum_i\Delta_i \hat p_i^2 \Bigr),
\label{eq:DW}
\end{equation}
where $D_W(t,\hat\vp)$ is the modified linear growth function for a Fourier mode along $\hat\vp$.
Eq.~\eqref{eq:DW} can be directly compared to the results derived with the standard
perturbation theory in the Einstein de-Sitter universe.
For an isotropic perturbation, $\Delta_i = - \Delta_0 / 3$, so $D^{(1)}_W = D^{(1)} (1 + 13
\Delta_0 / 21)$ and for pure tides, $\Delta_0=0$ and $\Delta_i = -\tau_i$, so $D^{(1)}_W = D^{(1)} (1 + 4 \sum_i\tau_i \hat p_i^2 / 7)$.

The above derivations have shown that the tidal effect on ZA is simpler in Fourier space,
and can be captured by a direction-dependent modulation on the linear growth function.
However, this is not the case for 2LPT, for which we find the correction more straightforward
in configuration space.
Again we define the second order displacement potential and its correction by
$\psi_{W}^{(2)} = \psi^{(2)} + \epsilon^{(2)}$.
For the matter dominated era, we find the following solution
\begin{align}
\sum_i\epsilon^{(2)}_{,ii}(t, \vq)
=&\frac14 \left[-\frac{16}{9}\sum_i\psi^{(2)}_{,ii}(t,\vq)\Delta_i + \frac89\sum_{ij}\psi^{(1)}_{,ij}(t,\vq)\psi^{(1)}_{,ji}(t,\vq)\Delta_i \right]
\nonumber\\
&+\frac16 \left[\sum_{i}\psi_{,ii}^{(2)} - \frac12\sum_{i}\left( \psi^{(1)}_{,ii}\right)^2- \frac12\sum_{ij}\psi^{(1)}_{,ij}\psi^{(1)}_{,ji}   \right]\Delta_0
\nonumber\\
&+\frac14\left[-\frac23\sum_{i}\psi^{(1)}_{,ii}(t,\vq)\sum_{j}\epsilon_{,jj}^{(1)}(t,\vq) + \frac{20}{9}\sum_{ij}\psi^{(1)}_{,ij}(t,\vq)\epsilon_{,ij}^{(1)}(t,\vq)  \right].
\end{align}
We present the general formula and its derivation in App.~\ref{app:2lpt}.

Having derived the leading order perturbations to ZA and 2LPT by the DC modes,
we can implement them numerically to generate initial conditions for our tidal simulations.
We modify the initial condition code \texttt{2LPTIC}~\cite{Crocce_etal:2006}
to include the $\epsilon^{(1)}$ and $\epsilon^{(2)}$ corrections to the displacements,
as well as the corresponding corrections to the initial velocities.
Since $\epsilon^{(1)}\propto D^2$ and $\epsilon^{(2)}\propto D^3$ during
matter domination, the corrections to ZA and 2LPT velocities are respectively given by
\begin{align}
    a\dot\epsilon^{(1)}_{,i} &= 2 H f_1 \epsilon^{(1)}_{,i},  \nonumber\\
    a\dot\epsilon^{(2)}_{,i} &= 3 H f_1 \epsilon^{(2)}_{,i}.
    \label{eq:vel_2LPT}
\end{align}
with $f_1\equiv \d \ln D/\d \ln a \simeq [\Om(a)]^{5/9}$.
Since we generate initial conditions at $a_\mathrm{i}$
deep in the matter dominated era, the above approximations should be accurate.

There is one other thing to note about the velocity modifications.
Here we derive the corrections to the peculiar velocity $a\dot \vx$.
However, the series of \texttt{Gadget} codes uses the canonical momentum
$\vP \equiv a^2m\dot \vx$ as the internal velocity variable,
and another velocity variable ${\bm u} \equiv \sqrt{a} \dot{\vx}$
in their data format when saving and loading data.
Therefore, given that we are dealing with anisotropic scale factors,
we need to be careful when converting among peculiar velocity,
momentum, and velocity variable on disks.
We multiplied $\sqrt{a}\dot\Psi_i$ by $\sqrt{1+\Delta_i}$ when generating
and saving initial conditions, and
then modified the conversion factor $a\sqrt{a}$ from ${\bm u}$ to $\vP$ as $a_i\sqrt{a_i}$,
which eventually results in the modified canonical momentum $P_i=a_i^2\dot x_i$.

\subsection{Particle-mesh and tree forces}
\label{sub:treepm}

Because the new effective background evolves anisotropically,
the gravitational force and the equation of motion,
isotropic in physical coordinates, need to be modified
and expressed in the local comoving coordinates.
We describe the modification of the force law here,
and explain the time integration in the next subsection.

We focus on the TreePM method which computes gravitational force efficiently
by splitting it into the long-range and short-range contributions,
computed by the particle mesh (PM) method and the tree algorithm \cite{Bagla02, BaglaRay03},
respectively.
The PM forces can be solved efficiently in Fourier space, and the tree forces
of nearby particles are summed with the help of a tree data structure.
In the absence of the DC modes,
\begin{align}
    \phi &= \phi^\PM + \phi^\Tree, \nonumber\\
    \phi^\PM(\vk) &= - 4\pi G \rhobarm a^2
        \frac{\delta(\vk)}{k^2} e^{- k^2 \xs^2}, \nonumber\\
    \phi^\Tree(\vx) &= - \frac{G m}a \sum_n \frac1{|\vx-\vx_n|}
    \erfc \Bigl( \frac{|\vx-\vx_n|}{2 \xs} \Bigr),
    \label{eq:phi_fiducial}
\end{align}
where $\vk$ is the wavevector, $\vx_n$ denotes the position of the $n$-th particle,
and the overdensity field is determined by the spatial distribution of the particles
\begin{equation}
    \rhobarm \bigl( 1 + \delta(\vx) \bigr)
    = \frac{m}{a^3} \sum_n \deltaD(\vx-\vx_n).
    \label{eq:density}
\end{equation}
The long- and short-range forces in \eqref{eq:phi_fiducial} are split
with a Gaussian kernel of comoving width $\xs$.
One can verify the above force splitting,
using the fact that $e^{- k^2 \xs^2} / k^2$ and ${\rm erf}(x / 2\xs) / 4\pi x$
are a 3D Fourier transform pair.
The acceleration due to the tree force is
\begin{equation}
    - \nabla_\vx \phi^\Tree = - \frac{G m}a \sum_n
    \frac{\vx - \vx_n}{|\vx - \vx_n|^3}
    \biggl[ \erfc \Bigl( \frac{|\vx - \vx_n|}{2 \xs} \Bigr)
            + \frac{|\vx - \vx_n|}{\xs \sqrt\pi}
            \exp \Bigl( - \frac{|\vx - \vx_n|^2}{4 \xs^2} \Bigr) \biggr].
\end{equation}

Now let us modify the above conventional TreePM algorithm
for an anisotropically expanding universe.
The modified Poisson equation \eqref{eq:poisson_local},
\begin{equation}
    (1+\Delta_i)^{-2} \frac{\partial^2}{\partial x_i^2} \phi
    = 4\pi G \rhobarm a^2 (1 + \Delta_0) \delta,
\end{equation}
implies that the TreePM potentials in the presence of the DC modes should be
\begin{align}
    \phi^\PM(\vp) &= - 4\pi G \rhobarm a^2 (1 + \Delta_0)
        \frac{\delta(\vp)}{k^2} e^{- p^2 \xs^2},  \nonumber\\
    \phi^\Tree(\vx) &= - \frac{Gm}{a}\sum_n \frac{a}{|\vr-\vr_n|}
    - \phi^{\rm PM}(\vx),
    \label{eq:phi_Fourier}
\end{align}
where we have introduced $\vp$ as the local comoving wavevector\footnote{Though
this Eulerian $\vp$ is technically different from
the Lagrangian one in Sec.~\ref{sub:lpt}, they are both Fourier conjugates
to the local comoving coordinates, therefore denoted by the same symbol.},
related to the global comoving wavevector by
\begin{equation}
    k_i = \frac{p_i}{1 + \Delta_i}.
    \label{eq:kp}
\end{equation}
And recall $r_i = a_i x_i$ is the physical coordinates.
As expected, both the PM and tree forces above are manifestly isotropic
in the physical or global comoving coordinates.
In App.~\ref{app:force} we provide the functional form of $\phi^{\rm PM}(\vx)$,
which involves integral and no longer has the simple form as in Eq.~\eqref{eq:phi_fiducial}.

Note that there are two choices on the force-splitting scale: 
isotropic in physical scales, i.e.\ anisotropic in local comoving scales,
or isotropic in local comoving scales, i.e.\ anisotropic in physical scales.
In the former case the equations become quite simple.
However, we have found this choice introduces numerical artifacts,
e.g.\ on the second order responses to tides of the halo abundance.
This is probably because in this case the force-splitting boundary
in local comoving scales (i.e., the simulation coordinates)
is neither isotropic nor constant in time,
and can interact with the anisotropic PM force artifacts at grid scales.
To avoid this problem we choose to split force isotropically in local comoving scales,
following Ref.~\cite{Stucker_etal:2020}.
The tree acceleration now is
\begin{equation}
    - \frac{\partial\phi^\Tree}{\partial x_i} = - \frac{G m }a \sum_n
    \frac{a a_i [\vr - \vr_n]_i}{|\vr - \vr_n|^3}
    +\pdv{\phi^{\rm PM}}{x_i}.
\end{equation}
Since $\pdv*{\phi^{\rm PM}}{x_i}$ is computationally expensive to exactly evaluate in simulations, we expanded this in terms of $\Delta_i$ and included up to the second order terms in $\Delta_i$ as in Ref.~\cite{Stucker_etal:2020}. Details are given in App.~\ref{app:force}.

\subsection{Time integration}

From Eq.~\eqref{eq:eom}, 
the equation of motion for the peculiar part takes a simple form of
\begin{equation}
     \frac{ \dot P_{i} }{m} = - \frac{\partial \phi}{\partial x^i},
    \label{eq:mod_EoM}
\end{equation}
where $\vP = a^2 m \dot \vx$ is the canonical momentum of the $N$-body Hamiltonian
\begin{equation}
    H = \sum_n \frac{\vP_n^2}{2 m a^2}
    + \sum_{n \neq n'} \frac{m^2 \varphi(\vx_n - \vx_{n'})}{2a},
\end{equation}
in which $\varphi$ is the potential of a unit-mass particle
in a box of comoving size $L$ at $a=1$:
\begin{equation}
    \nabla_\vx^2 \varphi = 4\pi G \Bigl(\deltaD(\vx-\vx') - \frac1{L^3} \Bigr).
\end{equation}
Note that we have made the time dependence in the potential explicit
by introducing $\varphi$, which otherwise is simply related to $\phi$
by Eq.~\eqref{eq:poisson_local} and Eq.~\eqref{eq:density}.

$N$-body simulations use the computed gravitational forces
to update the particle velocities and then them to evolve the particle positions in time.
This time integration is performed using the kick and drift leapfrog
operators \cite{QuinnKatzEtAl97, Springel05}
\begin{align}
    \mathrm{Kick} &: \quad \frac{\vP}m \to \frac{\vP}m
        - \nabla_\vx \phi \int_t^{t+\Delta t} \frac{\d t}a,  \\
    \mathrm{Drift} &: \quad \vx \to \vx
        + \frac{\vP}{m} \int_t^{t+\Delta t} \frac{\d t}{a^2}.
\end{align}

On the anisotropic background,
the Hamiltonian leading to the modified EoM Eq.~\eqref{eq:mod_EoM} is
\begin{equation}
    H = \sum_n \sum_{i=1}^3 \frac{P_{n,i}^2}{2 m a_i^2}
    + \sum_{n \neq n'} \frac{m^2 \varphi(\vx_n - \vx_{n'})}{2a},
\end{equation}
where $P_{n,i}\equiv a_i^2 m \dot x_{n,i}$ are the conjugate momenta of $x_{n,i}$.
Thus we have
\begin{align}
    \mathrm{Kick} &: \quad \frac{\vP}m \to \frac{\vP}m
        - \nabla_{\vx} \phi \int_t^{t+\Delta t} \frac{d t}a,  \\
    \mathrm{Drift} &: \quad x_i \to x_i
        + \frac{P_i}{m} \int_t^{t+\Delta t} \frac{d t}{a_i^2}.
\end{align}
Note that the drift operator changes due to the modified canonical momentum,
whereas the kick operator is unchanged because
the anisotropic effect on force calculations is already accounted
in Sec.~\ref{sub:treepm}.
We implemented these modifications of the TreePM forces and time integration
based on \texttt{L-Gadget2}~\cite{Springel05}.

\section{Simulations}
\label{sec:sim}

\begin{table}[htb]
  \centering
  \begin{tabular}{|c|c|c|c||c|c|c|} \hline 
    type of simulations 
    & $\Delta_0^{(1)}$ & $\Delta_{\rm p}$ & $\Delta_{\rm e}$ 
    & $L~[{\rm Mpc}/h]$ & $N_{\rm p}$ & realizations 
    \\ \hline \hline
        & & & & 250 & \multirow{12}{*}{$1024^3$} & \multirow{12}{*}{6} \\ \cline{5-5}
      fiducial &  0  &  0  &  0  & 1000 &  & \\ \cline{5-5}
        & & & & 3000 &  &  \\ \cline{1-5}
        & & & & 250 & &  \\ \cline{5-5}
      $\Delta_0$-type &  $\pm 0.09$  &  0  &  0  & 1000 &  & \\ \cline{5-5}
        & & & & 3000 &  &  \\ \cline{1-5}
        & & & & 250 &  &  \\ \cline{5-5}
      $\Delta_{\rm p}$-type &  0  &  $\pm 0.15$  &  0  & 1000 &  & \\ \cline{5-5}
        & & & & 3000 &  &  \\ \cline{1-5}
        & & & & 250 &  &  \\ \cline{5-5}
      $\Delta_{\rm e}$-type &  0  &  0  &  $\pm 0.1$  & 1000 &  & \\ \cline{5-5}
        & & & & 3000 &  &  \\ \hline
  \end{tabular}
    \caption{Summary of our 126 $N$-body simulations, of different types of mean strains (related to the DC density and tidal modes) in the box.
    The $\Delta_0$-type simulations have isotropic strain (the DC density perturbation);
    the $\Delta_\mathrm{p}$-type and $\Delta_\mathrm{e}$-type perturb
    the two DC tidal modes separately as decomposed in Eq.~\eqref{0ep};
    and the fiducial type are conventional simulations without
    any DC modes on global background cosmology.
    $\Delta_0^{(1)}$, $\Delta_\mathrm{p}$, and $\Delta_\mathrm{e}$ specify
    the strain amplitudes of their respective types;
    $L$ are the simulation box sizes;
    and $N_{\rm p}$ is the total number of particle.
    }
    \label{tab:simulations}
\end{table}

We perform $N$-body simulations in the tidal backgrounds
as described in the previous section,
using the Planck 2015 cosmology~\cite{Planck2015}:
$\Om = 0.3089$, $\Omega_\Lambda = 0.6911$, $h = 0.6774$,
$n_s = 0.9667$, and $\sigma_8 = 0.8159$.
We generate initial conditions at redshift $z_\mathrm{i}=49$ with $1024^3$ particles
using \texttt{CLASS} \cite{class} and our modified \texttt{2LPTIC}
\footnote{\url{https://cosmo.nyu.edu/roman/2LPT/}},
and then run the modified \texttt{L-Gadget2} \cite{Springel05}
with $N_{\rm mesh}=2048^3$ TreePM grid.
To cover a large range of halo masses, we use 3 different box sizes:
$L=250~\mathrm{Mpc}/h$, $L=1~\mathrm{Gpc}/h$, and $L=3~\mathrm{Gpc}/h$.
For each type of simulations, we set the force softening scales to $4\%$ of the mean particle distances,
which means $9.77~{\rm kpc}/h$, $39.1~{\rm kpc}/h$, and $117~{\rm kpc}/h$ for $L=250~\mathrm{Mpc}/h$, $L=1~\mathrm{Gpc}/h$, and $L=3~\mathrm{Gpc}/h$ simulations, respectively. About the accuracy of the time integration, \texttt{ErrTolIntAccuracy} is set to 0.05.
These parameters on the force accuracy potentially have an impact on the halo shape as recently presented in Ref.~\cite{Mansfield&Avestruz:2020}.

So far we have used $\Delta_1$, $\Delta_2$, and $\Delta_3$ to parametrize
the remaining three eigenvalues of the background strain.
Alternatively one can parametrize it by one isotropic dilation and two anisotropic scalings \cite{BondMyers96I},
with the former equal to the negative DC overdensity at linear order,
\begin{align}
    -\Delta^{(1)}_0 &= \Delta^{(1)}_1 + \Delta^{(1)}_2 + \Delta^{(1)}_3
    =-(\tau_1+\tau_2+\tau_3), \nonumber\\
    \Delta_\mathrm{e} &\equiv \frac{\Delta^{(1)}_1 - \Delta^{(1)}_2}2
    =-\frac{\tau_1-\tau_2}{2}, \nonumber\\
    \Delta_\mathrm{p} &\equiv \Delta^{(1)}_3 - \frac{\Delta^{(1)}_1 + \Delta^{(1)}_2}2
    =-\tau_3+\frac{\tau_1+\tau_2}{2}.
    \label{0ep}
\end{align}
The subscripts of the two anisotropic scaling modes stand for ellipticity and
prolaticity\footnote{We do not impose ordering on $\Delta_i$'s to compress the
parameter space (cf.\ \cite{BondMyers96I}).}.

To focus on tidal effects, we can remove the linear order overdensity mode
by setting $\Delta_0^{(1)}=0$, and run ``pure'' tidal simulations by
perturbing the remaining two degrees of freedom,
$\Delta_{\rm p}$ and $\Delta_{\rm e}$.
We also vary the two modes separately, further dividing those pure tidal simulations
into $\Delta_\mathrm{p}$- and $\Delta_\mathrm{e}$-type simulations.
For $\Delta_\mathrm{p}$-type simulations
we chose the background strain to satisfy
$\Delta^{(1)}_0=0$ and $\Delta_{\rm e}=0$ so that the remaining degree of freedom is only $\Delta_{\rm p} = 3\Delta^{(1)}_3/2$.
For $\Delta_\mathrm{e}$-type simulations
we chose the background strain to satisfy
$\Delta^{(1)}_0=0$ and $\Delta_{\rm p}=0$ so that the remaining degree of freedom is only $\Delta_{\rm e} = \Delta^{(1)}_1$.
This means the configuration of the tides and linear strain takes the form
$(\tau_1, \tau_2, \tau_3) = ( 1, 1, -2 )\Delta_{\rm p}/3$
for $\Delta_\mathrm{p}$-type simulations,
and $(\tau_1, \tau_2, \tau_3) = ( -1, 1, 0)\Delta_{\rm e}$
for $\Delta_\mathrm{e}$-type simulations.

Our modification for the tidal background also works for the isotropic configuration,
where $\Delta_{\rm p}=\Delta_{\rm e}=0$ and only $\Delta_0^{(1)}$ is varied.
In fact this configuration is equivalent to the usual isotropic separate universe
simulation at the nonlinear level.
We validate our simulation pipeline by running the $\Delta_0$-type simulations 
and comparing them to the usual separate universe simulations.
We find their power spectrum responses and halo biases are fully consistent
with each other, as summarized in App.~\ref{app:SU_comp}.

As reference and normalization for the response estimations,
we also run some conventional fiducial simulations without
any variation in the DC density or tidal modes.
For each of type of simulations, each box size, and each sign
(positive, negative, or fiducial) of the DC modes,
we have performed six simulations, and in total used 126 simulations
to produce our results.
Simulations with different DC modes shares the same random phases in the 
initial conditions so that the sample variances in the response functions are suppressed.
We summarize all our simulations in Table~\ref{tab:simulations}.

\section{Results}
\label{sec:results}

In this section, we present analytic expressions and numerical calibrations
of the power spectrum tidal response using 2LPT and $N$-body simulations for high and low redshift, respectively;
we also present measurements of the halo shape response
to the tidal field.
For the latter, we study its dependence on the halo mass and other (secondary)
halo properties, including the concentration and axis ratio.
This is analogous to the halo assembly bias, the dependence of halo abundance on variables beyond the halo mass.

\subsection{Power spectrum responses}
\label{sub:response}

The long-wavelength modes modulate the evolution of the short-wavelength ones,
in their amplitudes and scales,
known as the growth and dilation effects~\cite{Li_etal:2014a}, respectively.
The growth effect focuses on the changes in the short mode amplitude
in the local comoving coordinate system, as considered in Sec.~\ref{sub:lpt}.
On the other hand, the dilation effect arises from the (anisotropic) expansion
of the local comoving coordinate system with respect to the global one,
as described in Sec.~\ref{sub:background}.
In addition to the growth and dilation effects,
the change in the reference mean density by $(1 + \Delta_0)$
in power spectrum estimation also contributes to its responses, as shown below.

Let us derive analytically this separation of the growth and dilation effects
in the power spectrum response functions.
As in Sec.~\ref{sub:treepm}, we use $\vp$ to denote a wavevector in
the local comoving coordinates, which is related to the global comoving wavevector $\vk$
by Eq.~\eqref{eq:kp}.
At linear order, the density and tidal perturbations induce
anisotropic responses in the monopole and quadrupole.
Therefore we define the total response functions $R^\Eul$,
including both growth and dilation effects,
and the growth only response functions $R^\Lag$, by
\begin{align}
    - \frac{\partial\ln P}{\partial\Delta^{(1)}_i} \bigg|_{\vk, \vDelta=0}
    &\equiv R^\Eul_0(k) + \L_2(\hat k_i) R^\Eul_2(k), \nonumber\\
    - \frac{\partial\ln P_W}{\partial\Delta^{(1)}_i} \bigg|_{\vp, \vDelta=0}
    &\equiv R^\Lag_0(p) + \L_2(\hat p_i) R^\Lag_2(p),
    \label{eq:responses}
\end{align}
where $P_W$ is the power spectrum measured in local comoving space.
The derivatives with respect to $\Delta_i$ is taken
holding the other two $\Delta_j$'s ($j \neq i$) fixed,
at $\vDelta=0$ with $\vDelta$ being $\Delta_0$ and $\Delta_i$'s.
The function $\L_2$ is the second order Legendre polynomial.

Our notation for the power spectrum responses bear a resemblance
to that for the halo biases.
The linear halo bias can also be measured as responses of the halo abundance
with respect to the long-wavelength density mode.
One measures the Eulerian bias $b^\Eul$ or Lagrangian bias $b^\Lag$,
depending on whether the measurement is carried out in either
global or local comoving space (see e.g.\ \cite{Li_etal:2015}), respectively.
Therefore this ``Lagrangian'' superscript in $R^\Lag$ is not to be confused with
\emph{Lagrangian} perturbation theory in Sec.~\ref{sub:lpt}.

Because the variances should be conserved when transforming
between Fourier-space volume elements,
\begin{equation}
    P (\vk;\vDelta)\,\d^3\vk = (1 + \Delta_0)^2 P_W(\vp;\vDelta) \,\d^3\vp,
\end{equation}
where the $(1+\Delta_0)^2$ factor on the right hand side is due to a change
of the reference density in $P_W$.
Therefore, for the dimensionless power spectra we have
\begin{equation}
    k^3 P (\vk;\vDelta)= (1 + \Delta_0)^2 p^3 P_W(\vp;\vDelta).
    \label{eq:P_PW_relation}
\end{equation}
Using Eq.~\eqref{eq:kp} and the chain rule we can relate
the total and growth responses in Eq.~\eqref{eq:responses} by
\begin{align}
    - \frac{\partial\ln k^3 P}{\partial\Delta^{(1)}_i} \bigg|_\vk
    = 2 - \frac{\partial\ln p^3 P_W}{\partial\Delta^{(1)}_i} \bigg|_\vk
    &= 2 - \frac{\partial\ln P_W}{\partial\Delta^{(1)}_i} \bigg|_\vp
    - \frac{\partial\ln p^3 P_W}{\partial p_j}\bigg|_{\vDelta}
        \frac{\partial p_j}{\partial \Delta^{(1)}_i} \bigg|_{\vk}
    \nonumber\\
    &= 2 - \frac{\partial\ln P_W}{\partial\Delta^{(1)}_i} \bigg|_\vp
    - \hat k_i^2 \frac{\d\ln k^3 P}{\d\ln k}, \nonumber\\
    &= 2 - \frac{\partial\ln P_W}{\partial\Delta^{(1)}_i} \bigg|_\vp
    - \frac13 \frac{\d\ln k^3 P}{\d\ln k}
    - \L_2(\hat k_i) \frac23 \frac{\d\ln k^3 P}{\d\ln k},
\end{align}
where $\hat k_i = k_i / k$ and $\vDelta = 0$ is assumed.

Hence, the total (Eulerian) and growth (Lagrangian) responses
are related with each other through
\begin{align}
    R^\Eul_0 &= 2 + R^\Lag_0 - \frac13 \frac{\d\ln k^3 P}{\d\ln k},
    \nonumber\\
    R^\Eul_2 &= 0 + R^\Lag_2 - \frac23 \frac{\d\ln k^3 P}{\d\ln k}.
    \label{eq:RE_RL}
\end{align}
As explained in the beginning of this subsection, we have decomposed
the total responses $R^\Eul$ into three contributions: the constant two
due to the change of reference density,
the growth responses $R^\Lag$, and the dilation term proportional to the
slope of the dimensionless power spectrum.
Note the quadrupole response is not affected by the reference density,
which is emphasized above with 0.

In the linear regime $P_W \propto D_W^2$
and we have calculated the modified growth factor $D_W$ in Sec.~\ref{sub:lpt}.
Rewriting Eq.~\eqref{eq:DW} in Legendre polynomial,
\begin{equation}
    D_W = D \Bigl( 1 + \frac{13}{21} \Delta_0
    - \frac8{21} \L_2(\hat p_i) \Delta^{(1)}_i \Bigr),
\end{equation}
so that the tree-level response functions are
\begin{align}
    R^\Lag_0 = \frac{26}{21}, \qquad
    & R^\Eul_0 
    = \frac{68}{21} - \frac13 \frac{\d\ln k^3 P}{\d\ln k};
    \nonumber\\
    R^\Lag_2 = \frac{16}{21}, \qquad
    & R^\Eul_2 
    = \frac{58}{21} - \frac23 \frac{\d\ln k^3 P}{\d\ln k}.
    \label{eq:RL_treept}
\end{align}
These results are consistent with the standard perturbation theory calculations,
e.g.~\cite{Akitsu_Takada:2017,LiSchmittfullSeljak17}.

In the nonlinear regime, we need to numerically calibrate the response functions
with simulations.
We focus on the growth response because the other contributions in Eq.~\eqref{eq:RE_RL}
are well understood, with the dilation term readily computable from
the nonlinear power spectrum.
A pure long tidal mode has only two degrees of freedom, as parametrized
in Eq.~\eqref{0ep} by $\Delta_\mathrm{p}$ and $\Delta_\mathrm{e}$.
One can show that they modulate the power spectrum respectively by
\begin{align}
    \delta P_W(\vp; \Delta_\mathrm{p})
    &= - P(p) R^\Lag_2(p) \Delta_\mathrm{p} \L_2(\hat p_3),  \nonumber\\
    \delta P_W(\vp; \Delta_\mathrm{e})
    &= - P(p) R^\Lag_2(p) \Delta_\mathrm{e} \bigl[ \L_2(\hat p_1) - \L_2(\hat p_2) \bigr].
    \label{eq:Rp_Re}
\end{align}
In practice, we measure $R_2^\Lag$ from those $\Delta_\mathrm{p}$-type
and $\Delta_\mathrm{e}$-type of simulations.

It is more straightforward to extract $R_2^\Lag$ from the
$\Delta_\mathrm{p}$-type simulations.
Eq.~\eqref{eq:Rp_Re} implies that the relevant changes in $P_W$ lie in its
quadrupole along the $z$-axis (denoted by $\ell_\mathrm{p}$ below),
\begin{equation}
P^{\ell_\mathrm{p}=2}_W(p;\Delta_{\rm p}) \equiv
5\int \frac{\d^2\hat \vp}{4\pi} P_W(\vp;\Delta_{\rm p}) \L_2(\hat p_3)
= - P(p)R_2^{\rm L}(p)\Delta_{\rm p},
\end{equation}
due to the orthonormality of the Legendre polynomials and the fact that
$P(p)$ is istropic.
Therefore, the estimator for $R_2^{\rm L}(p)$ can be constructed as
\begin{equation}
R_2^{\rm L}(p,z) = - \frac{P_W^{\ell_\mathrm{p}=2}(p,z;\Delta_\mathrm{p}=+\epsilon) - P_W^{\ell_\mathrm{p}=2}(p,z;\Delta_\mathrm{p}=-\epsilon)}{2\epsilon D(z) P(p,z)}.
\label{eq:RL2_from_p}
\end{equation}

Similarly, we can also estimate the growth response from
the power spectrum quadrupoles along both $x$ and $y$ axes, from the $\Delta_\mathrm{e}$-type simulations,
\begin{equation}
P^{\ell_\mathrm{e}=2}_W(p;\Delta_{\rm e}) \equiv
5\int \frac{\d^2\hat \vp}{4\pi} P_W(\vp;\Delta_{\rm e})
\bigl[ \L_2(\hat p_1) - \L_2(\hat p_2) \bigr]
= - P(p)R_2^{\rm L}(p)\Delta_{\rm e},
\end{equation}
where we have used the orthonormality of the Legendre polynomials and
\begin{equation}
\int \frac{\d^2\hat \vp}{4\pi} \L_2(\hat p_1)\L_2(\hat p_2)
=-\frac{1}{10}.
\end{equation}
Thus, we can estimate $R_2^{\rm L}(p)$ by
\begin{equation}
R_2^{\rm L}(p,z) = - \frac{P_W^{\ell_\mathrm{e}=2}(p,z;\Delta_\mathrm{e}=+\epsilon) - P_W^{\ell_\mathrm{e}=2}(p,z;\Delta_\mathrm{e}=-\epsilon)}{2\epsilon D(z) P(p)}.
\label{eq:RL2_from_e}
\end{equation}

\subsubsection{Responses at high redshifts from 2LPT}
\label{subsub:response_2LPT}

At high redshifts 2LPT works accurately with more modes in the linear regime.
Using results derived in Sec.~\ref{sub:lpt}, we have modified the initial condition
generator to incorporate the leading order impact of the long modes
to 2LPT.
This allows us to measure the power spectrum responses reliably
at high redshifts.
For this purpose, we generated eight pairs of 2LPT realizations
with $\Delta_{\rm p}$-type which contain $1024^3$ particles
in $100~{\rm Mpc}/h$ boxes at $z=49, 15, 10, \textrm{and } 7$.

We first compare the isotropic response $R^\Lag_0$ measured from our 2LPT to
that from the usual separate universe simulations, in order to find out the 
scale above which our 2LPT responses converge.
Fig.~\ref{fig:RL0_SU_2LPT} in App.~\ref{app:SU_comp} shows the results of this convergence test.
The isotropic 2LPT responses are accurate to within $5\%$ of the separate
universe results, for $k \leq 9~h/{\rm Mpc}$ at $z=15$,
$k\leq 4~h/{\rm Mpc}$ at $z=10$,
and $k\leq 2~h/{\rm Mpc}$ at $z=7$.

\begin{figure}[tb]
\centering
\includegraphics[width=0.67\textwidth]{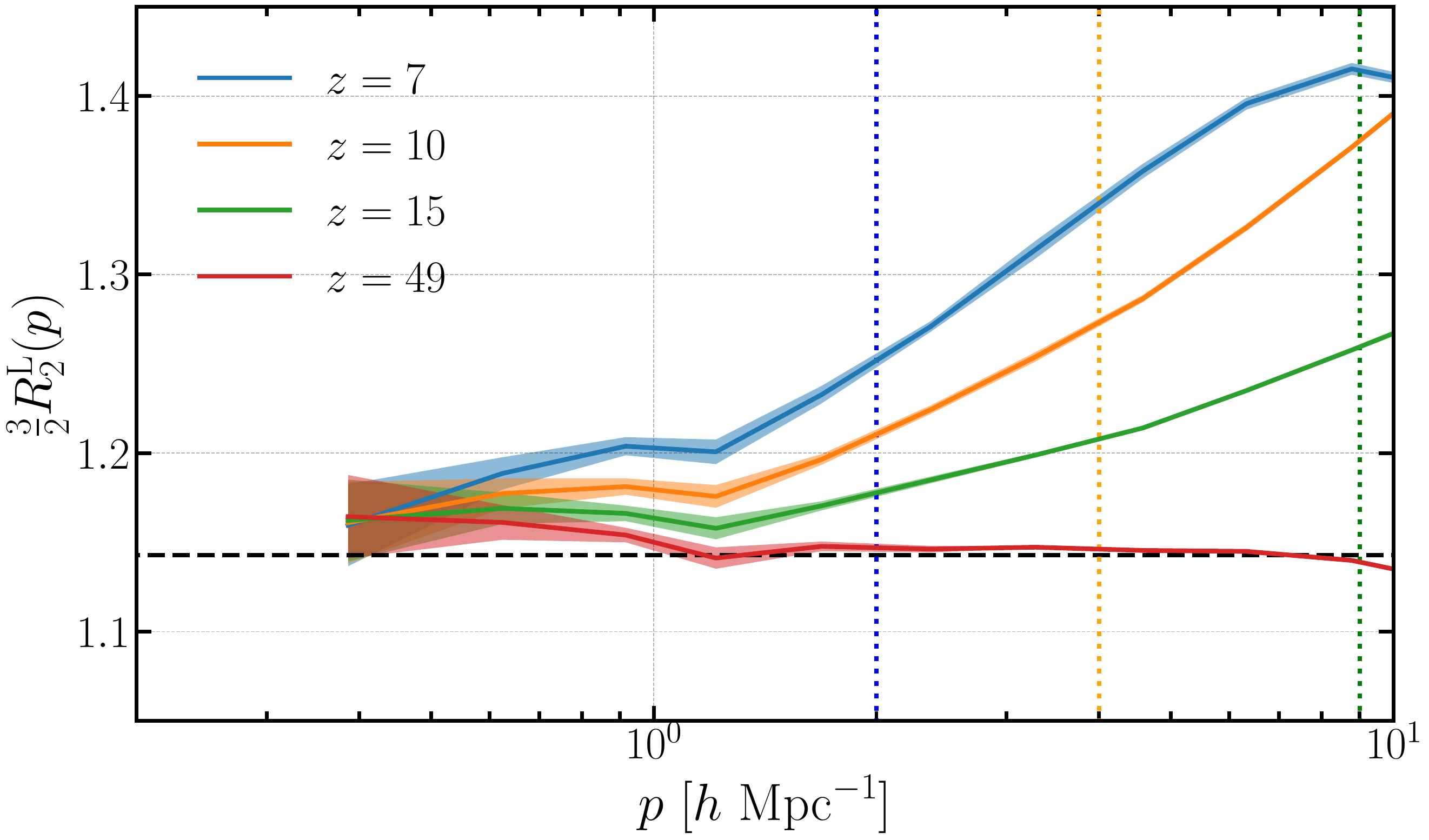}
\caption{
Power spectrum growth response function to the large-scale tidal field,
$R_2^\Lag$,
measured at high redshifts from 2LPT realizations.
The shaded region shows the $1\sigma$ error.
At $z=49$, the response agrees well with the linear perturbation theory
prediction on all scales, which is shown by the horizontal dashed line, while at later times the responses are
enhanced on small scales.
Vertical dotted lines mark for each redshift the
convergence scale, above which isotropic 2LPT responses $R_0^\Lag$
converge to within $5\%$ of the validation separate universe simulations.
Though of similar trend, our results show quantitatively less enhancement
than that in Ref.~\cite{Masaki:2020b},
even after taking into account the possible error of 2LPT.
}
\label{fig:RL2_2LPT}
\end{figure}

We measure the 2LPT response $R_2^\Lag$ at each redshift,
and show the results in Fig.~\ref{fig:RL2_2LPT}.
The valid scale obtained from the comparison of $R_0^{\rm L}$ are shown as vertical dotted lines.
Note that we have scaled the vertical axis for easy comparison
with previous studies, $3 R^{\rm L}_2(p)/2 = G(k)$
used in Refs.~\cite{Stucker_etal:2020,Masaki:2020a,Masaki:2020b}.
It is clear that at these high redshifts
$R_{2}^{\rm L}(p)$ grows more than the prediction of the tree-level
perturbation theory on small scales for $k\geq2~h/{\rm Mpc}$.
While the overall trend is in agreement with
the results of Ref.~\cite{Masaki:2020b},
our responses have quantitatively less enhancement than 
that in Ref.~\cite{Masaki:2020b},
even after taking into account the possible error of 2LPT.
This may be attributed to the difference on details of the implementations,
e.g.\ generating initial conditions at different orders,
and needs the further examination.

\subsubsection{Responses at low redshifts from $N$-body}
\label{subsub:response_nbody}

Although Refs.~\cite{Stucker_etal:2020,Masaki:2020a} already investigated the linear tidal response of a matter power spectrum from their simulations,
we measure the growth response $R_{2}^{\rm L}(p)$ at low redshift from our $N$-body simulations
to check the validity of our numerical implementation.
In this subsection we show results from $1~{\rm Gpc}/h$ simulations to cover both linear and nonlinear regimes.

\begin{figure}[tb]
\centering
\includegraphics[width=0.49\textwidth]{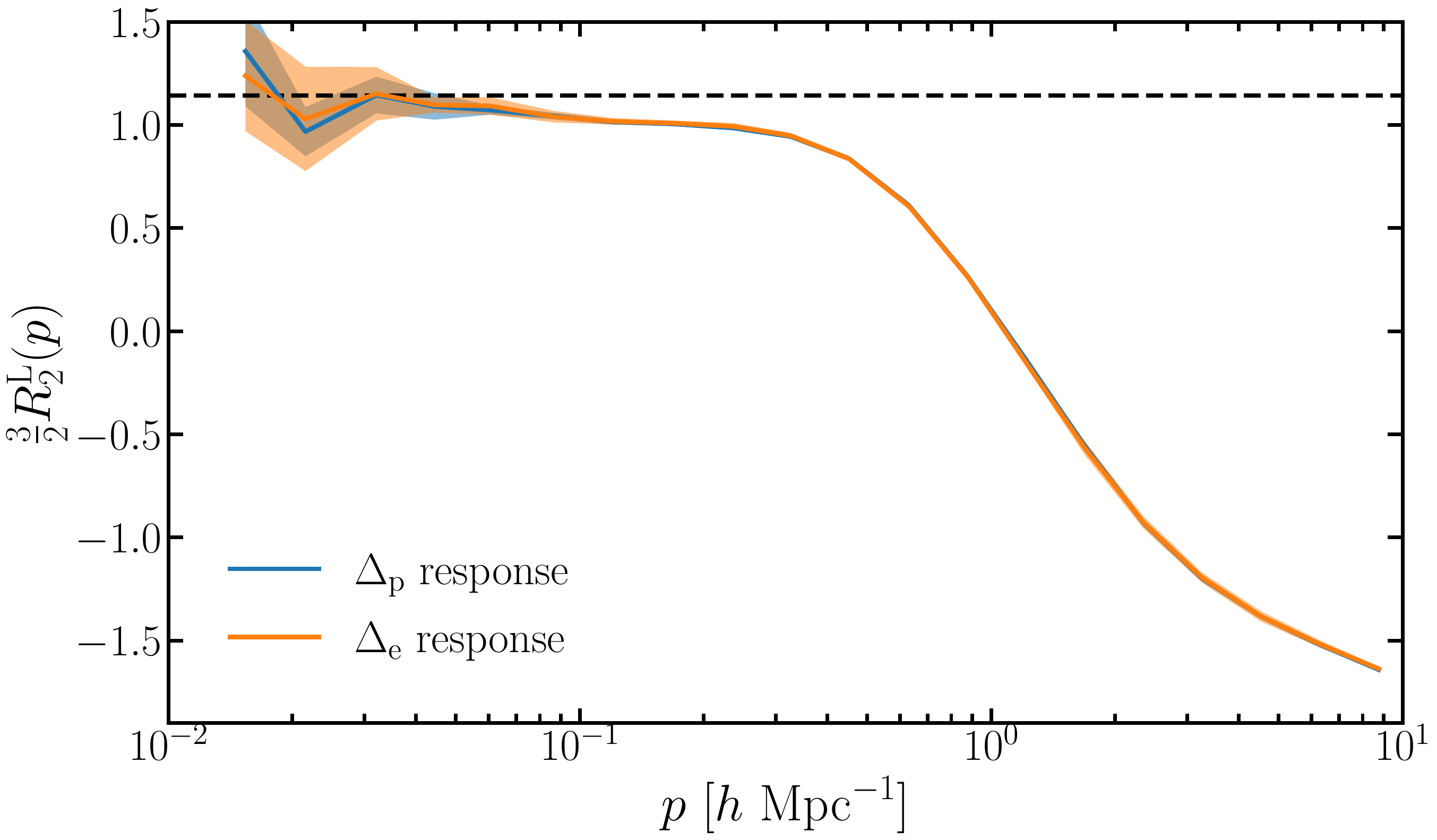}
\hfill
\includegraphics[width=0.49\textwidth]{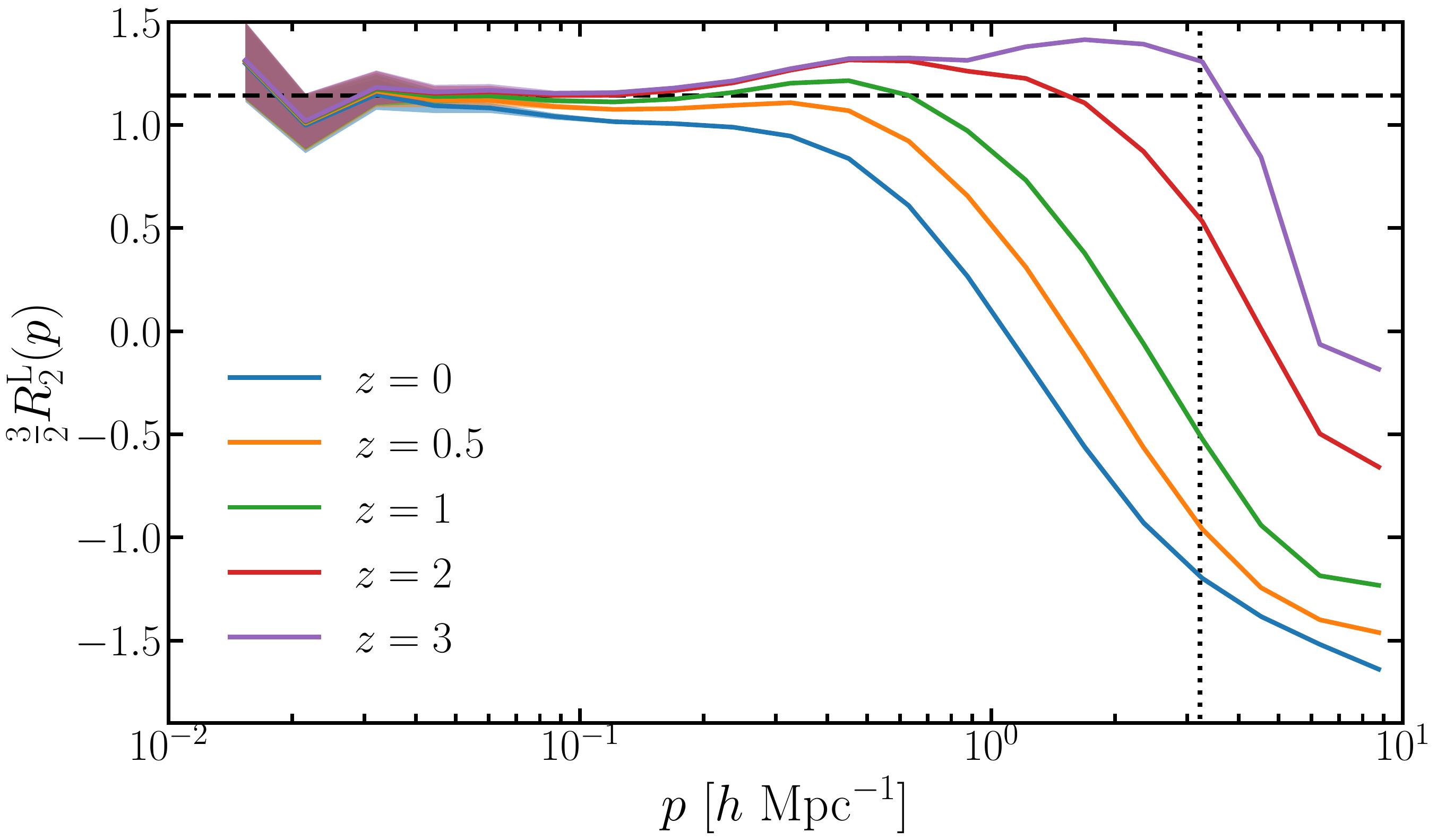}
\caption{Power spectrum responses as in the previous figure,
measured at low redshifts from $N$-body simulations with $L=1~{\rm Gpc}/h$.
The left plot shows measurements
from $\Delta_{\rm p}$- and $\Delta_{\rm e}$-type simulations at $z=0$,
and the right one combines $\Delta_\mathrm{p}$ and $\Delta_\mathrm{e}$ types
at multiple low redshifts.
The shaded region shows the $1\sigma$ error.
The horizontal dashed line is the tree-level perturbation theory prediction.
The vertical dotted line on the right panel depicts the particle Nyquist wavenumber for $L=1~{\rm Gpc}/h$ simulations.
}
\label{fig:RL2}
\end{figure}

First, since we ran two different types of the tidal simulations ($\Delta_{\rm p}$- and $\Delta_{\rm e}$-type), we test whether the simulations with these different types give converging results.
In the left panel of Fig.~\ref{fig:RL2} we present $R_{2}^{\rm L}(p)$ from both $\Delta_{\rm p}$-type simulations (Eq.~\eqref{eq:RL2_from_p}) and $\Delta_{\rm e}$-type simulations (Eq.~\eqref{eq:RL2_from_e}). 
Both results agree well up to $k=10~h/{\rm Mpc}$.
Therefore we combine $\Delta_{\rm e}$- and $\Delta_{\rm p}$-type simulations to estimate $R_2^{\rm L}(p)$ in the results below.

The right panel of Fig.~\ref{fig:RL2} shows $R_2^{\rm L}(p)$ for several redshifts: $z=3,~2,~1,~0.5,$ and $0$.
At these low redshifts, $R_2^{\rm L}(p)$ decreases over all the scales as the redshift decreases.
This is likely because strong nonlinearity tends to erase the memory of large-scale tidal field gradually.
Although these features are in general consistent with previous studies,
quantitatively there is a small difference.
For instance, at $z=3$ tidal response from our simulations takes a maximum $3R_2^{\rm L}/2\sim 1.5$ while 
in Ref.~\cite{Masaki:2020b} the maximum at $z=3$ is less than $1.5$. 
In addition, Ref.~\cite{Stucker_etal:2020} reports the different behaviour of $R_2^{\rm L}$ at $z=2$
from ours.
These disagreements may arise from the halo sample variance in the nonlinear regime, which is also seen in the separate universe simulations (e.g. Ref.~\cite{Li_etal:2014a}) or the difference in details of the numerical implementations and require further studies.

\subsection{Halo shape response}

The tidal field is known to cause intrinsic alignment
of galaxy shapes.
Likewise, halo shapes respond to the large-scale tidal modes,
with the sensitivity captured by the shape bias.
This is analogous to the halo bias which is a response of the halo
abundance to the large-scale density mode.
In this section, we present measurements of the shape bias
using our tidal simulations.
We show its universal behavior as a function of the linear bias,
and its dependence on properties beyond mass, which we call the shape assembly bias.

We identify dark matter halos from our simulations with \texttt{AHF}~\cite{AHF}.
It uses adaptively refined meshes to find local density peaks as centers
of prospective halos.
It then defines the halos as spherical overdensity (SO) regions
$\Delta_\mathrm{h}$ times denser than the mean matter density $\rhobarm$.
We choose $\Delta_\mathrm{h} = 200$, and disable the
gravitational unbinding procedure to find the host halos with more than 400 particles.
While we need to identify SO halos in the global comoving coordinates,
\texttt{AHF} by default uses the simulation coordinates
that are local comoving, so that halos
are not identified as the spherical overdensity in the global ones. 
Therefore, we modify the \texttt{AHF} code to use the Euclidean metric for 
the halo identification in the global coordinates.

Given an SO halo, we define its quadrupole shape in two ways,
the inertia tensor and the reduced inertia tensor.
The former is defined as
\begin{equation}
  I_{ij} \equiv \sum_{i=1}^{N_\mathrm{p}} m_\mathrm{p} x_i x_j,
  \label{inertial_tensor}
\end{equation}
summing over all $N_\mathrm{p}$ particles of the halo.
$m_\mathrm{p}$ is the particle mass and $x_i$ is $i$-th components of
the particle location with respect to the halo center.
In the literature $I_{ij}$ is sometimes normalized by the halo mass,
which however does not affect our response measurement presented below.

The reduced inertia tensor is defined similarly but with additional
radius weighting
\begin{equation}
  J_{ij} \equiv \sum_{i=1}^{N_\mathrm{p}} m_\mathrm{p}
  \frac{x_i x_j}{x^2},
  \label{reduced_inertial_tensor}
\end{equation}
where $x$ is the distance of a particle to the halo center.
This definition uses a dimensionless ratio and therefore weight each mass
equally only by angular position regardless of radial distance $x$.
Compared to Eq.~\eqref{inertial_tensor}, $J_{ij}$ upweights the inner masses 
and thus should be more strongly correlated with properties of galaxies which reside in the halo.\footnote{See, e.g., Refs. \cite{Okumura2009,Faltenbacher2009,Okumura2009a} for the detection of large misalignments ($\sim 30 ~{\rm deg}$) between the major axes of central galaxies and their host halos when $I_{ij}$ is used to define halo shapes.} 
Therefore, in the following main text, we use $J_{ij}$ to estimate halo shapes. The results from $I_{ij}$ are summarized in App.~\ref{app:nonreduced}. 
Ref.~\cite{Zemp_etal:2011} gives detailed discussion on the dependence of the halo shape on its definition.
Ref.~\cite{Shi2020} also presented comparisons of $I_{ij}$ and $J_{ij}$ in the context of intrinsic alignments of galaxies (see their Appendix B).

According to the linear alignment model, at leading order halo shapes responds to the external tidal field as 
\begin{equation}
    J_{ij} = J_0 \left[\frac13 \delta_{ij}^{\rm K} +  b_K K_{ij}\right],
    \label{3D_IA}
\end{equation}
where $J_0$ is the trace component of the shape tensor: $J_0={\rm Tr}[J_{ij}]$, $K_{ij}$ is the DC tidal field: $K_{ij} = (\partial_i\partial_j/\partial^2-\delta^{\rm K}_{ij}/3)\delta$ and $b_K$ is the dimensionless linear shape bias parameter, which is related to the conventionally used linear alignment coefficient $C_1$ through $b_K = - a^3 C_1 \bar{\rho}(a)/D(a)$. 
The shape bias $b_K$ represents the strength of the response or alignment and thus an analogous parameter to the linear bias $b_1$, which describes the response of the number density of halos to the spherically symmetric long-wavelength perturbation: $\delta_{\rm h} = b_1 \delta$.

Following the decomposition of the traceless components of the background strain into $\Delta_{\rm p}$ and $\Delta_{\rm e}$, it is convenient to define the following two quantities
\begin{align}
    J_{\rm p} &\equiv J_{33} - \frac{ J_{11} + J_{22}}{2}, \\
    J_{\rm e} &\equiv \frac{J_{11} - J_{22}}{2}.
\end{align}
Then, Eq.~\eqref{3D_IA} implies
\begin{align}
    b_K(M,z) &= -\frac{J_{\rm p}(M,z;\Delta_{\rm p} = +\epsilon)-J_{\rm p}(M,z;\Delta_{\rm p} = -\epsilon)}{ 2\epsilon D(z) J_0(M,z)}, \\
    b_K(M,z) &= -\frac{J_{\rm e}(M,z;\Delta_{\rm e} = +\epsilon)-J_{\rm e}(M,z;\Delta_{\rm e} = -\epsilon)}{ 2\epsilon D(z) J_0(M,z)},
\end{align}
for $\Delta_{\rm p}$-type and $\Delta_{\rm e}$-type simulations, respectively.
Having these relations, we can estimate $b_K$ from our tidal simulations by measuring 
the averaged trace of the shape tensor, $J_0$, and the averaged traceless components $J_{\rm p}$ and $J_{\rm e}$ from fiducial, $\Delta_{\rm p}$-type, and $\Delta_{\rm e}$-type simulations respectively.

Note that the linear shape bias $b_K$ measured in this way should be regarded as the Lagrangian shape bias since we do not take into account the volume distortion due to the background strain. For linear shape bias, however, there is no difference between the Lagrangian shape bias $b_K^{\rm L}$ and the Eulerian one $b_K^{\rm E}$ unlike the number density bias where the Lagrangian linear bias $b^{\rm L}_1$ is related to the Eulerian one $b^{\rm E}_1$ through $b_1^{\rm E} = b_1^{\rm L}+1$.
This is because the pure tidal field does not induce the volume distortion at linear order of the tides and can be explicitly shown by considering the conservation laws: $J^{\rm E}_{ij}(\vx)d^3\vx = J^{\rm L}_{ij}(\vq)d^3\vq$ and $(1+\delta^{\rm E}(\vx))d^3\vx=d^3\vq$\footnote{
For the second order shape bias, the Lagrangian shape bias is no longer identical to the Eulerian one. See the discussion in \cite{Schmitz_etal:2018}}.
Thus, in this paper we do not distinguish $b_K^{\rm L}$ from $b_K^{\rm E}$ and the linear shape bias is just written as $b_K$.

\subsubsection{Convergence on the resolution and external tides}

\begin{figure}[tb]
\centering
\includegraphics[width=0.49\textwidth]{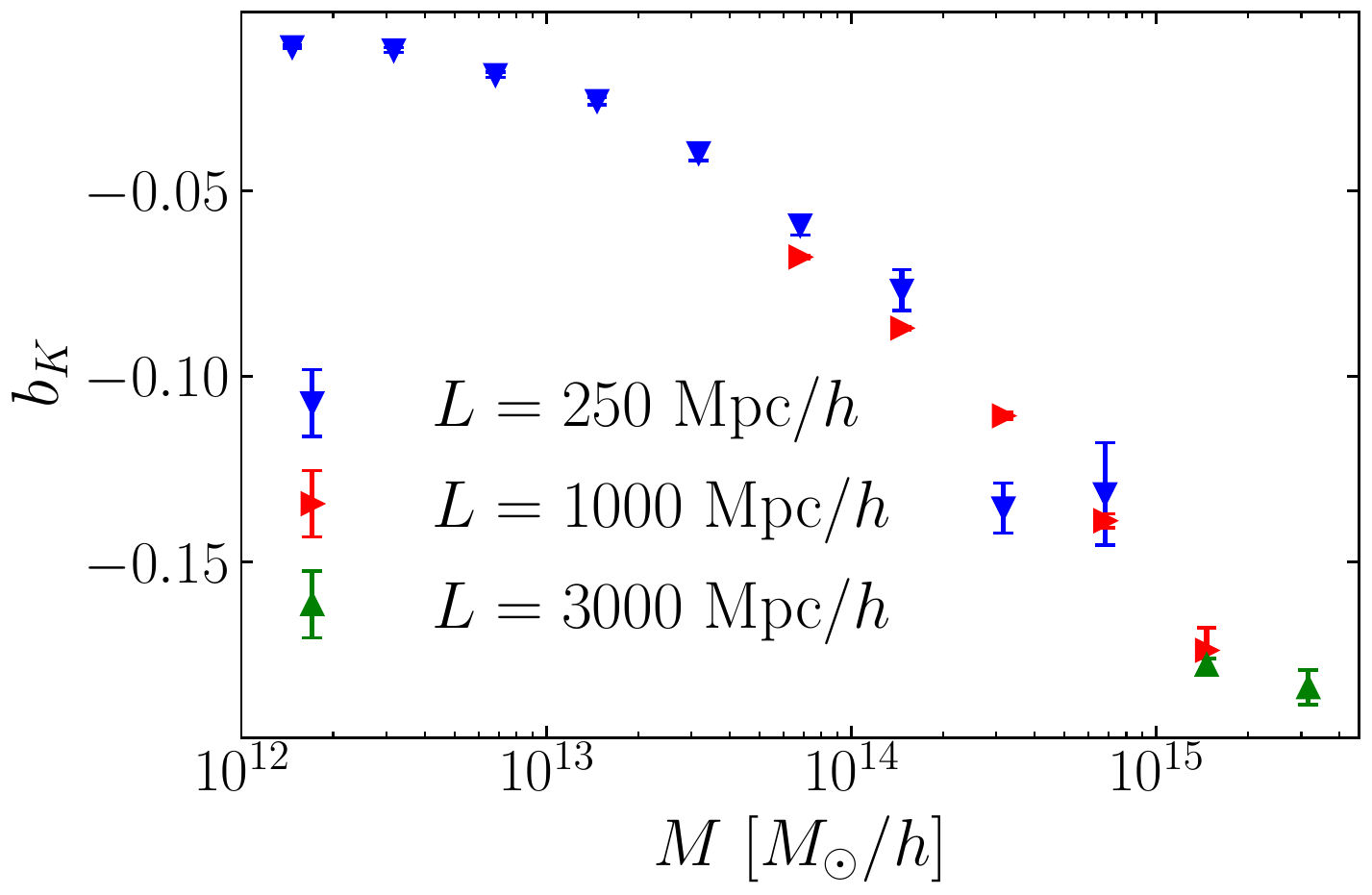}
\hfill
\includegraphics[width=0.49\textwidth]{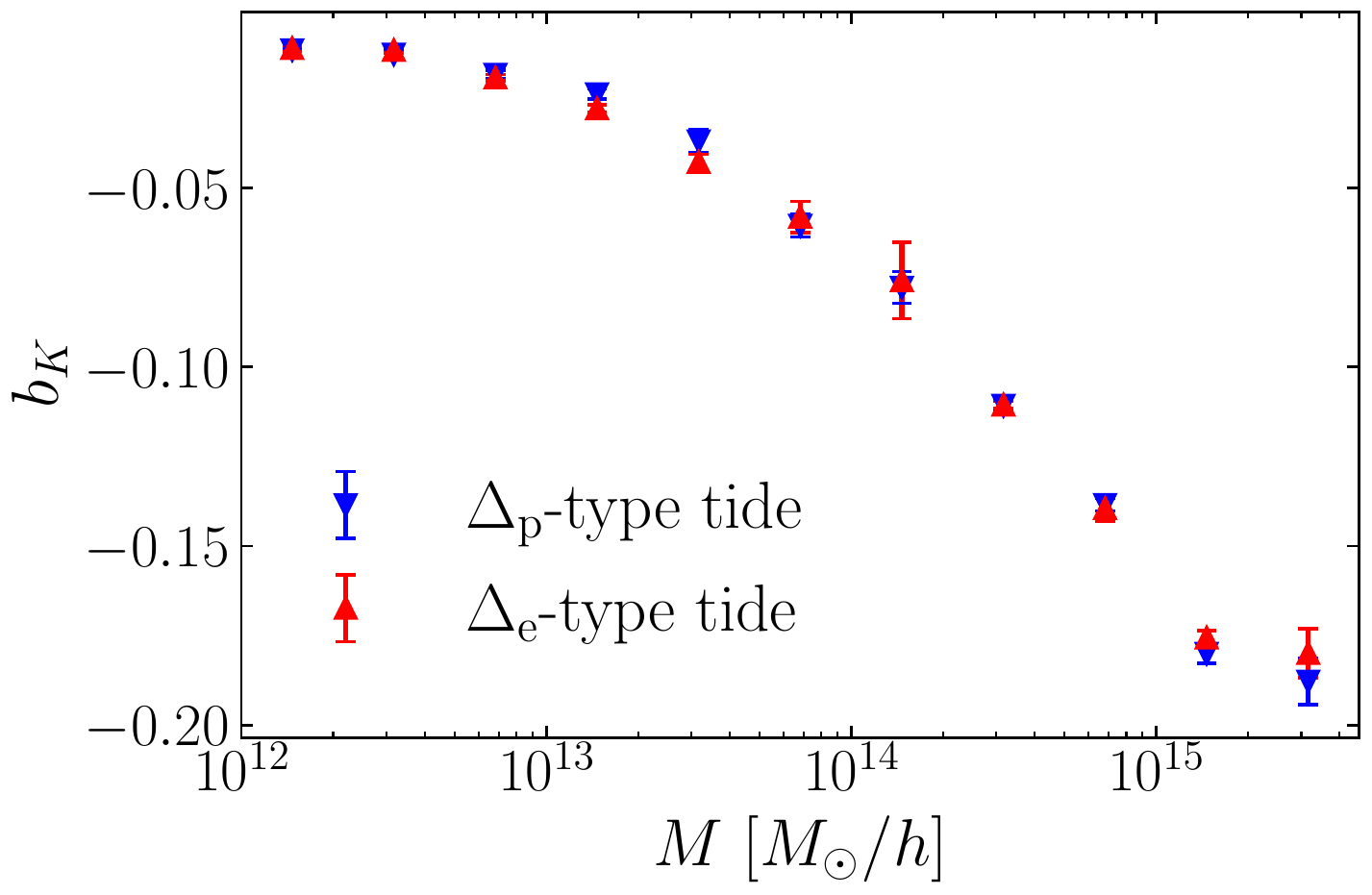}
\caption{Linear alignment coefficient, $b_K$, for the reduced inertial tensor, $J_{ij}$, at $z=0$.
The left plot shows measurements from simulations of different box sizes,
and the right one shows measurements from different tidal types.
}
\label{fig:bK_reduced}
\end{figure}

Before showing the redshift- and environment-dependence of $b_K$, 
here we discuss the convergence of measured $b_K$ for different resolutions and different kinds of tides.

The left panel of Figure~\ref{fig:bK_reduced} shows $b_K$ from different boxsize simulations, meaning the different resolutions since we fix the number of particles.
The results are in agreement with each other over all the mass range except for the 6th-8th mass bins where the $250~{\rm Mpc}/h$ and $1~{\rm Gpc}/h$ simulations give slightly different results. 
For the 6th and 7th mass bins, these differences can be attributed to the insufficient number of particles in the inner regime of halos in $1~{\rm Gpc}/h$ simulation to determine halo shapes,
given that the results from $I_{ij}$ are converged at these mass bins (see Fig.~\ref{fig:bK_nonreduced} in App.~\ref{app:nonreduced}).
On the other hand, at the 8th mass bin $b_K$ from $250~{\rm Mpc}/h$ simulations are not in agreement with that from $1~{\rm Gpc}/h$ for both $J_{ij}$ and $I_{ij}$ results.
This could happen due to the small number of halos at this mass bin in $250~{\rm Mpc}/h$ simulations.
Considering these results, in the following we use $250~{\rm Mpc}/h$, $1~{\rm Gpc}/h$, and $3~{\rm Gpc}/h$ simulations for the 1st-7th, 8th-9th, and 10th-11th mass bins, respectively.

In the right panel of Fig.~\ref{fig:bK_reduced} we show $b_K$ from different kinds of tides, namely $\Delta_{\rm p}$-type and $\Delta_{\rm e}$-type tides. They are in good agreement with each other, which implies that the validity of the linear alignment model is irrelevant to the substructure of the cosmic web such as knots, filaments, or pancakes.
Because the results from the two different tides are converged over all the mass range, in the following we combine two kinds of simulations to estimate $b_K$.

\subsubsection{Redshift-dependence: the relation between $b_K$ and $b_1^{\rm E}$}
\label{subsub:redshifts}

\begin{figure}[tb]
\centering
\includegraphics[width=0.67\textwidth]{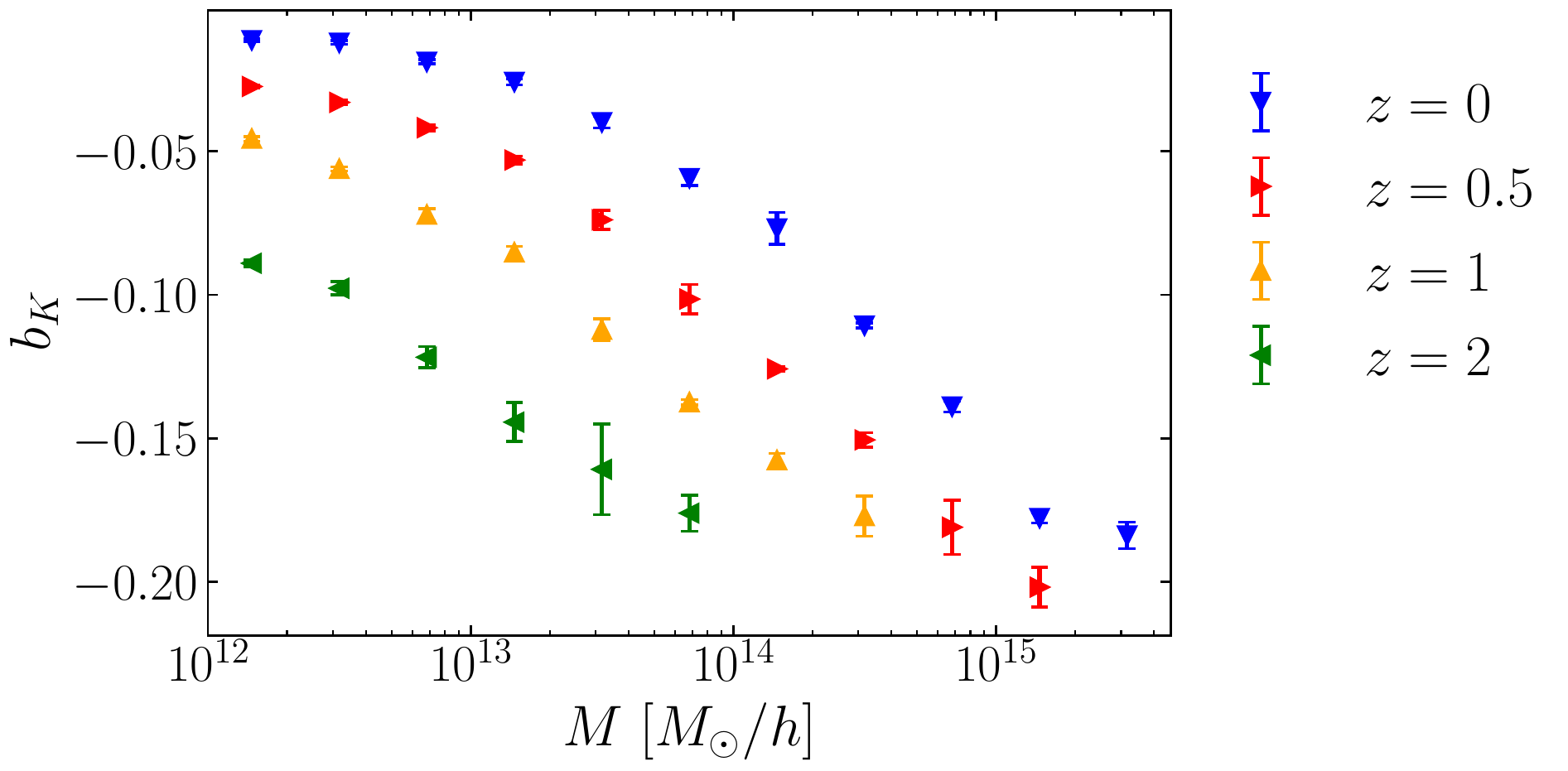}
\caption{Linear alignment coefficient, $b_K$, for the reduced inertial tensor, $J_{ij}$, at various redshifts.
Here we have combined results from difference box sizes and different tides.
}
\label{fig:bK_reduced_redshifts}
\end{figure}

Here we discuss the redshift-dependence of the linear alignment coefficient and show that there seems an universal relation between $b_K$ and $b_1$.

Fig.~\ref{fig:bK_reduced_redshifts} shows the linear alignment coefficient, $b_K$, for different redshifts.
It is clear that the absolute value of $b_K$ is greater at more massive halos and at higher redshift.
This means that more massive halos align stronger than less massive ones and the strength of the alignment becomes larger as redshift increases for all mass range.
These trends are supposed to originate from the fact that the alignment of halo shape is also affected by the surrounding matter distribution of each halos; less massive halos are susceptible to their surroundings and as time evolves the impact of their surroundings becomes greater.

\begin{figure}[tb]
\centering
\includegraphics[width=0.6\textwidth]{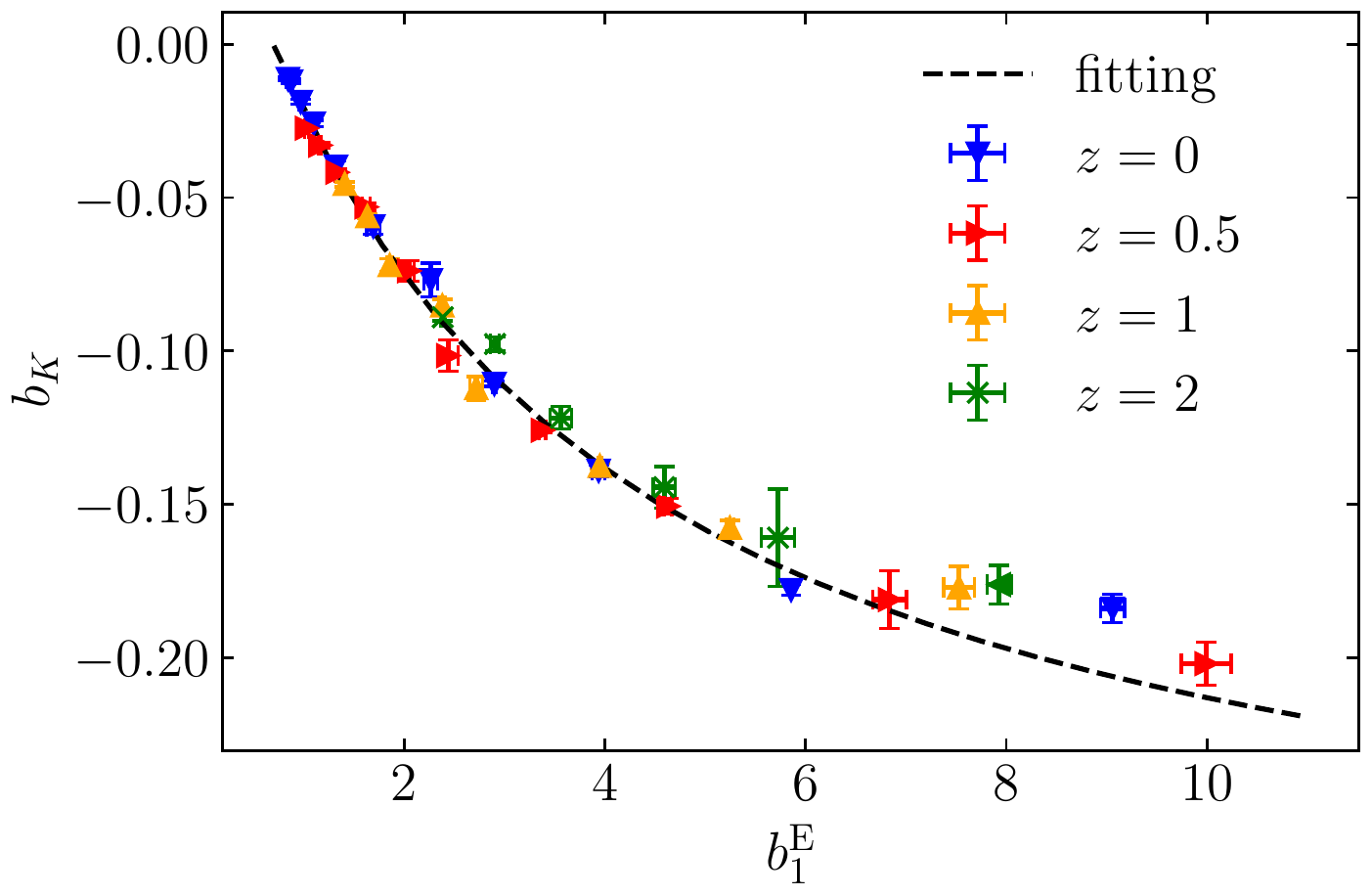}
\caption{Linear alignment coefficient, $b_K$, as a function of the Eulerian linear bias, $b_1^{E}$, combining all redshift and mass information.
The dashed curve is the fitting given in Eq.~\eqref{eq:bE1_bK_fit_J}.
}
\label{fig:b1bK_reduced}
\end{figure}

These trends are similar to the linear bias $b_1$ 
so it is interesting to explore the relation between $b_K$ and $b_1$.
In Fig.~\ref{fig:b1bK_reduced} we plot $b_K$ as a function of the Eulerian linear bias $b_1^{\rm E}$ for various redshifts. 
$b_1^{\rm E}$ is estimated from the Lagrangian linear bias $b_1^{\rm L}$, which is directly measured as the response of the halo number in our $\Delta_0$-type simulations (see also App.~\ref{app:SU_comp}), using $b_1^{\rm E} = b_1^{\rm L}+1$. 
We find the relation between $b_K$ and $b_1^{\rm E}$ shows an universal behaviour over the range $z=0\sim 2$.
This universal relation is also found when using $I_{ij}$ (see App.~\ref{app:nonreduced}) and thus is not relevant to how to measure the halo shapes.
This strongly suggests that $b_K$ is also uniquely determined by some quantity depending mass as $b_1^{\rm E}$ does by the variance of the dark matter density field.
Since our simulations enable both $b_K$ and $b_1$ to be measured very accurately, here we provide the fitting formula in the form of $b_K(b_1^{\rm E})$ for convenience.
We combine results from all redshifts and obtain a fitting formula
of the $b_K$-$b_1$ relation using
a very simple rational function:
\begin{equation}
    b_K = \frac{0.06461 - 0.09322 b_1^\Eul}{1 + 0.3073 b_1^\Eul}
    \label{eq:bE1_bK_fit_J}
\end{equation}
This fitting function is shown as dashed curve in Fig.~\ref{fig:b1bK_reduced}.

\subsubsection{Secondary halo shape responses}
\label{subsub:secondary}

Studies of the halo assembly bias show that the halo bias depends on properties
other than the mass.
Likewise, it is natural that the halo shape response also possesses rich dependences
beyond the halo mass.
Here we study dependences of $b_K$ on the halo concentration and the eccentricity of inertial tensor of halos.

We use the \texttt{AHF} halo finder to measure the halo concentration parameter $c$.
Instead of fitting a NFW halo profile to each halo, \texttt{AHF} measures the ratio
$v_\mathrm{max} / v_{200}$. $v_\mathrm{max}$ is the maximum circular velocity,
$v_\mathrm{max} \equiv \max\sqrt{G M(<R) / R}$,
and $v_{200}$ is the circular velocity at virial radius
$v_\mathrm{200} \equiv \sqrt{G M_{200} / R_{200}}$.
Assuming a NFW halo profile, $v_\mathrm{max} / v_{200}$ is related to $c$
as given by \cite{prada2012},
and thus is used in \texttt{AHF} to determine the concentration.

\begin{figure}[tb]
\centering
\includegraphics[width=0.4\textwidth]{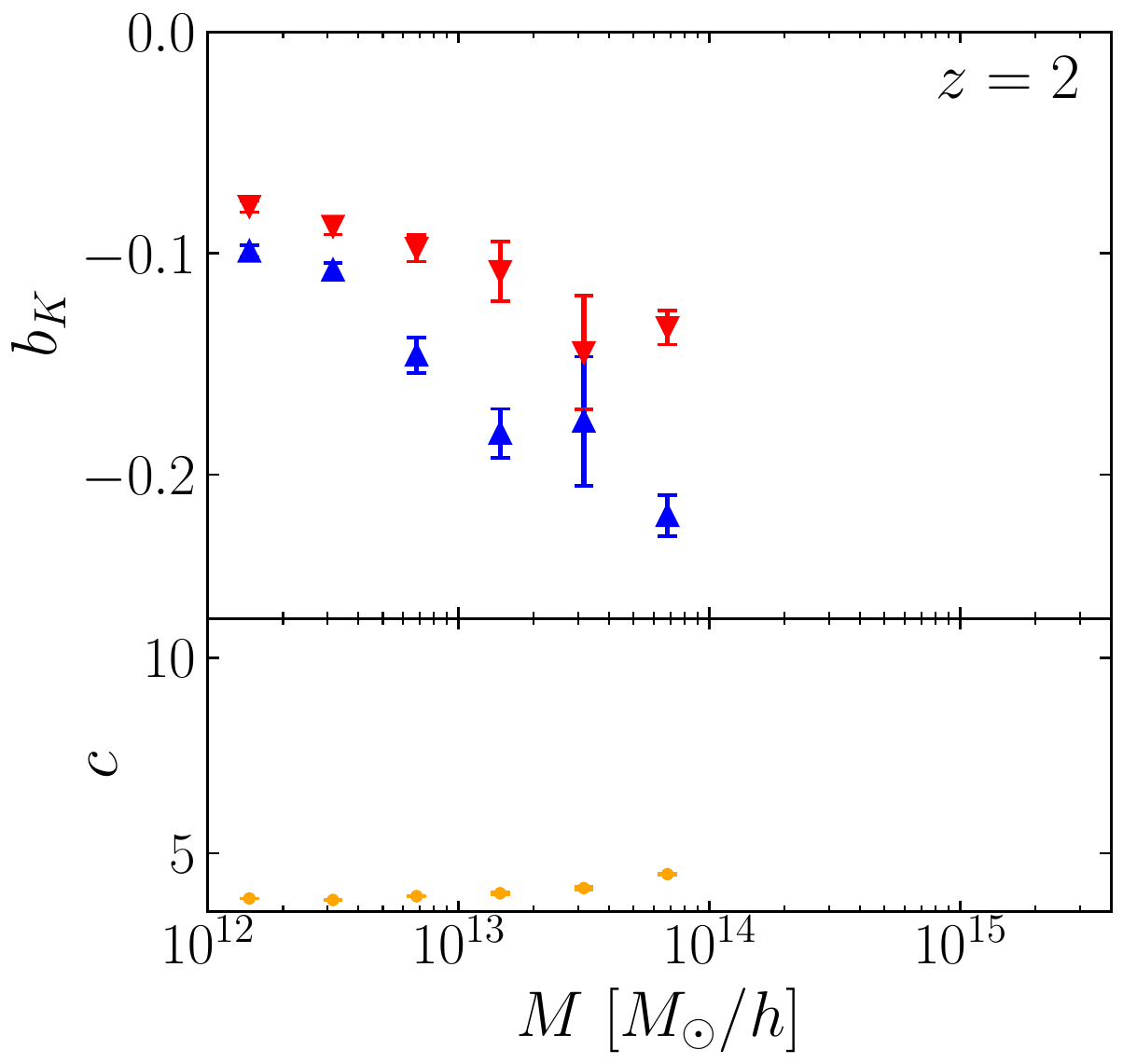}
\hspace{1em}
\includegraphics[width=0.52\textwidth]{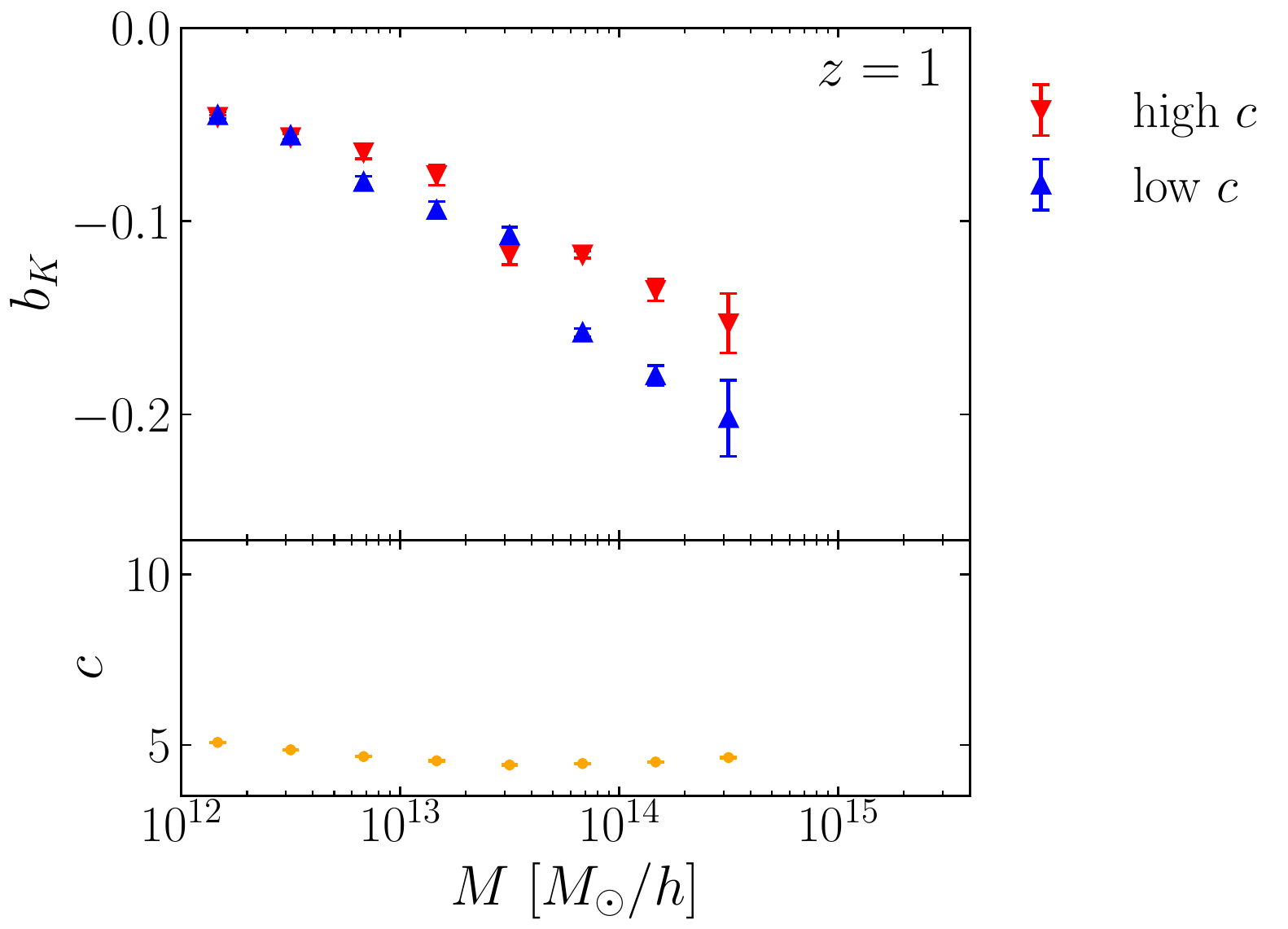}

\vspace{1em}
\includegraphics[width=0.4\textwidth]{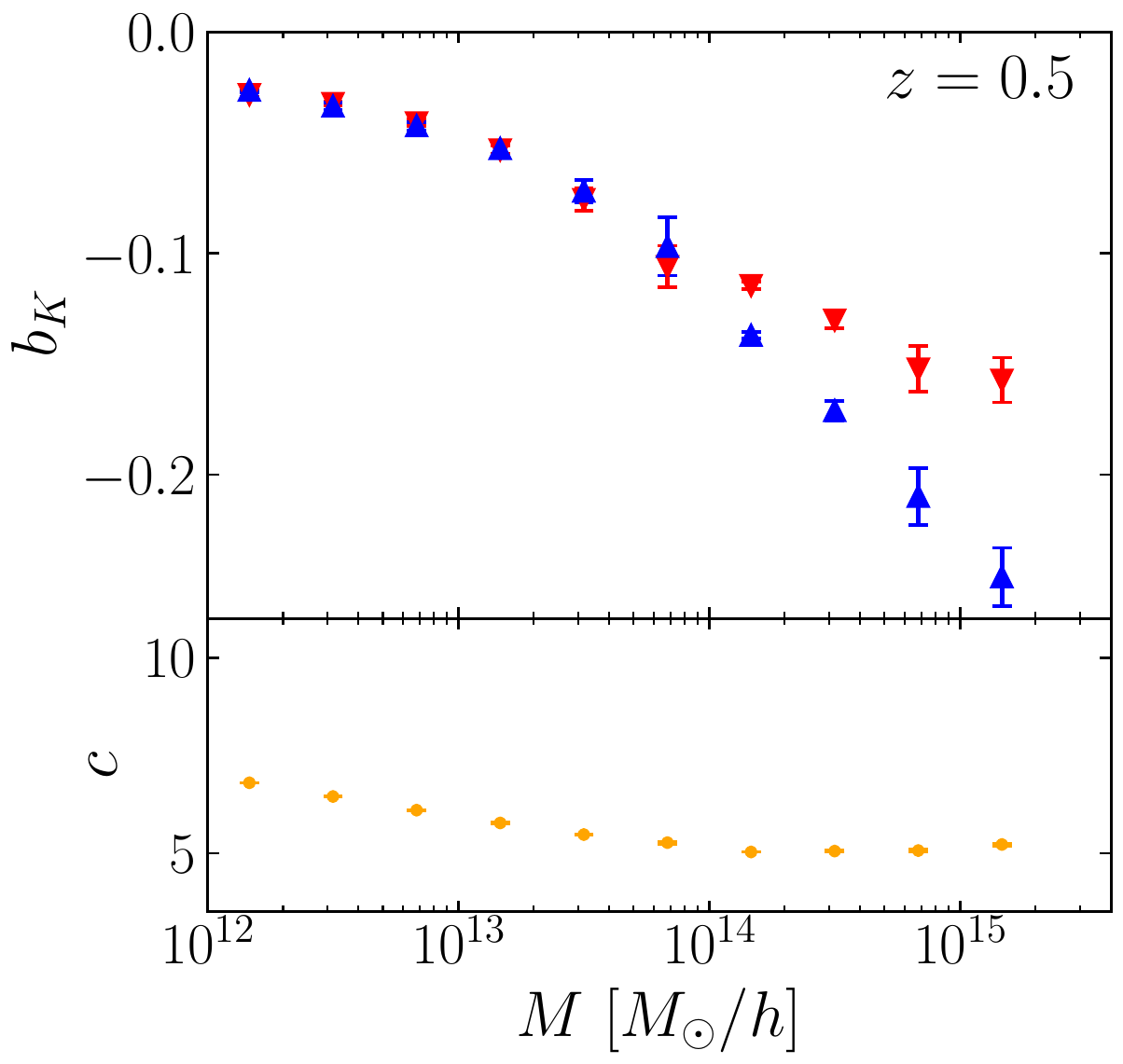}
\hspace{1em}
\includegraphics[width=0.52\textwidth]{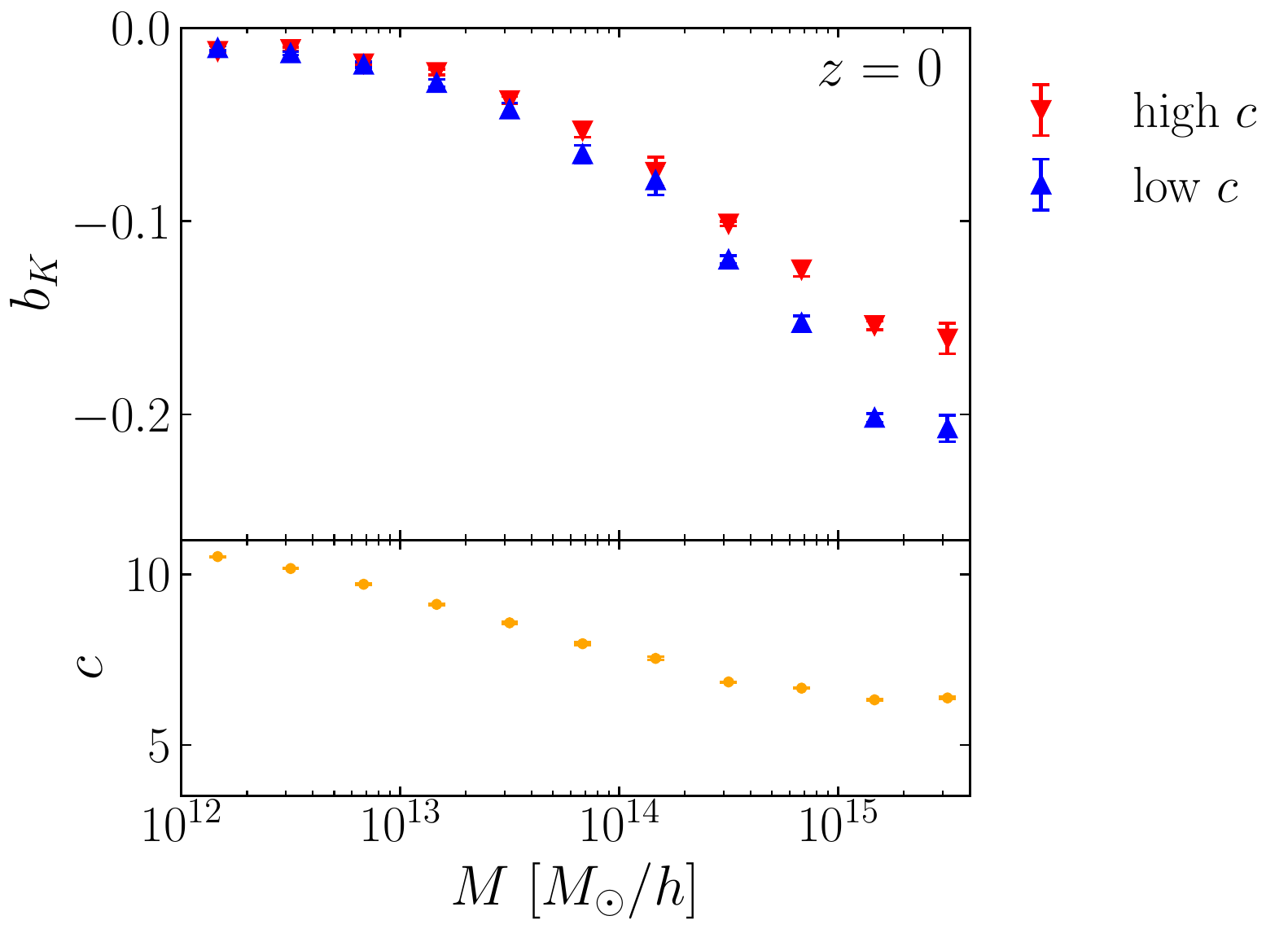}
\caption{{\it Upper panels:} The halo concentration dependence of the linear alignment coefficient, $b_K$, at various redshifts and masses:
$b_K$ from high concentration (red) and low concentration (blue).
{\it Lower panels:} The median concentration by which we divided halo samples.
}
\label{fig:bK_reduced_c}
\end{figure}

At each redshift and mass-bin we divided halo samples into those with greater than the median of the concentration and those with smaller concentration. 
Then we measured $b_K$ from each group. 
The results are shown in Fig.~\ref{fig:bK_reduced_c}.
For all redshifts halos with lower concentration tend to have a large amplitude of $|b_K|$ at high mass, while the difference is likely to become small at low mass.
Since halos with the high concentration are expected to be formed from highly curved peaks~\cite{Dalal_etal:2008},
the process of collapse into halo is not expected to be much affected by large-scale tidal field.
Further, since they were formed earlier they tend to lose their memory on large-scale tidal field
through interacting with the local surroundings for a long time.

\begin{figure}[tb]
\centering
\includegraphics[width=0.4\textwidth]{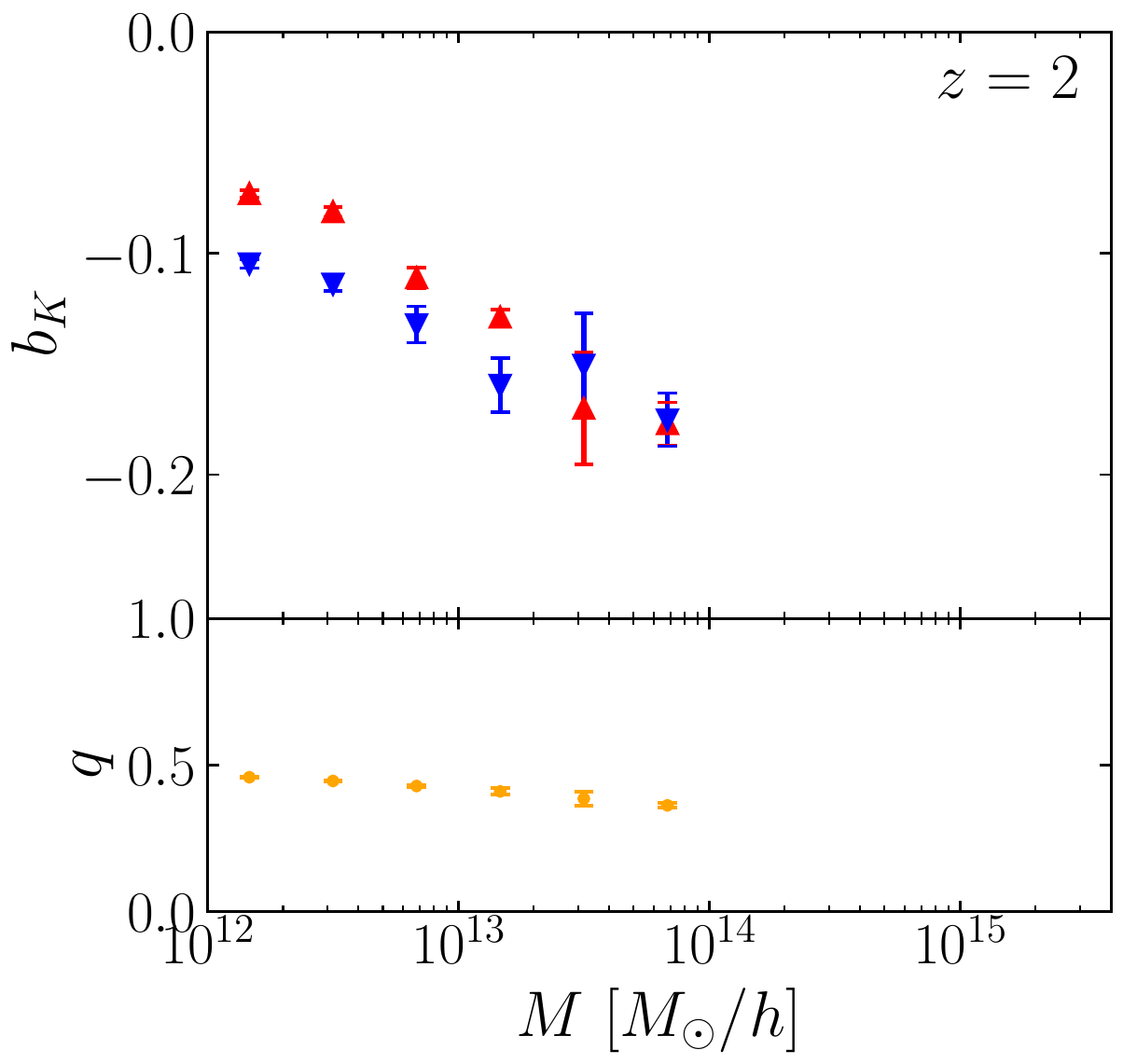}
\hspace{1em}
\includegraphics[width=0.52\textwidth]{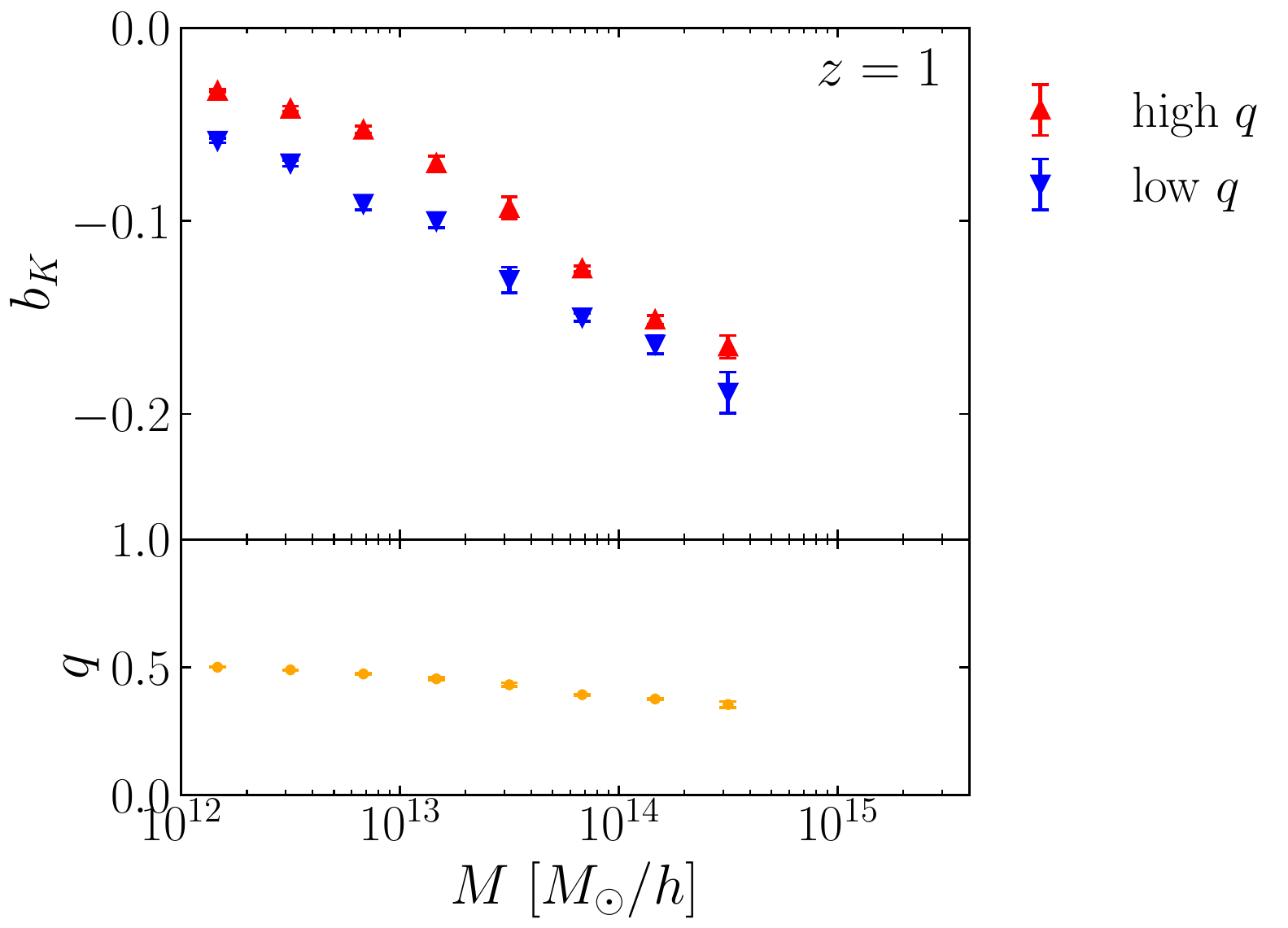}

\vspace{1em}
\includegraphics[width=0.4\textwidth]{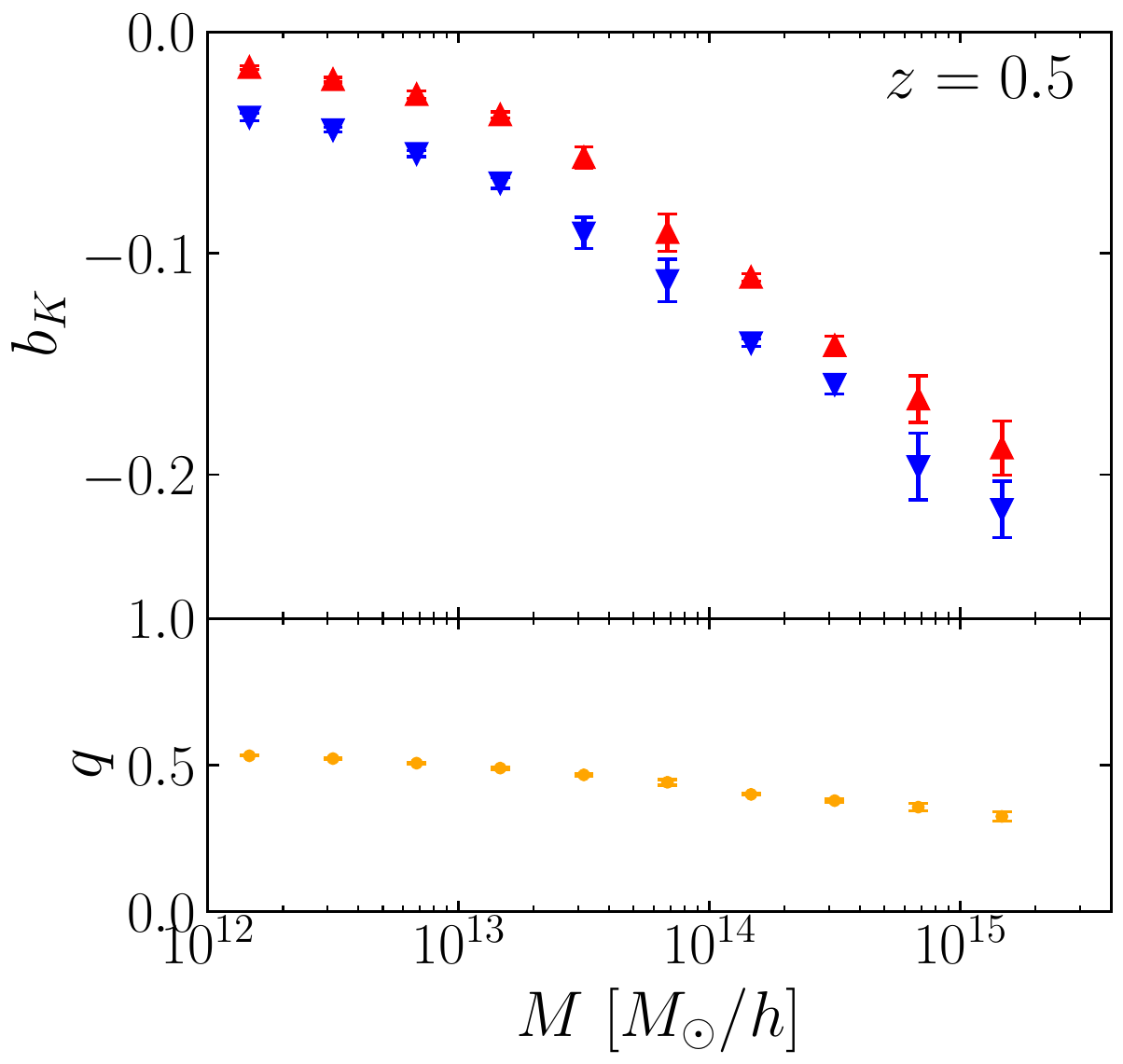}
\hspace{1em}
\includegraphics[width=0.52\textwidth]{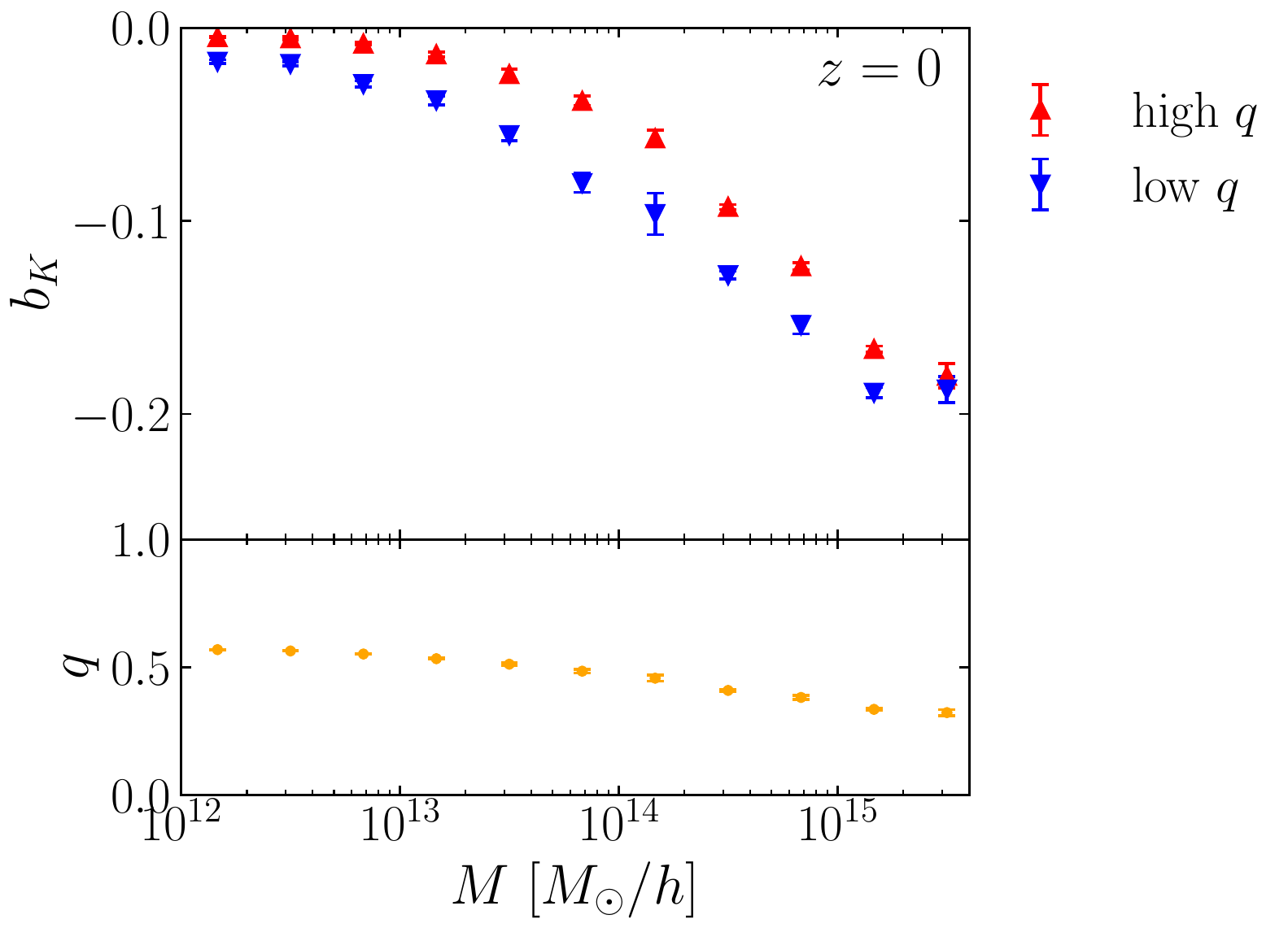}
\caption{
{\it Upper panels:} The axis-ratio ($q=J_3/J_1$ with $ J_1 \geq J_3$) dependence of the linear alignment coefficient, $b_K$, at various redshifts and masses: 
$b_K$ from high $q$ (low ellipticity, blue) and low $q$ (high ellipticity, red).
{\it Lower panels:} The median axis-ratio on which we divided halo samples.
}
\label{fig:bK_reduced_e}
\end{figure}

Next we discuss a dependence of the axis-ratio of halo shapes on $b_K$. We introduce the axis-ratio as the ratio of the major axis to the minor axis of the shape tensor: $q\equiv J_3/J_1$ where $J_1$, $J_2$, and $J_3$ are the eigenvalues of $J_{ij}$ satisfying $J_1 \geq J_2 \geq J_3$. As is done in the concentration, we divided halo samples into two groups: those above the median of the axis-ratio and those below the median, and then measured $b_K$ from each group.
Fig.~\ref{fig:bK_reduced_e} presents the results. 
For all redshifts and mass bins |$b_K$| from lower $q$ samples is greater than higher $q$ samples,
which means that halos with rounder shapes do not respond to the large-scale tidal field as strongly as halos with greater ellipticity.
This implies that distortion of halo shapes is indeed accelerated by the large-scale tidal field.
The same trend is found for a study of intrinsic alignments in Ref.~\cite{Okumura2009a}: more elongated halos are more tightly aligned with the surrounding matter distribution. Thus, the existence of the DC mode would not only bias measurements of the cosmic shear power spectrum in weak lensing surveys but also affect the cosmological application of intrinsic alignment itself.  

\section{Discussion}
\label{sec:discuss}

In this paper, we have implemented mean (DC) tidal and density fluctuations
into cosmological $N$-body simulations by absorbing them in effective 
anisotropic background expansion.
We have improved upon previous works \cite{Schmidt_etal:2018, 
Stucker_etal:2020, Masaki:2020a, Masaki:2020b}
on generating initial conditions for the tidal simulations,
with full second order Lagrangian dynamics properly solved in
general anisotropic background. 

The 2LPT in anisotropic background can be used to investigate
the linear tidal response of the matter power spectrum,
$R_2^{\rm L}$, on quasi-nonlinear scales. 
Since in simulations $R_2^{\rm L}$ at high redshifts suffer from
various  numerical artifacts \cite{Stucker_etal:2020},
it is of particular interest to measure $R_2^{\rm L}$ on small scales
at high redshifts from the 2LPT.
We have found $R_2^\Lag$ is enhanced upon the
tree-level perturbation theory prediction at $z=7\sim15$.
Our 2LPT results should be more robust since they are not affected
by high-redshift numerical artifacts that affect $N$-body simulations.
Though our findings are qualitatively consistent with
the result of Ref.~\cite{Masaki:2020b},
our enhancement is quantitatively weaker than that in Ref.~\cite{Masaki:2020b}.

We have also investigated the effect of large-scale tidal field on three-dimensional halo shapes using our simulations.
The linear alignment model predicts that 
the halo shape responds to the large-scale tidal field and thus linearly related with each other: $J_{ij} = b_K K_{ij}$.
Our tidal simulations allow us to directly test this relation and measure the proportional coefficient $b_K$ accurately.
We have found that the dependence of $b_K$ on redshifts and halo mass is similar to that of the linear halo bias $b^{\rm E}_1$; i.e., at the same halo mass $|b_K|$ is getting smaller as redshift decrease and more massive halos have greater $|b_K|$.
Furthermore, we have noticed that the relation between $b_K$ and $b_1^{\rm E}$ shows the universal behaviour
over a wide range of redshifts and masses; $z=0\sim 2$ and $M_{\rm h} = 10^{12}\sim 10^{15}~M_\odot/h$.
This implies that we can construct an analytical, physical model that can properly describe the mass- and redshift-dependence of $b_K$
as done for the linear bias using the peak theory and excursion set approach.
This kind of the theoretical prediction on $b_K$
is quite useful especially when treating the intrinsic alignments as the signal.
In particular, such a model is of crucial importance for exploring the angular-dependent (quadrupolar) primordial non-Gaussianity (PNG) from observations of the intrinsic alignments.
It is pointed out that the intrinsic alignments can uniquely probe the quadrupolar PNG in Refs.~\cite{Schmidt_etal:2015,Akitsu_etal:2020}.
However, since the bias induced by the quarupolar PNG on halo or galaxy shapes is completely degenerated 
with the quadrupolar PNG signal,
the lack of such a model makes it impossible to extract the information on the quadrupolar PNG
from measurements of the intrinsic alignments.
Therefore to develop a theory for the linear alignment coefficient $b_K$, analogous to the linear bias $b_1$ case, is an urgent issue and worth exploring in future works.

In addition, we have measured for the first time the secondary dependence of $b_K$ on halo 
properties other than the mass; it also depends on the halo concentration 
and axis-ratio.
This can be seen as the shape ``assembly bias'' as in the case of the number density bias
\footnote{Refs.~\cite{Obuljen_etal:2019,Obuljen_etal:2020} discuss the impact of the intrinsic alignment or galaxy shape on the density tracer as the assembly bias. This should be distinguished with ours that is the assembly bias \textit{of} the intrinsic alignment.}.
These findings will help to understand how halo shapes are determined in the hierarchical structure formation.

\acknowledgments

KA acknowledge support from JSPS Research Fellowship for
Young Scientists and JSPS KAKENHI Grant Numbers JP19J12254 and JP19H00677.
YL acknowledge support from Fellowships at Simons Foundation, the Kavli
IPMU established by World Premier International Research Center
Initiative (WPI) of the MEXT Japan, and at the Berkeley Center for
Cosmological Physics.
TO acknowledges support from the Ministry of Science and Technology of Taiwan under Grants No. MOST 109- 2112-M-001-027- and the Career Development Award, Academia Sinica (AS-CDA-108-M02) for the period of 2019 to 2023.

\appendix

\section{Solving DC density mode at second order}
\label{app:2spt}

At second order, Eq.~\eqref{eq:Deltai_eq} gives
\begin{equation}
    \ddot\Delta_i^{(2)} + 2H \dot\Delta_i^{(2)}
    = 4\pi G \rhobarm \Bigl( {\Delta_i^{(1)}}^2 - \frac13 \Delta_0^{(2)} \Bigr).
\end{equation}
Also expand Eq.~\eqref{eq:D0123} at second order and use Eq.~\eqref{eq:D0123_linear}
to derive
\begin{equation}
    \Delta_0^{(2)} = - \sum_i \Delta_i^{(2)}
    + \sum_{i<j} \Delta_i^{(1)} \Delta_j^{(1)} + \sum_i {\Delta_i^{(1)}}^2
    = - \sum_i \Delta_i^{(2)} + \frac12 {\Delta_0^{(1)}}^2
        + \frac12 \sum_i {\Delta_i^{(1)}}^2.
\end{equation}
Sum over $i$ of the first equation and plug in the second one,
one can show that at matter dominated era
\begin{equation}
    \sum_i \Delta_i^{(2)} = \frac3{14} \sum_i {\Delta_i^{(1)}}^2
    - \frac3{14} {\Delta_0^{(1)}}^2.
\end{equation}
Thus
\begin{equation}
    \Delta_0^{(2)} = \frac57 {\Delta_0^{(1)}}^2
    + \frac27 \sum_i {\Delta_i^{(1)}}^2
    = \frac{17}{21} {\Delta_0^{(1)}}^2
    + \frac27 \sum_i \tau_i^2,
\end{equation}
where the last equality follows from Eq.~\eqref{eq:Di_linear}.

\section{Second order Lagrangian perturbation theory in an anisotropic background}
\label{app:2lpt}

For 2LPT, the Jacobian determinant and matrix inverse in the master equation Eq.~\eqref{eq:master}
can be expanded as
\begin{align}
&\Bigl| \frac{\partial\vx}{\partial\vq} \Bigr| \simeq
    1 + \sum_i\Psi_{i,i}^{(1)} + \sum_i\Psi_{i,i}^{(2)} + \frac12 \left[\sum_i\left(\Psi_{i,i}^{(1)} \right)^2 - \sum_{ij}\Psi_{i,j}^{(1)}\Psi_{j,i}^{(1)} \right]
    \\
&\left[ \delta_{ij} + \Psi_{i,j} \right]^{-1} \simeq \delta_{ij} - \Psi_{i,j}^{(1)},
\end{align}
leading to the second-order equation
\begin{multline}
\sum_i\ddot \Psi^{(2)}_{i,i} + 2\sum_i H_i\dot \Psi^{(2)}_{i,i} -\frac32 H^2\Om(a)(1+\Delta_0)\sum_i\Psi_{i,i}^{(2)}
    \\
    = -\sum_i\Psi_{i,i}^{(1)}\sum_j\left[\ddot \Psi^{(1)}_{j,j} + 2H_j\dot \Psi_{j,j}^{(1)} \right]
    + \sum_{ij}\Psi_{i,j}^{(1)}\left[\ddot \Psi^{(1)}_{i,j} + 2H_i\dot \Psi_{i,j}^{(1)} \right]
    \\
    + \frac32 H^2 \Om(a) (1+\Delta_0)
    \left[\frac12\sum_i\left(\Psi_{i,i}^{(1)}\right)^2 - \frac12\sum_{ij}\Psi_{i,j}^{(1)}\Psi_{j,i}^{(1)}  \right].
\end{multline}

Similar to the linear order, we introduce the second order displacement potential
through $\Psi_i^{(2)} = \partial\psi_W^{(2)}/\partial q_i \equiv \psi^{(2)}_{W,i}$.
In the absence of the long modes, the equation for $\psi^{(2)}$ reduces to
\begin{equation}
\sum_i\ddot \psi^{(2)}_{,ii} + 2H\sum_i\dot \psi^{(2)}_{,ii}-\frac32 H^2 \Om(a)\sum_i\psi^{(2)}_{,ii}
 = -\frac32 H^2 \Om(a)\left[\frac12 \left( \sum_i\psi^{(1)}_{,ii} \right)^2 -\frac12 \sum_{ij}\psi^{(1)}_{,ij}\psi^{(1)}_{,ji} \right],
\end{equation}
where we used the linear equation Eq.~\eqref{eq:psi_1st}.
In this usual case, we denote the time-dependent part of $\psi^{(2)}$ as $D^{(2)}(t)$, which obeys 
\begin{align}
\ddot D^{(2)} + 2H \dot D^{(2)} - \frac32 H^2 \Om(a) D^{(2)} =
-\frac32 H^2\Omega_{\rm m} {D^{(1)}}^2.
\end{align}
In the matter-domination, we have $D^{(2)} = 3 {D^{(1)}}^2 / 7$.
The correction induced by the long modes, which is expressed by $\epsilon^{(2)}(t,\vq) \equiv \psi_W^{(2)}(t,\vq) - \psi(t,\vq)$, follows
\begin{align}
\sum_i\ddot{\epsilon}_{,ii}^{(2)} +& 2H\sum_i\dot{\epsilon}^{(2)}_{,ii}-\frac32 H^2 \Om(a) \sum_i\epsilon_{,ii}^{(2)}=
\nonumber\\
&-2\sum_i\dot\psi^{(2)}_{,ii} \dot\Delta_i + 2\sum_{ij}\dot\Delta_i\dot\psi_{,ij}^{(1)}\psi_{,ji}^{(1)}
-\frac32 H^2 \Om(a) \sum_{j}\epsilon^{(1)}_{,jj}\sum_{i}\psi^{(1)}_{,ii} 
\nonumber\\
&+ \sum_{ij}\psi^{(1)}_{,ij} \left[ \ddot\epsilon_{,ij}^{(1)} + 2H\dot\epsilon_{,ij}^{(1)} \right]
\nonumber\\
&+\frac32 H^2 \Om(a)\Delta_0\left[ \sum_{i}\psi_{,ii}^{(2)} - \frac12\sum_{i}\left( \psi^{(1)}_{,ii}\right)^2- \frac12\sum_{ij}\psi^{(1)}_{,ij}\psi^{(1)}_{,ji} \right],
\label{eq:2nd_epsilon}
\end{align}
where we have neglected $\mathcal{O}(\Delta_i^2)$ terms.
Notice that $\epsilon^{(1)}$ is $\mathcal{O}(\Delta_i)$.
For the matter dominated era,
the solution for Eq.~\eqref{eq:2nd_epsilon} is given by
\begin{align}
\sum_{i}\epsilon^{(2)}_{,ii}(t, \vq)
=&\frac14 \sum_{i}\left[-\frac{16}{9}\psi^{(2)}_{,ii}(t,\vq) + \frac89\sum_{j}\psi^{(1)}_{,ij}(t,\vq)\psi^{(1)}_{,ji}(t,\vq) \right]\Delta_i
\nonumber\\
&+\frac16 \left[\sum_{i}\psi_{,ii}^{(2)} - \frac12\sum_{i}\left( \psi^{(1)}_{,ii}\right)^2- \frac12\sum_{ij}\psi^{(1)}_{,ij}\psi^{(1)}_{,ji}   \right]\Delta_0
\nonumber\\
&+\frac14\left[-\frac23\sum_{i}\psi^{(1)}_{,ii}(t,\vq)\sum_{j}\epsilon_{,jj}^{(1)}(t,\vq) + \frac{20}{9}\sum_{ij}\psi^{(1)}_{,ij}(t,\vq)\epsilon_{,ij}^{(1)}(t,\vq)  \right].
\end{align}
Although the modified second order growth factor, $D^{(2)}_W$, due to the long modes can be identified as $D^{(2)}_W (t, \vp) = D^{(2)}(1 + \Delta_0/6 - 4\hat p_i^2\Delta_i/9 )$, the local gravitational tides cannot be neglected at second order.

\section{Force computation}
\label{app:force}
In this appendix, we review how to evaluate the tree force, especially the real-space counterparts of the PM force. 
$\phi^{\rm PM}(\vx)$ satisfies
\begin{align}
    \sum_i(1+\Delta_i)^{-2}\pdv{x_i}^2 \phi^{\rm PM}(\vx) 
    = - 4\pi G \bar{\rho}_m a^2 (1+\Delta_0)
    \int \dd^3\vx'\delta(\vx')\cdot
    \left[ \frac{1}{8 \pi \sqrt{\pi} x^3_s } \exp\left(-\frac{|\vx-\vx'|^2}{4x_s^2} \right) \right],
\end{align}
where the function inside the last bracket corresponds to the Fourier transform of $\exp(-p^2x_s^2)$,
which is the Gaussian smoothing kernel used in Eq.~\eqref{eq:phi_Fourier} to split force.
The PM potential is related to the tree potential as $\phi^{\rm PM} = \phi - \phi^{\rm T}$ and 
the solution for the PM potential is found to be 
\begin{align}
    \phi^{\rm PM}(\vx) 
    =-\frac{G m }{2a\sqrt{\pi}x_s}
    \int_0^\infty 
    \frac{\exp[-\frac{1}{4x_s^2}
    \left(\frac{ (1+\Delta_1)^2 x^2}{(1+\Delta_1)^2+\lambda}
    +\frac{ (1+\Delta_2)^2 y^2}{(1+\Delta_2)^2+\lambda}
    +\frac{ (1+\Delta_3)^2 z^2}{(1+\Delta_3)^2+\lambda}\right)]}
    {\sqrt{((1+\Delta_1)^2+\lambda)((1+\Delta_2)^2+\lambda)((1+\Delta_3)^2+\lambda)}}
    ~\dd \lambda,
    \label{eq:phi_PM_Real}
\end{align}
where we used $\bar{\rho}_{\rm m}(1+\Delta_0)\left[1+\delta(\vx) \right] = m(1+\Delta_0)/a^3\cdot\sum_n\delta^{\rm D}(\vx - \vx_n)$.
Although \eqref{eq:phi_PM_Real} has no closed analytic form, 
we can approximate this potential by Taylor expansion in $\Delta_i$
\footnote{Our expansion is different from that in Ref.~\cite{Stucker_etal:2020}, where $\phi^{\rm PM}(\vx)$ is expanded
in $1 + \Delta_i - \sqrt[3]{1+\Delta_0}$.}.
Using the following identity
\begin{equation}
    I_k(r) \equiv
    \int_0^\infty
    \frac{\exp[ - \frac{1}{4x_s^2}
    \frac{r^2}{a^2(1+\lambda)}]}
    {\sqrt{(1+\lambda)^k}}
    ~\dd \lambda
    =
    \biggl( \frac{r}{2 a x_s} \biggr)^{2-k}
    \gamma\biggl( \frac{k-2}{2}, \frac{r^2}{4a^2 x_s^2 }\biggr)
\end{equation}
where $\gamma$ is the lower incomplete gamma function.
We can express the approximated $\phi^{\rm PM}(\vx)$ up to the second order of $\Delta_i$ as
\begin{align}
    \phi^{\rm PM}({\bm r}) 
    =&- \frac{G m }{2a\sqrt{\pi}x_s}
    \left[
    I_3(r) + 
    \sum_i\left(-I_5(r)+\frac{r_i^2}{2a^2x_s^2}I_7(r) \right)\Delta_i
    \right.
    \nonumber\\
    &+\frac12\sum_i
    \left( 
    -I_5(r) 
    + \left(3 + \frac{r_i^2}{2a^2x_s^2} \right)I_7(r)
    -\frac{3r_i^2}{a^2x_s^2}I_9(r)
    +\frac{r_i^4}{4a^4x_s^4}I_{11}(r)
    \right)\Delta_i^2
    \nonumber\\
    &\left.+\frac12\sum_{i\neq j}
    \left(
    I_7(r) - \frac{r_i^2+r_j^2}{2a^2x_s^2}I_9(r) + \frac{r_i^2r_j^2}{4a^4x_s^4}I_{11}(r)
    \right)\Delta_i\Delta_j
    \right]
    +{\cal O}(\Delta_i^3).
\end{align}
In order to derive the force from this potential, we must be careful 
that the derivative should be taken with respect to the local comoving coordinate $\vx$, not to ${\bm r}$.
Thus, the force from the PM potential is computed as
\begin{align}
    \frac{\partial \phi^{\rm PM}}{\partial x^k}
    =a(1+\Delta_k)\frac{\partial \phi^{\rm PM}}{\partial r^k},
\end{align}
and
\begin{align}
    \pdv{\phi^{\rm PM}}{r^k}
    =&-\frac{G m }{2a\sqrt{\pi}x_s}
    \left[
    \frac{r_k}{r} I'_3(r)
    +\frac{r_k}{r}\sum_i\left( 
    -I'_5(r) + \frac{r_i^2}{2a^2x_s^2}I_7'(r)
    \right)\Delta_i
    + \frac{r_k}{a^2x_s^2} I_7(r) \Delta_k
    \right.
    \nonumber\\
    & + \frac{1}{2}\frac{r_k}{r}\sum_i
    \left(
    -I_5'(r) + \left(3+ \frac{r_i^2}{2a^2x_s^2} \right)I_7'(r)
    -\frac{3r_i^2}{a^2x_s^2}I'_9(r)
    +\frac{r_i^4}{4a^4x_s^{4}}
    I'_{11}(r)
    \right)
    \Delta_i^2
    \nonumber\\
    &\hspace{0.8cm}+\left(
    \frac{r_k}{2a^2x_s^2}I_7(r) - \frac{3r_k}{a^2 x_s^2} I_9(r) + \frac{r_k^3}{2a^4x_s^4} I_{11}(r)
    \right)\Delta_k^2
    \nonumber\\
    &+\frac{1}{2}\frac{r_k}{r}\sum_{i\neq j}
    \left(
    I'_7(r)
     - \frac{r_i^2+r_j^2}{2a^2x_s^2}I'_9(r)
     +\frac{r_i^2r_j^2}{4a^4x_s^4} I'_{11}(r)
    \right)\Delta_i\Delta_j
    \nonumber\\
    &\left.\hspace{0.8cm}+
    \Delta_k\sum_{i\neq k}\left(
    -\frac{r_k}{2a^2x_s^2}I_9(r) 
    + \frac{r^2_i r_k}{4a^4x_s^4} I_{11}(r)
    \right)\Delta_i
    \right],
\end{align}
where $I'_k(r)$ is the derivative of $I_{k}(r)$ given by 
\begin{equation}
    I_{k}^{\prime}(r) = \pdv{I_k(r)}{r}
    = \frac{2}{r} \biggl[
    \exp\biggl( - \frac{r^2}{4a^2x_s^2}\biggr)
    - \frac{k-2}2 \biggl(\frac{r}{2 a x_s}\biggr)^{2-k}
    \gamma\biggl( \frac{k-2}{2}, \frac{r^2}{4a^2 x_s^2 }\biggr)
    \biggr].
\end{equation}

\section{Comparison between our simulation and the conventional separate universe simulation}
\label{app:SU_comp}

In this appendix, we show the result of the convergence test for the isotropic background
by comparing our simulations with the conventional separate universe simulations.
\subsection{Recap of the usual separate universe simulations}
One way to incorporate the isotropic super-box mode is to change the background parameters according to its value.
This technique is based on the fact that
the flat FLRW universe with the spherically homogenious density perturbation $\Delta_0$ is
equivalent to the curved FLRW universe without $\Delta_0$.
This means $\Delta_0$ can be absorbed into the background parameters in cosmological simulations.
The relation of cosmological parameters between the global and local universe can be characterized
by the ratio of Hubble parameters,
\begin{equation}
\delta h \equiv \frac{h_W}{h} = \sqrt{1- \frac35 \frac{\Om}{D(t_0)}\Delta_0^{(1)}(t_0) },
\end{equation}
where $h_W$ is the local Hubble parameter and hereafter the subscript $_W$ denotes local quantities.
In terms of $\delta h$, other cosmological paramters are given by
\begin{align}
\Omega_{{\rm m}W} &= \Om  \delta h^{-2},\\
\Omega_{\Lambda W} &= \Omega_{\Lambda} \delta h^{-2},\\
\Omega_{KW} &= 1 - \delta h^{-2}.
\end{align}

In general, the time and comoving coordinates are also different among the global and local universes.
Each cosmology has its own expansion history and thus $a_W\neq 1$ when $a=1$.
Therefore we need to find the relation between the global and local scale factors at the same physical time $t$.
We can compute the difference between $a_W$ and $a$ by numerically solving Eq.~\eqref{eq:Deltai_eq} with $\tau_i=0$,
since the difference of the Friedmann equations in the two cosmology solves the spherical collapse.
We have to be careful to this mapping of time in generating the initial conditions and determining the output time.

As for the comoving length, it is common to use the unit of $\mathrm{Mpc}/h$ so the simulation box are given by
$L~{\rm Mpc}/h$ and $L_W~{\rm Mpc}/h_W$ in each cosmology.
The choice of $L_W$ depends on what one wants to measure directly from separate universe simulations.
If one needs to obtain the Eulerian response directly, $L_W$ is set to follow $a_W(t_{\rm out})L_W/h_W = a(t_{\rm out})L/h$
with $t_{\rm out}$ being the output physical tim.
In order to get the Lagrangian response directly, one have to set $L_W/h_W = L/h$.
The former and the latter is called as the total derivative method and the growth-dilation method respectively in Ref.~\cite{Li_etal:2014a}.
We employ the growth-dilation method where we set $L_W/h_W = L/h$ at all times to share the ramdom fluctuations
in the comoving scale in Mpc.

In this comparison study, we ran 6 pairs of separate universe simulations with $\Delta_0^{(1)}=\pm 0.09$. The boxsize is $L=250~{\rm Mpc}/h$ and the number of particles is $N_{\rm p} = 1024^3$, which are the same as our high-resolution simulations.

\begin{figure}[tb]
\centering
\includegraphics[width=0.4\textwidth]{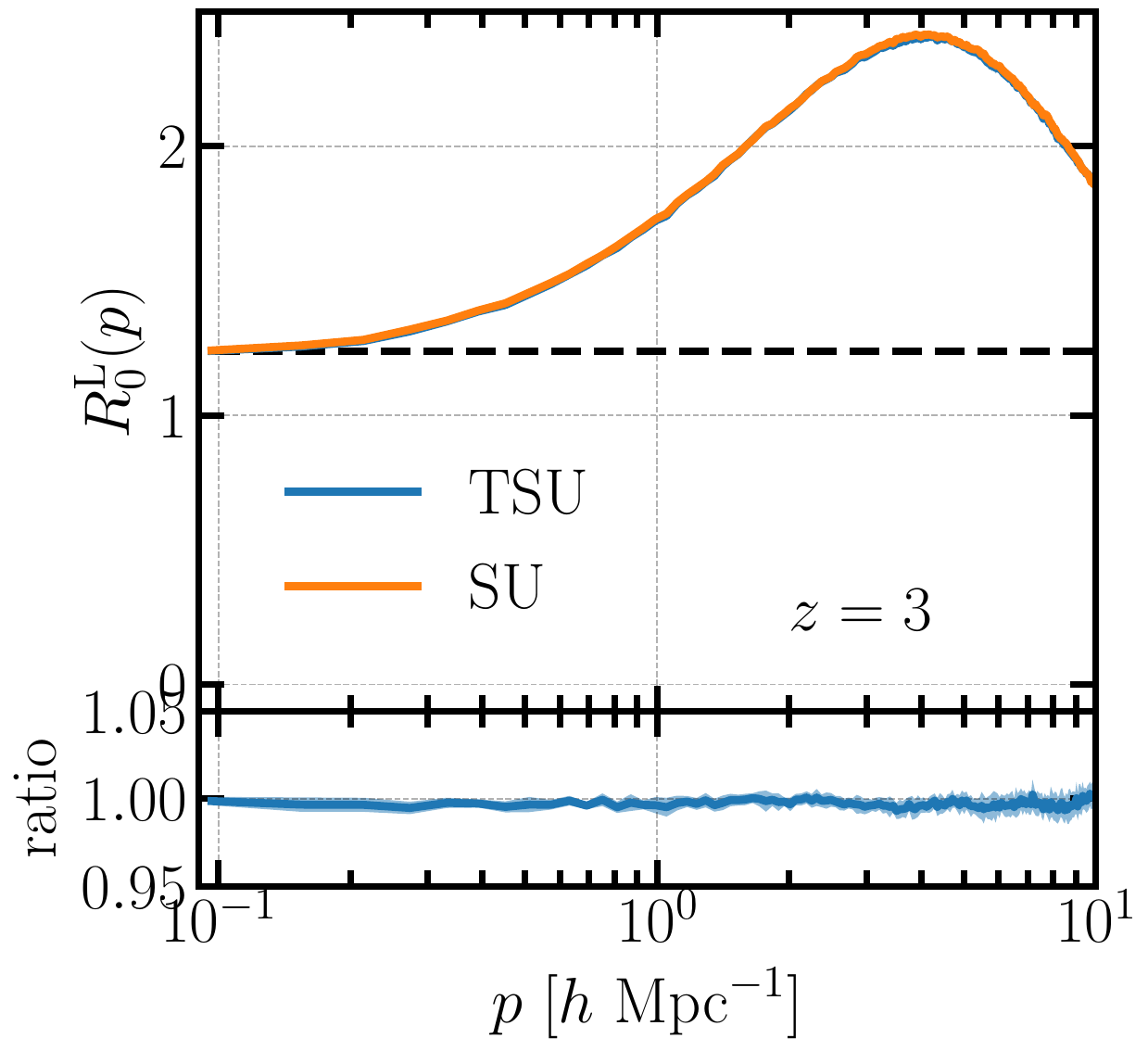}
\hspace{1em}
\includegraphics[width=0.4\textwidth]{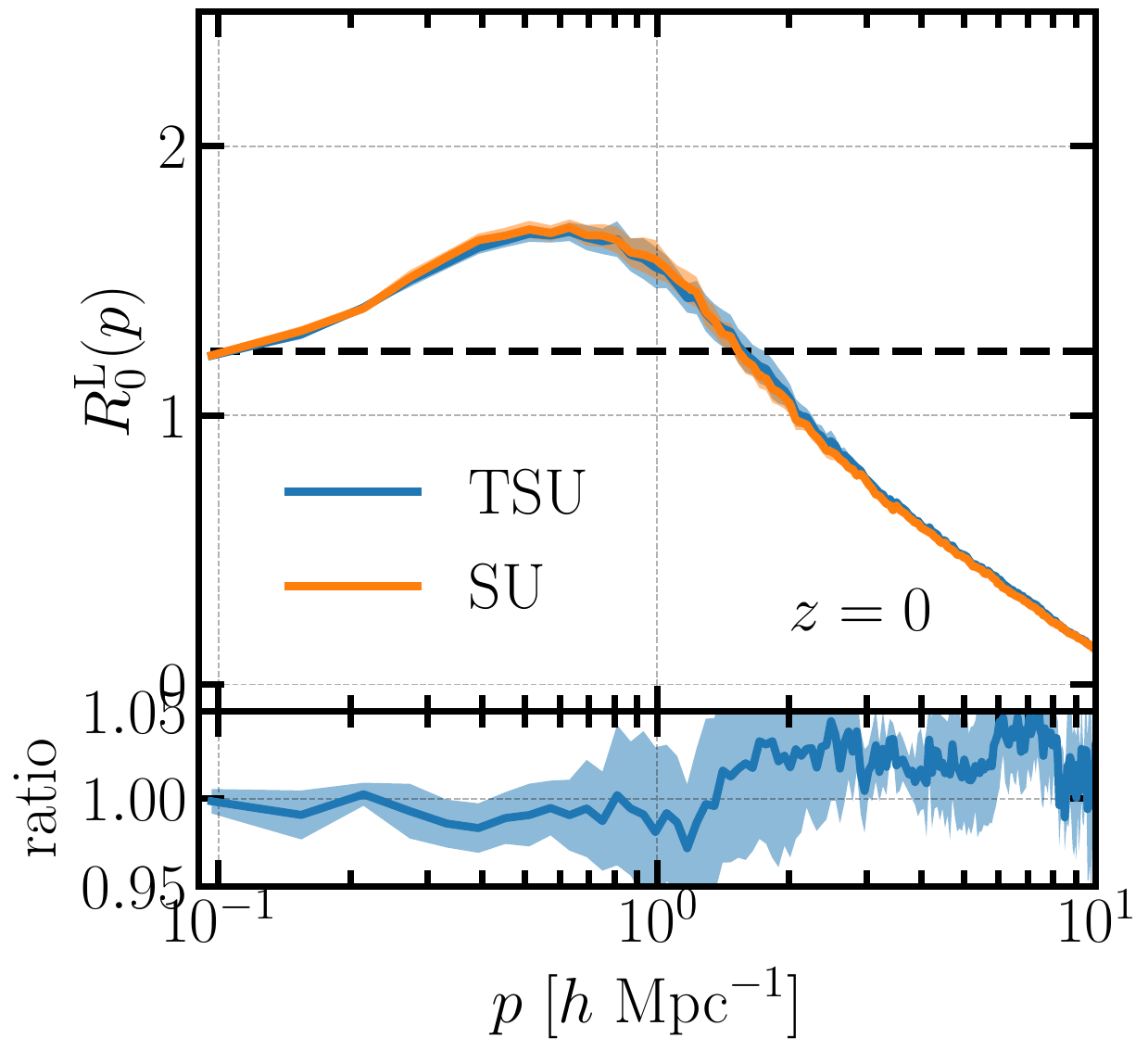}
\caption{
{\it Upper panels:}
Power spectrum responses to the large-scale overdensity ($\Delta_{\rm 0}$) from our tidal separate universe simulations (TSU, blue) and usual separate universe simulations (SU, orange)} at $z=0$ and 3.
The horizontal dashed line presents the tree-level prediction of the perturbation theory (Eq.~\eqref{eq:RL_treept}).
{\it Lower panels:} the ratio of the two: TSU/SU.

\label{fig:RL0_SU_ours}
\end{figure}

\subsection{Power spectrum response}
This subsection presents the comparison of the power spectrum response to the density perturbation $\Delta_0$ from both our $N$-body simulation and 2LPT with that from the usual separate universe simulations.
We estimate $R_0^{\rm L}(p)$ as
\begin{align}
    R_0^{\rm L}(p,z)
    =\frac{P_W(p,z;\Delta_0^{(1)}=+\epsilon)-P_W(p,z;\Delta_0^{(1)}=-\epsilon)}{2\epsilon D(z) P(p)}.
\end{align}

\subsubsection{Convergence of $N-$body results}
Fig.~\ref{fig:RL0_SU_ours} shows the $R_0^{\rm L}(p)$ responses from our simulations and the usual separate universe simulations.
For both $z=3$ and $z=0$, our results are in good agreement with the usual separate universe one, down to $k\simeq10~h{\rm /Mpc}$.

\subsubsection{On the valid scale of our 2LPT at high redshifts}
\begin{figure}[tb]
\centering
\includegraphics[width=0.35\textwidth]{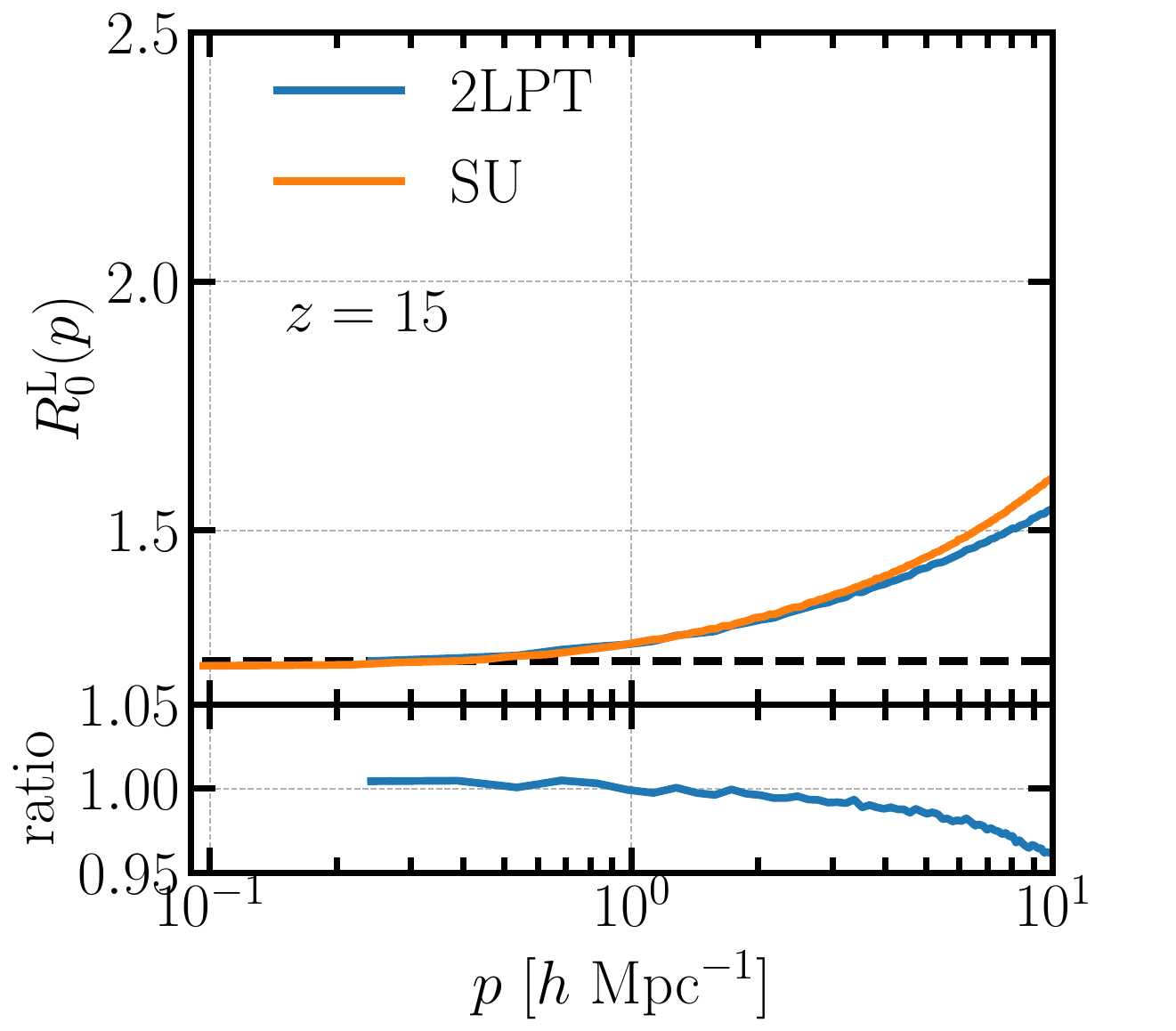}
\hspace{-2em}
\includegraphics[width=0.35\textwidth]{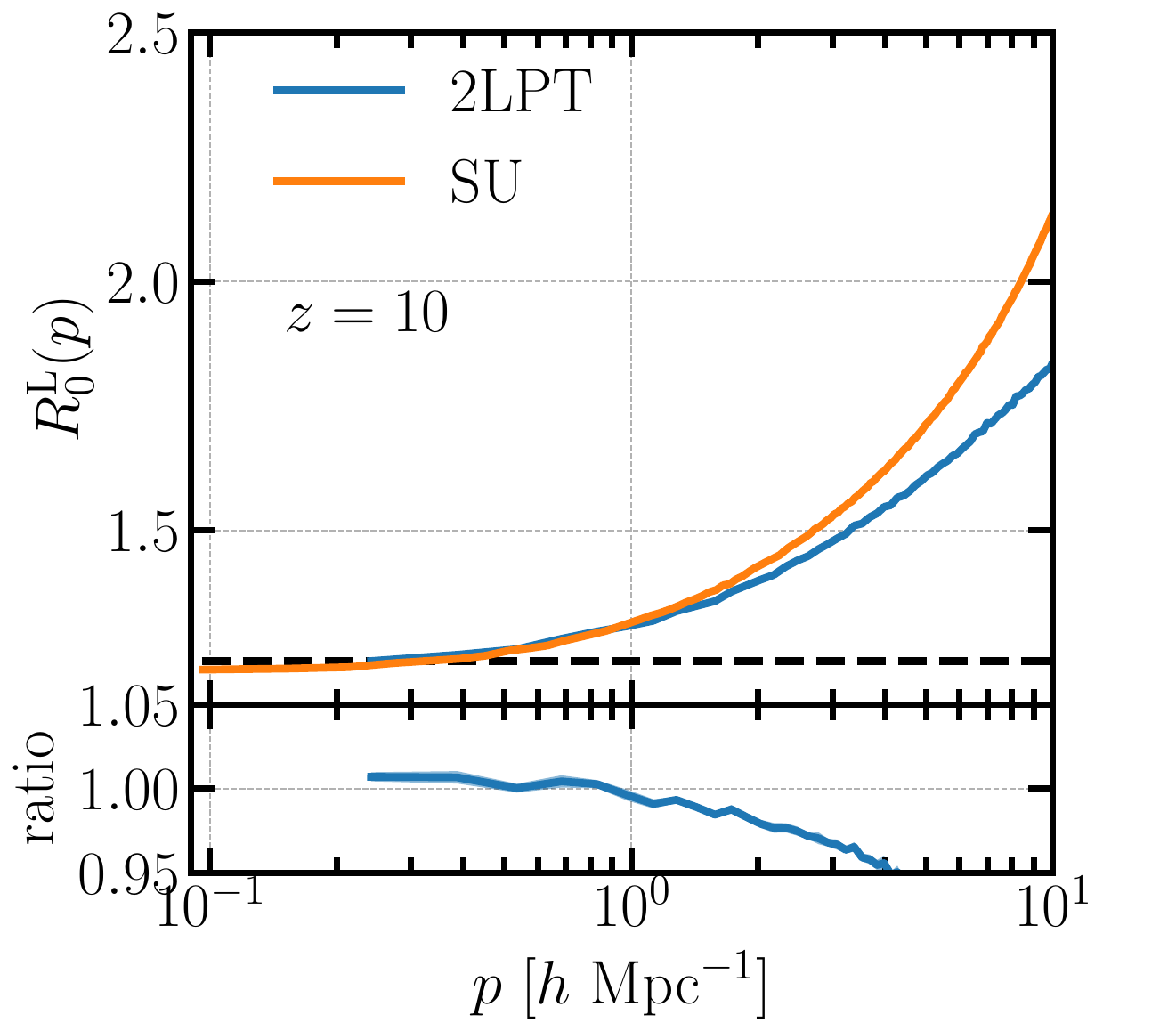}
\hspace{-2em}
\includegraphics[width=0.35\textwidth]{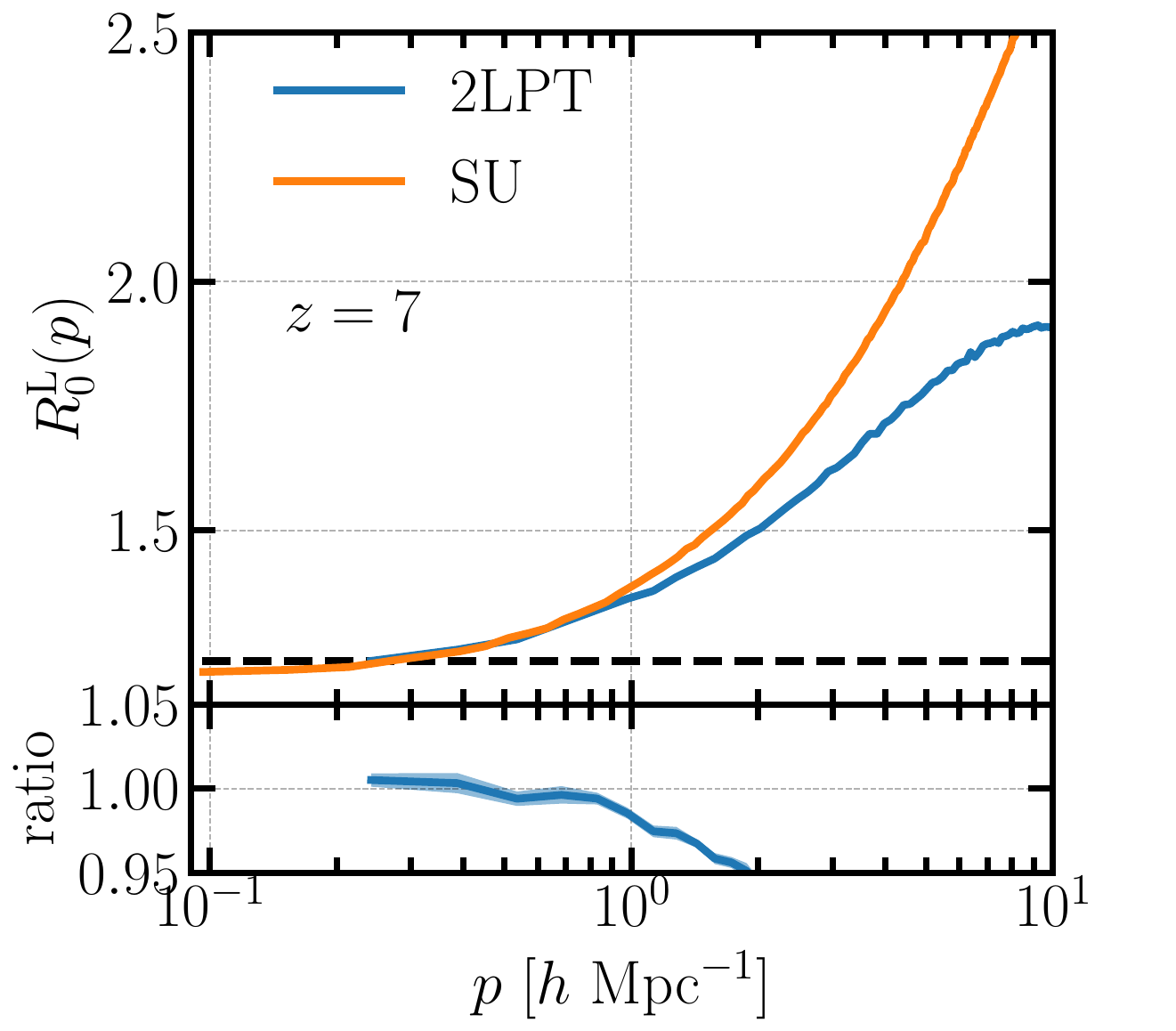}
\caption{
{\it Upper panels:}
Power spectrum responses to the large-scale overdensity ($\Delta_{\rm 0}$) from our modified 2LPT (blue) and usual separate universe (SU) simulations (orange) at $z=15$, 10, and 7.
{\it Lower panels:} the ratio of the two: 2LPT/SU.
}
\label{fig:RL0_SU_2LPT}
\end{figure}

Here we discuss the valid scales of our 2LPT by comparing it with the separate univserse $N$-body results.
Fig.~\ref{fig:RL0_SU_2LPT} presents the $R_0^{\rm L}(p)$ responses at $z=15,~10,$ and $7$ from our 2LPT and the usual separate universe simulations.
Note that the boxsize of our 2LPT is $L=100~{\rm Mpc}/h$ with $N_{\rm p}=1024^3$ while the usual separate universe simulations have $L=250~{\rm Mpc}/h$ with $N_{\rm p}=1024^3$.
Setting the criteria to be $5\%$ difference between our 2LPT and the separate universe simulations, we conclude the responses from our 2LPT are reliable up to $k=~9,~4,~{\rm and}~ 2h/{\rm Mpc}$ at $z=15,~ 10,$ and $7$, respectively.

\subsection{Linear bias}
\begin{figure}[tb]
\centering
\includegraphics[width=0.4\textwidth]{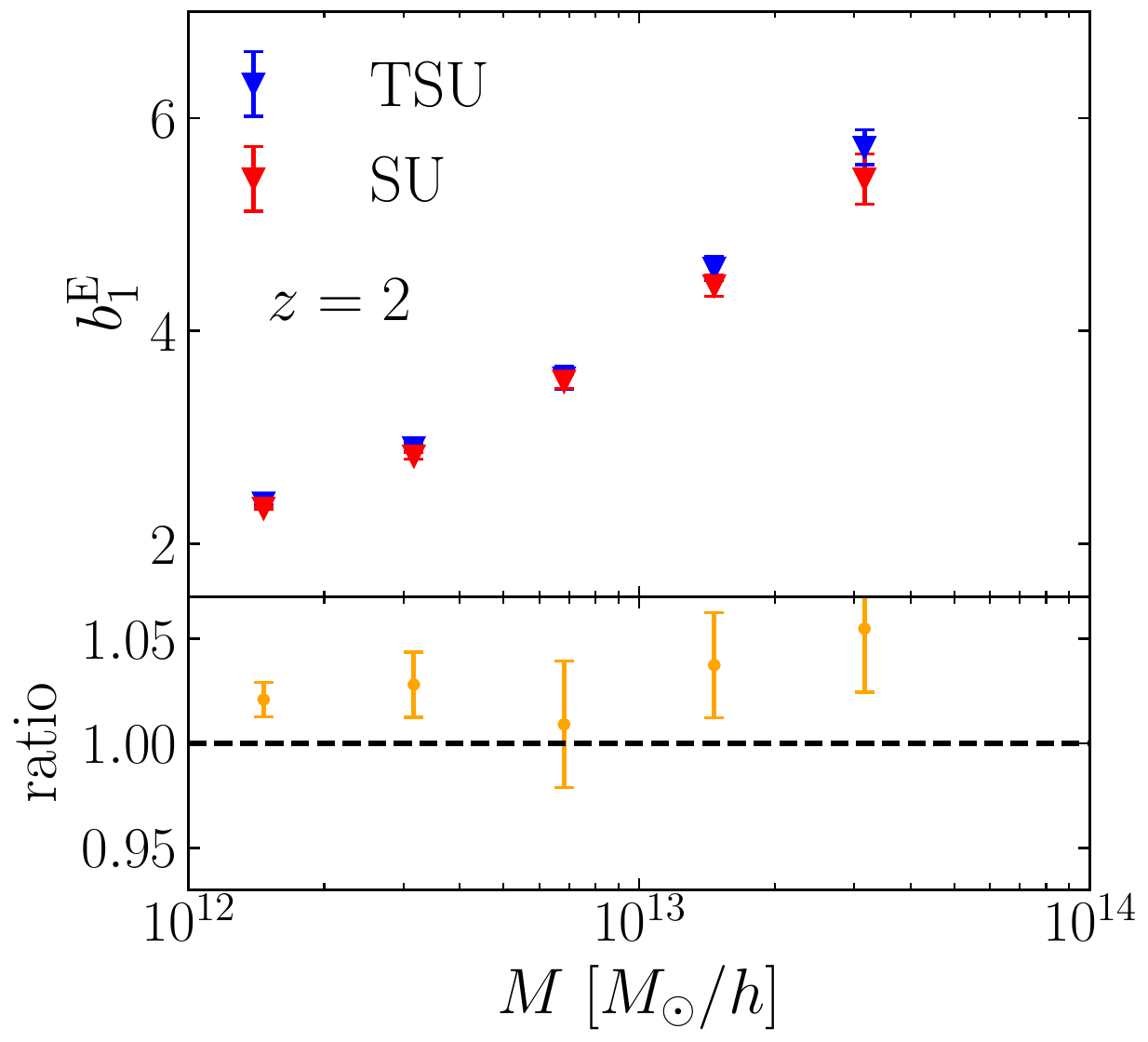}
\hspace{1em}
\includegraphics[width=0.4\textwidth]{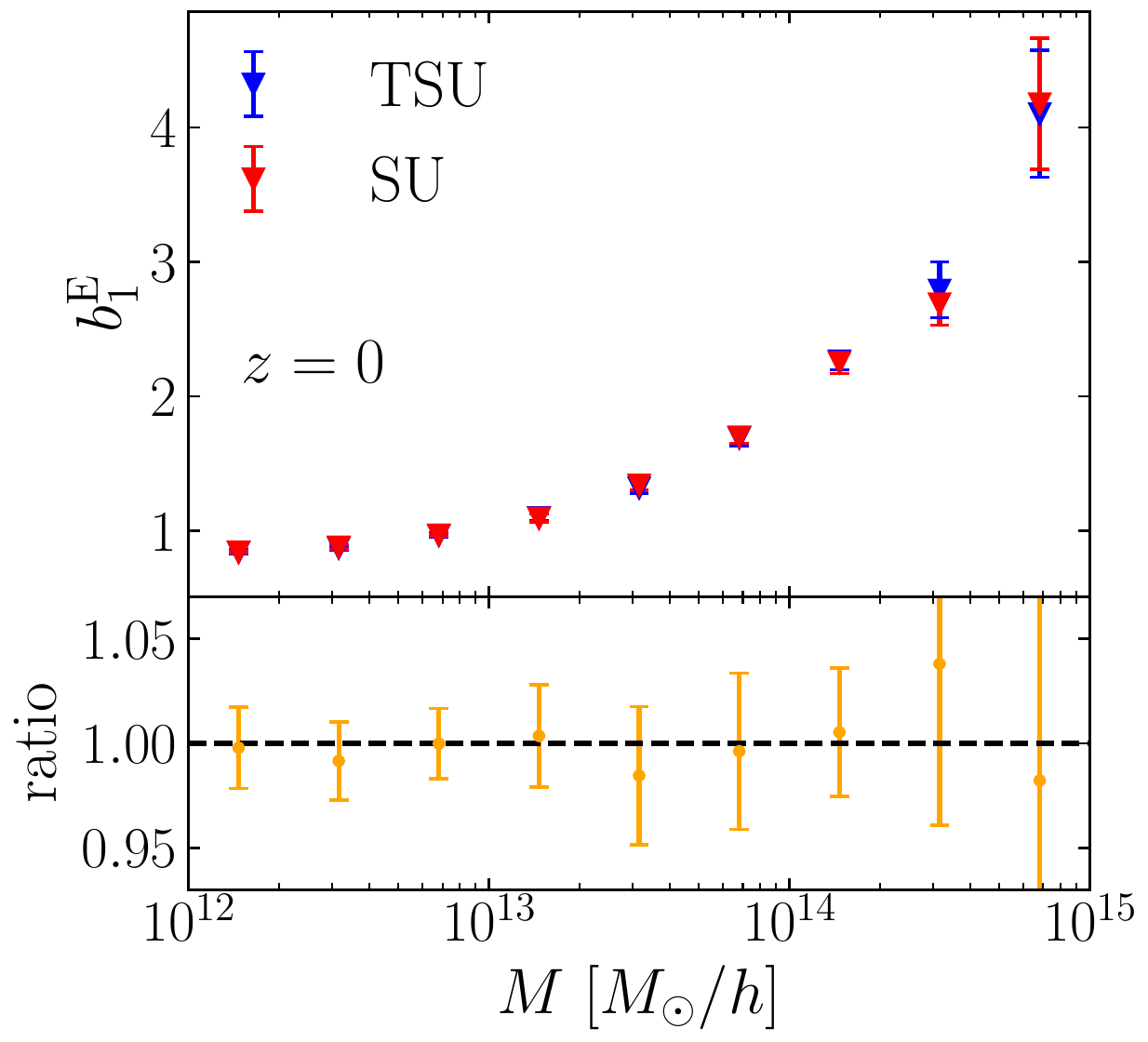}
\caption{
{\it Upper panels:}
$b_1^{\rm E}$ from our tidal separate universe simulations (TSU, blue) and usual separate universe simulations (SU, red).
{\it Lower panels:} the ratio of the two: TSU/SU.
}
\label{fig:bE1_SU_ours}
\end{figure}

Finally we also compare results on the linear bias measured from our simulations and the separate universe simulations.
Fist, we directly measure the Lagrangian linear bias $b_1^{\rm L}$ as
\begin{align}
    b_1^{\rm L}(M,z) = \frac{N_{\rm h}(M,z;\Delta_0^{\rm (1)}=+\epsilon)-N_{\rm h}(M,z;\Delta_0^{\rm (1)}=-\epsilon)}{2\epsilon D(z)N_{\rm h}(M,z;\Delta_0^{\rm (1)}=0)},
\end{align}
where $N_{\rm h}(M,z)$ is the total number of halos at mass $M$
in the local comoving volume.
Then the Eulerian linear bias is computed as $b_1^{\rm E} = b_1^{\rm L}+1$.
Fig.~\ref{fig:bE1_SU_ours} shows $b_1^{\rm E}$ at $z=0$ and $2$.
For all mass range our results agree with those from the usual separate universe simulations.
Together with the results about $R_0^{\rm L}(p)$, this suggests that 
our implementation is correctly working.

\section{Results from the inertial tensor $I_{ij}$}
\label{app:nonreduced}

\begin{figure}[tb]
\centering
\includegraphics[width=0.49\textwidth]{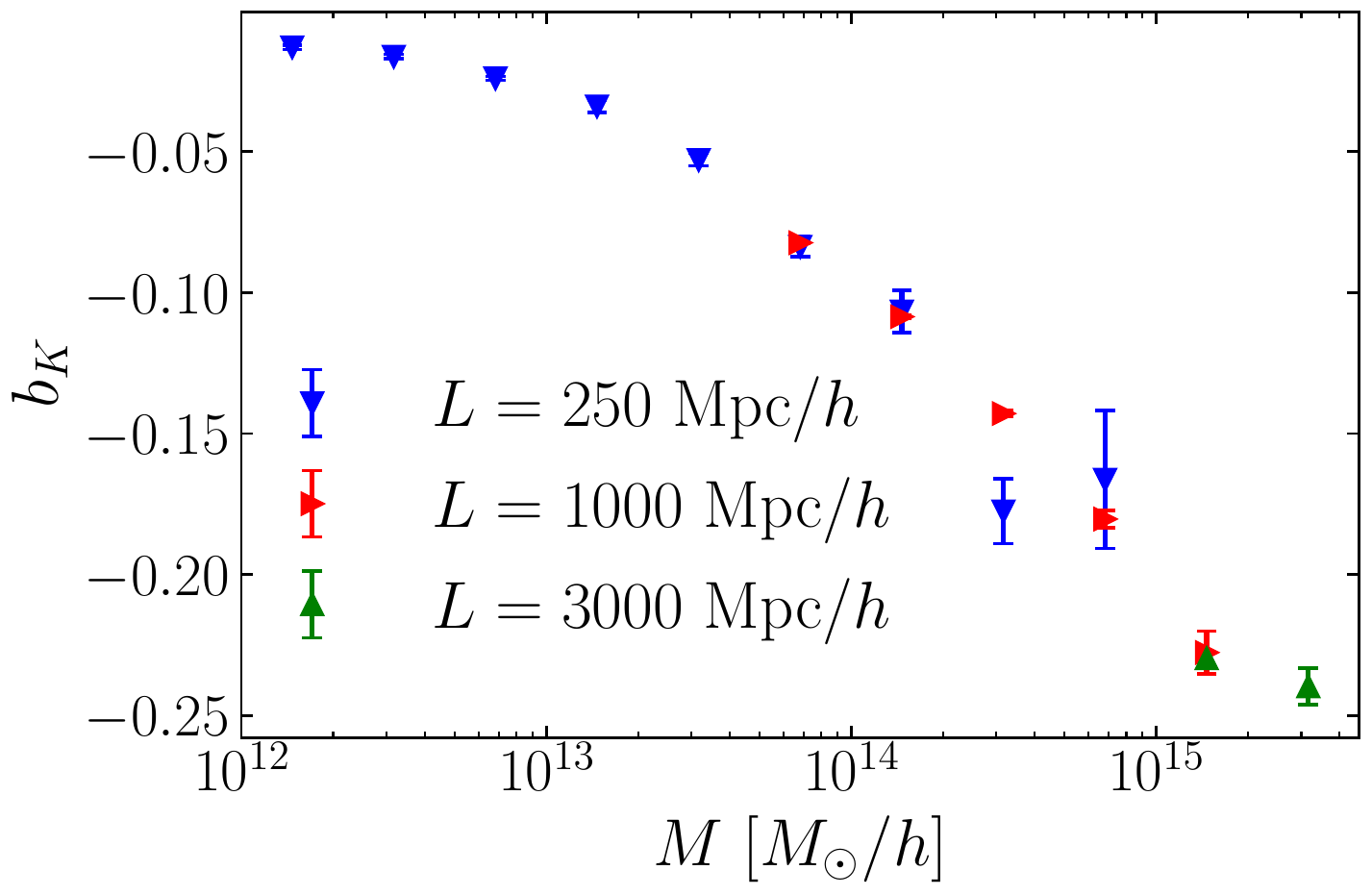}
\hfill
\includegraphics[width=0.49\textwidth]{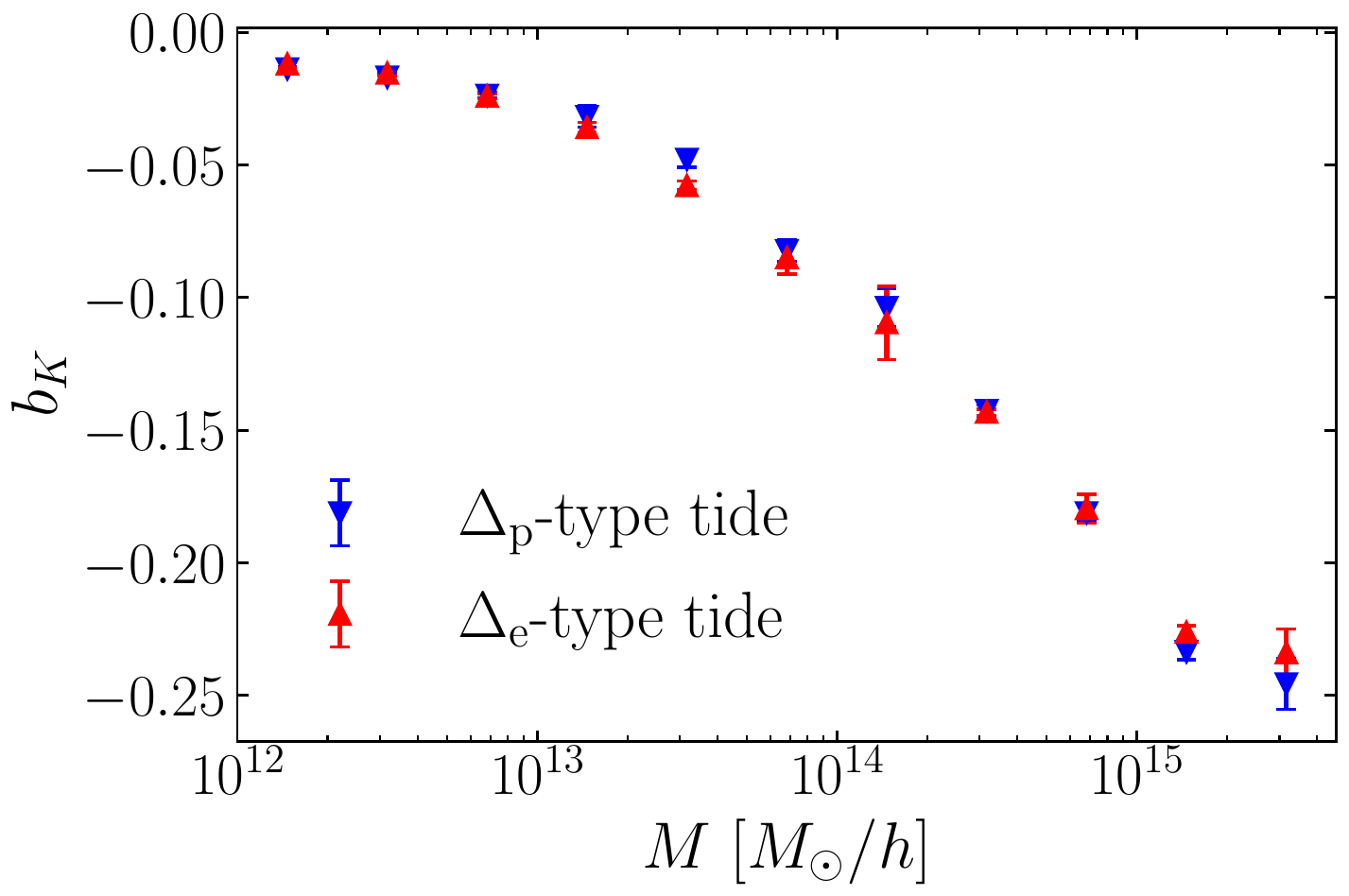}
\caption{Linear alignment coefficient, $b_K$, for the reduced inertial tensor, $I_{ij}$, at $z=0$.
The left plot shows measurements from simulations of different box sizes,
and the right one shows measurements from different tidal types.
}
\label{fig:bK_nonreduced}
\end{figure}

In this appendix, we summarize the shape response results when using the inertial tensor $I_{ij}$.
Fig.~\ref{fig:bK_nonreduced} shows the boxsize or equivalently resolution dependence of $b_K$.
Unlike using $J_{ij}$, 
there is no difference between $b_K$ at 6th and 7th mass-bin from $L=250~{\rm Mpc}/h$ and $L=1~{\rm Gpc}/h$.
This can be explained by the enough number of particle to determine the halo shape when using $J_{ij}$ in $1~{\rm Gpc}/h$ simulations since the number of particles used to define the halo shape is effectively higher in $I_{ij}$ than in $J_{ij}$.
Given these results, in the following in this appendix we use $250~{\rm Mpc}/h$, $1~{\rm Gpc}/h$, and $3~{\rm Gpc}/h$ simulations for 1st-6th, 7th-9th, and 10th-11th mass-bin, respectively.

\begin{figure}[tb]
\centering
\includegraphics[width=0.7\textwidth]{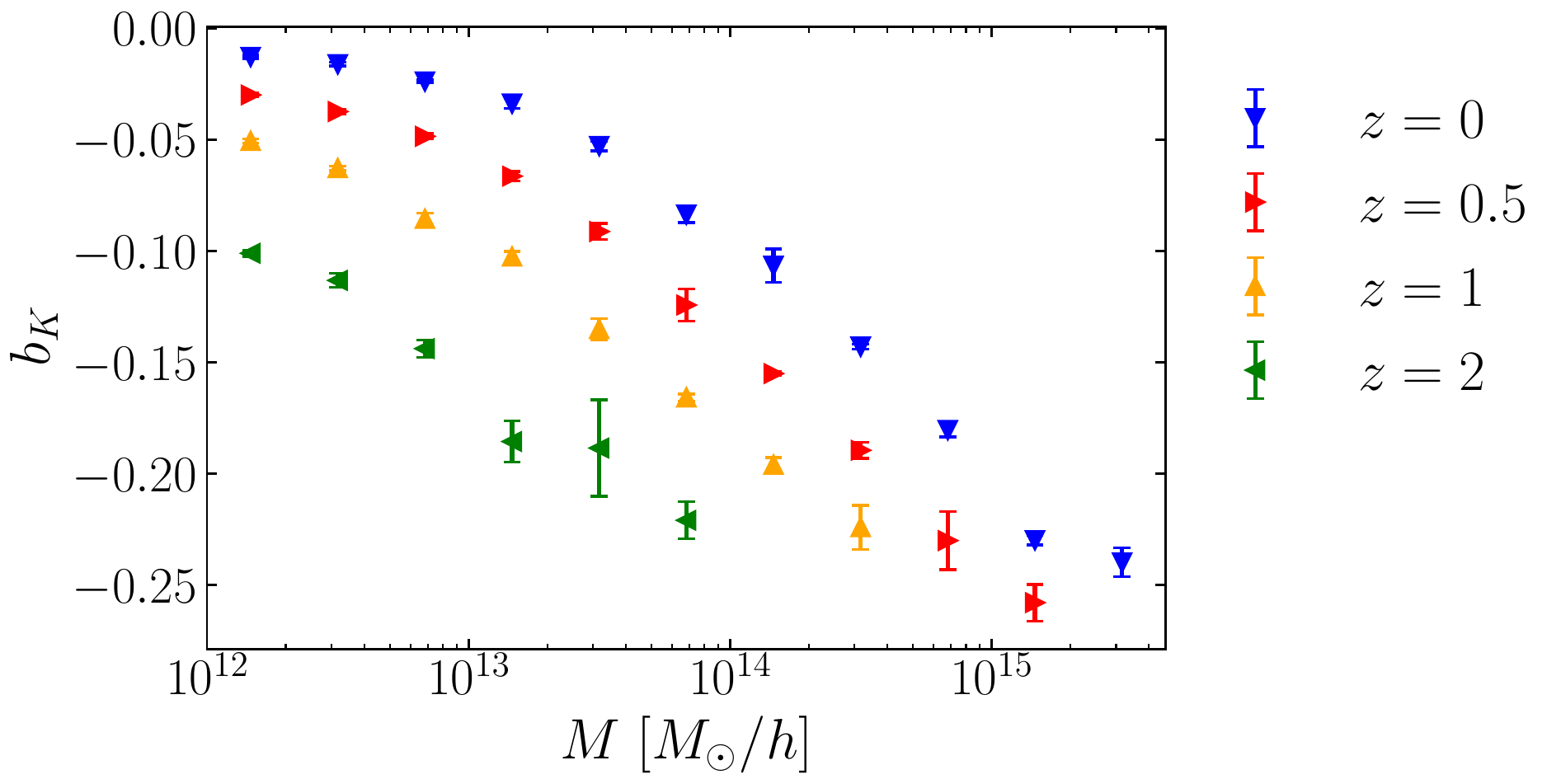}
\caption{Linear alignment coefficient, $b_K$, for the reduced inertial tensor, $I_{ij}$, at various redshifts.
Here we combine the results from difference box sizes and different tides.
}
\label{fig:bK_nonreduced_redshifts}
\end{figure}

\begin{figure}[tb]
\centering
\includegraphics[width=0.5\textwidth]{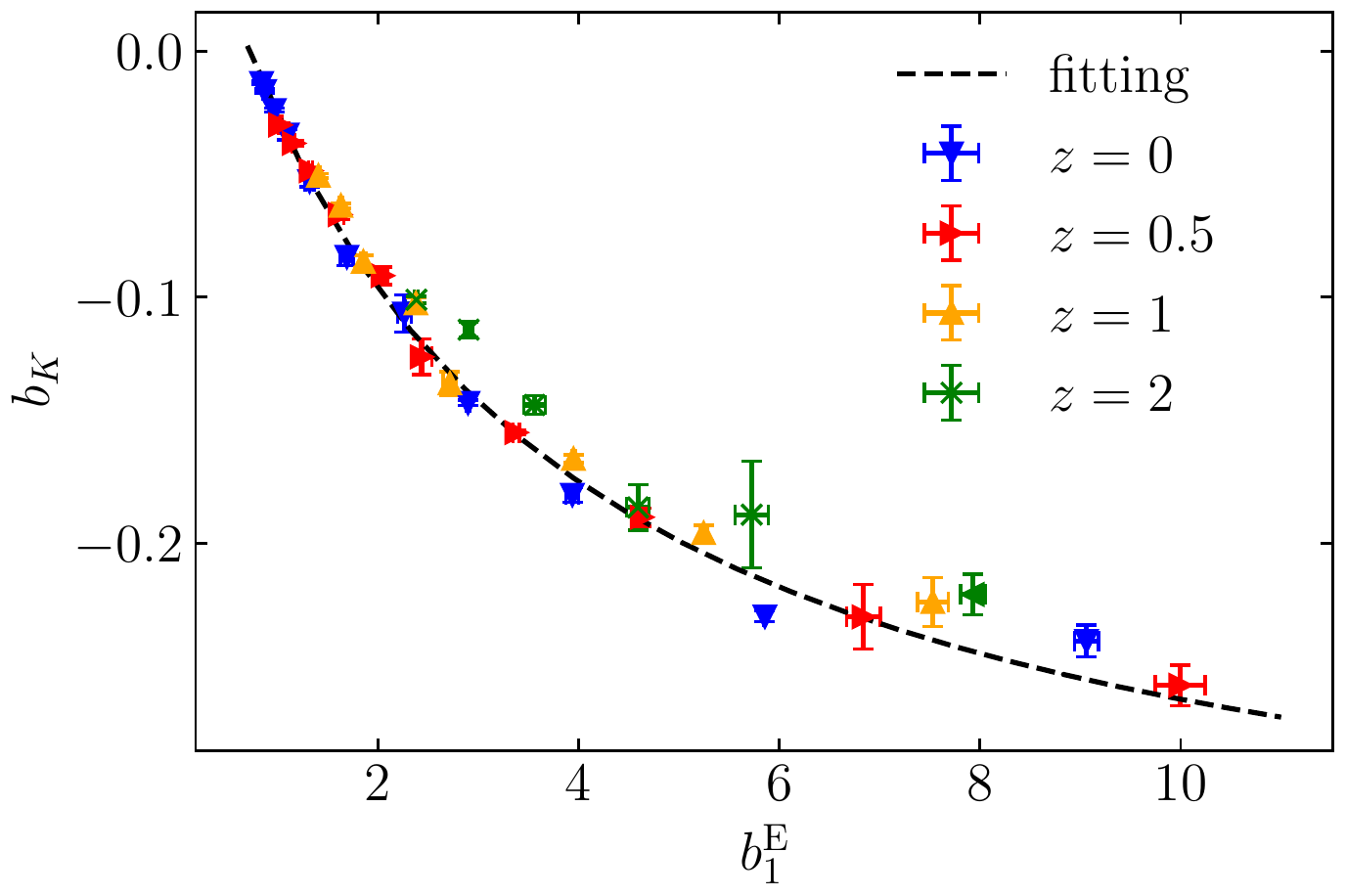}
\caption{Linear alignment coefficient, $b_K$, from $I_{ij}$, as a function of the Eulerian linear bias, $b_1^\mathrm{E}$, combining all redshift and mass information.
The dashed curve is the fitting given in Eq.~\eqref{eq:bE1_bK_fit_I}.
}
\label{fig:b1bK_nonreduced}
\end{figure}

Fig.~\ref{fig:bK_nonreduced} also compares $b_K$ from different kinds of tides and 
Fig.~\ref{fig:bK_nonreduced_redshifts} shows the time evolution of $b_K$ when using $I_{ij}$.
While the amplitude of $b_K$ from $I_{ij}$ differs from $J_{ij}$,
$I_{ij}$ results show the same trend as $J_{ij}$.
The universal behavior between $b_1^{\rm E}$ and $b_K$ is also found from $I_{ij}$ results as shown in Fig.~\ref{fig:b1bK_nonreduced}.
This implies that this relation is indeed ``universal'', regardless of the definition of shapes.
In $I_{ij}$ case we can fit this relation as
\begin{align}
    b_K = \frac{0.09302 - 0.1289 b_1^\Eul}{1 + 0.3541 b_1^\Eul}
    \label{eq:bE1_bK_fit_I}
\end{align}

\begin{figure}[tb]
\centering
\includegraphics[width=0.4\textwidth]{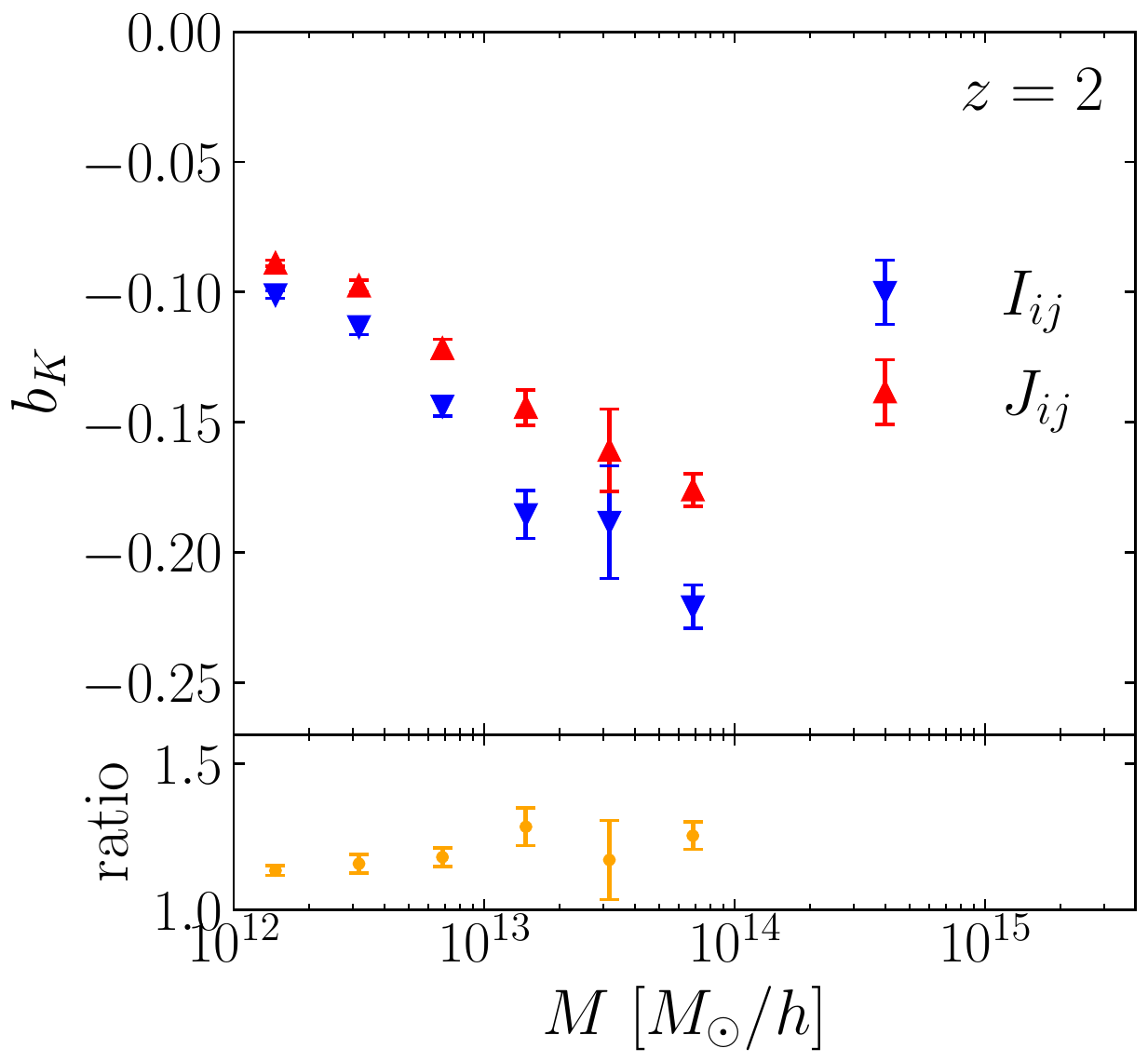}
\hspace{1em}
\includegraphics[width=0.4\textwidth]{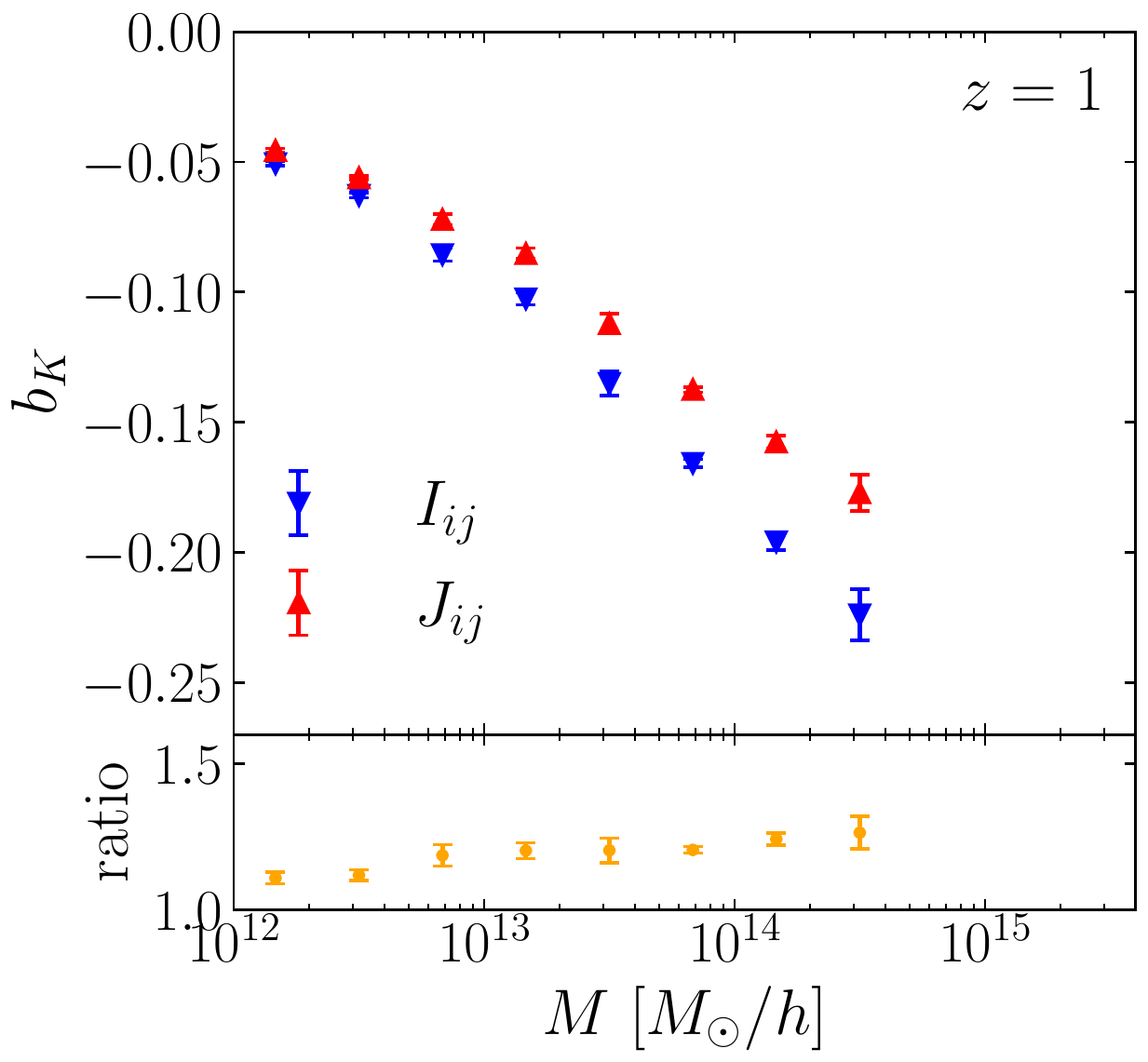}

\vspace{1em}
\includegraphics[width=0.4\textwidth]{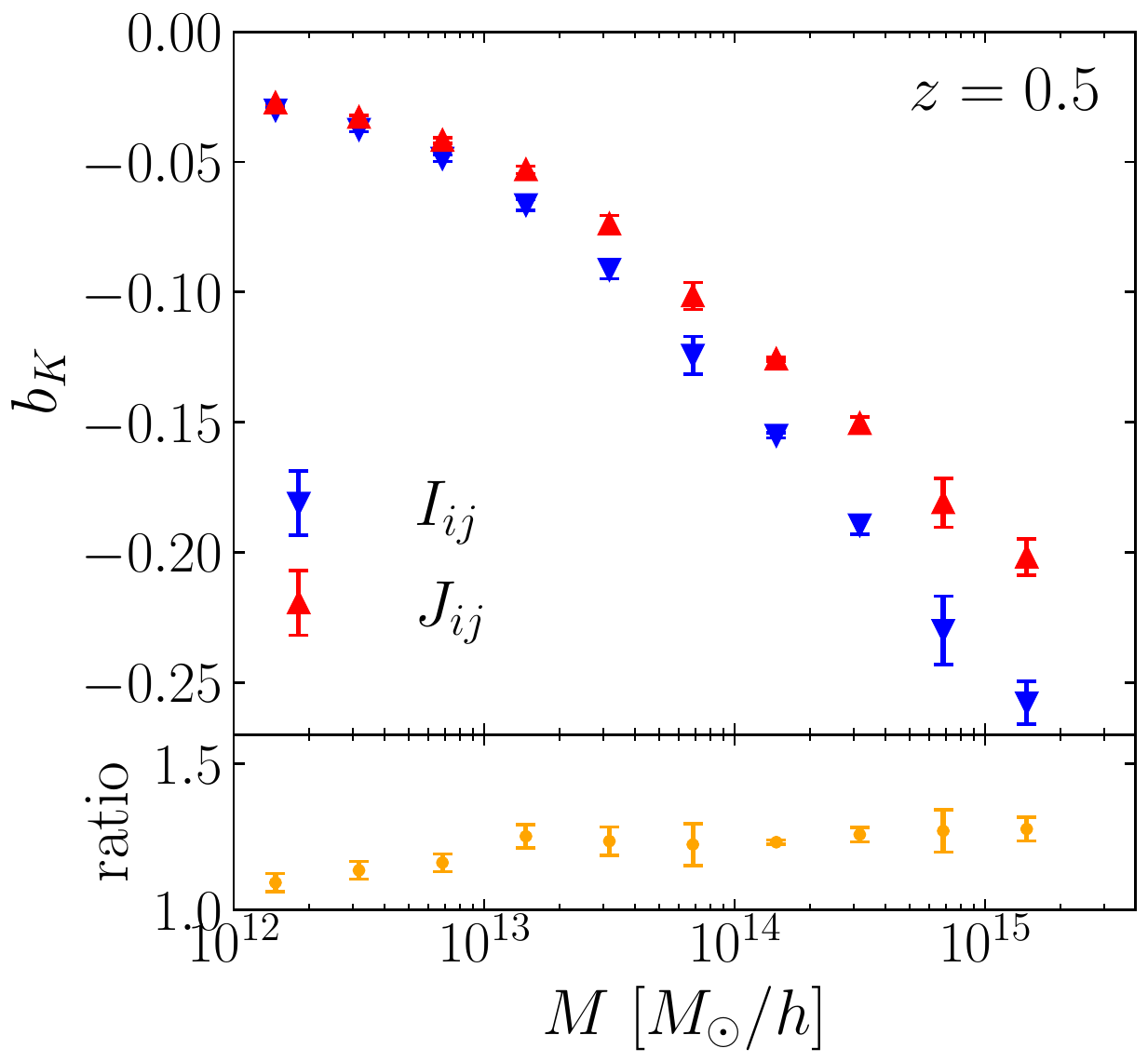}
\hspace{1em}
\includegraphics[width=0.4\textwidth]{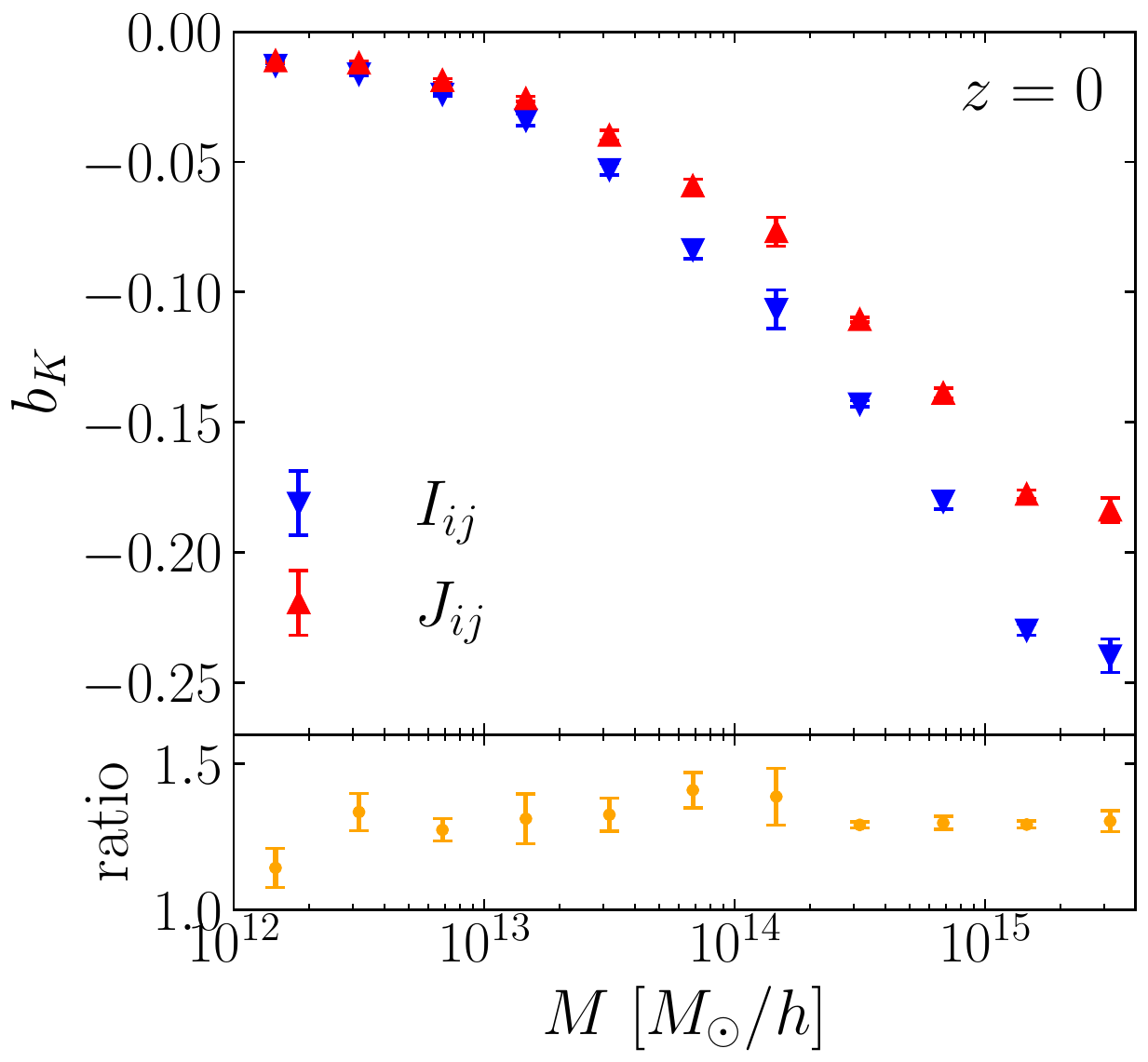}
\caption{
{\it Upper panels:} The shape definition dependence of the linear alignment coefficient, $b_K$, at various redshifts and masses:
$b_K$ from $I_{ij}$ (blue) and $J_{ij}$ (red).
{\it Lower panels:} The ratio of the two: $I_{ij}/J_{ij}$.}
\label{fig:bK_comp}
\end{figure}

In Fig.~\ref{fig:bK_comp} we provide the comparison of $b_K$ measured from $I_{ij}$ and $J_{ij}$ for various redshifts and mass-bins.
For all points, the amplitude of $|b_K|$ is greater when using $I_{ij}$ than when using $J_{ij}$ and its ratio does not change significantly over all the redshift and mass range.
Thus the choice of the definition of shapes does not change the dependence of $b_K$ on mass or redshift as already seen in Fig.~\ref{fig:bK_nonreduced_redshifts}.

\begin{figure}[tb]
\centering
\includegraphics[width=0.4\textwidth]{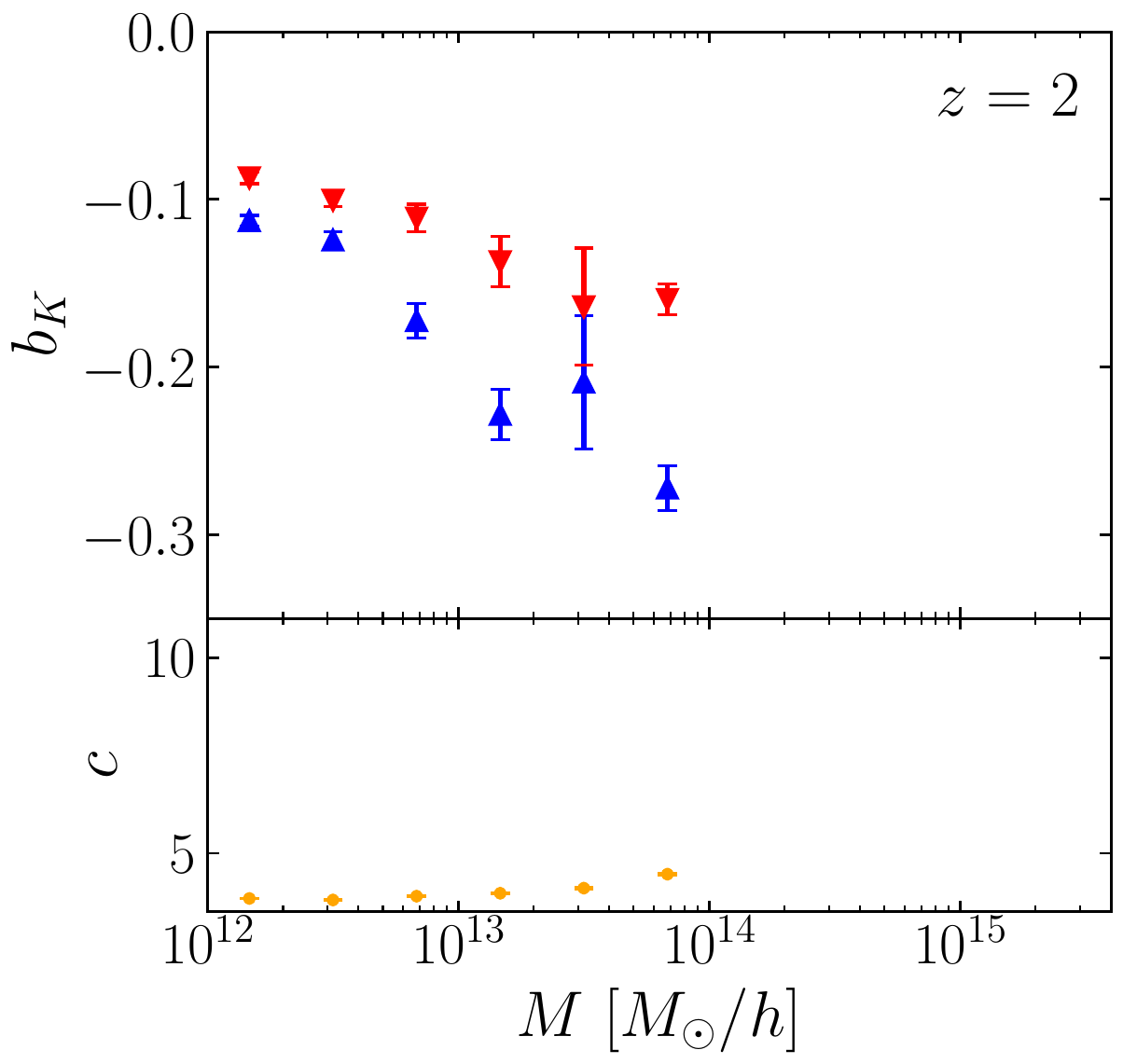}
\hspace{1em}
\includegraphics[width=0.52\textwidth]{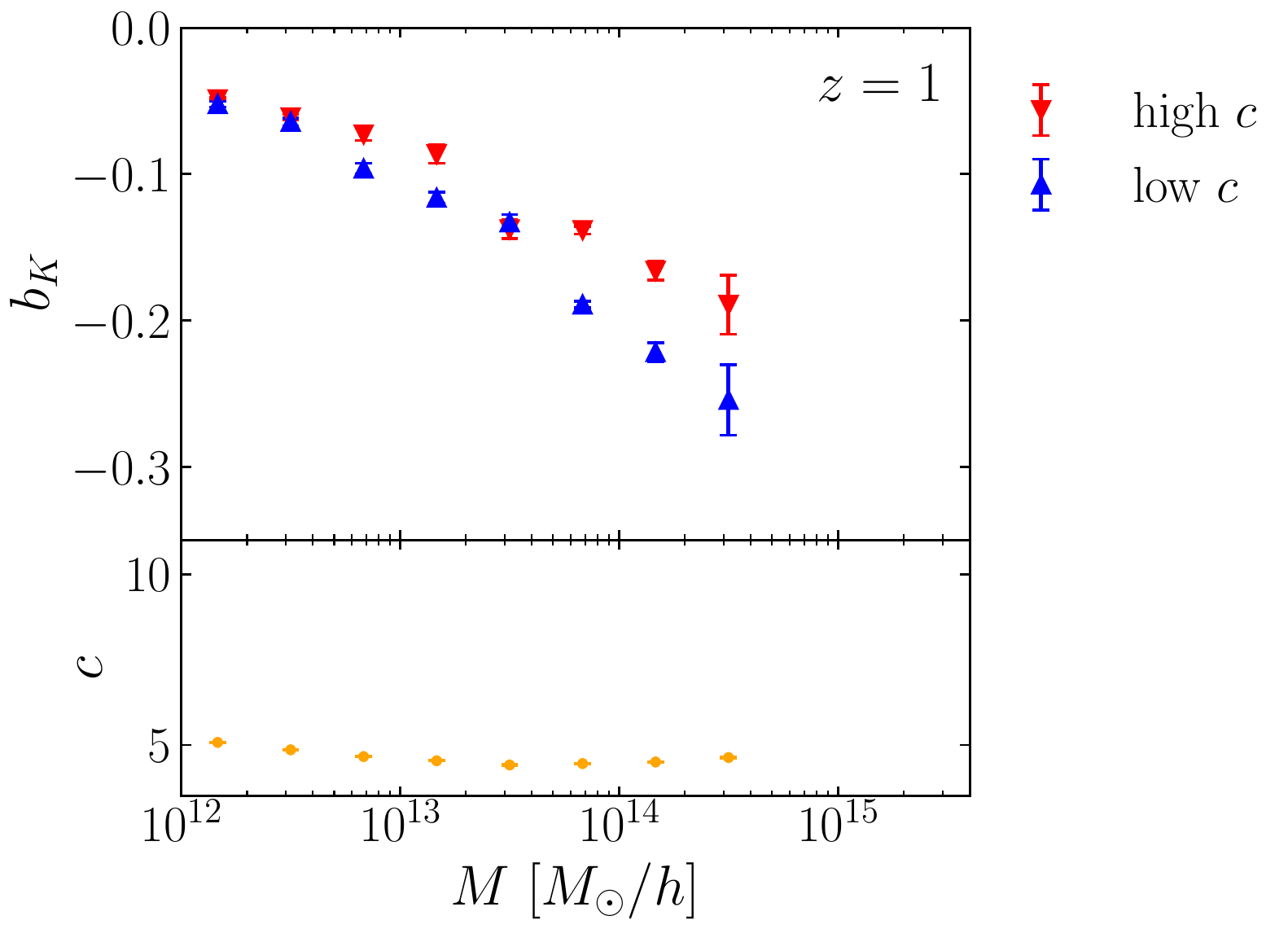}

\vspace{1em}
\includegraphics[width=0.4\textwidth]{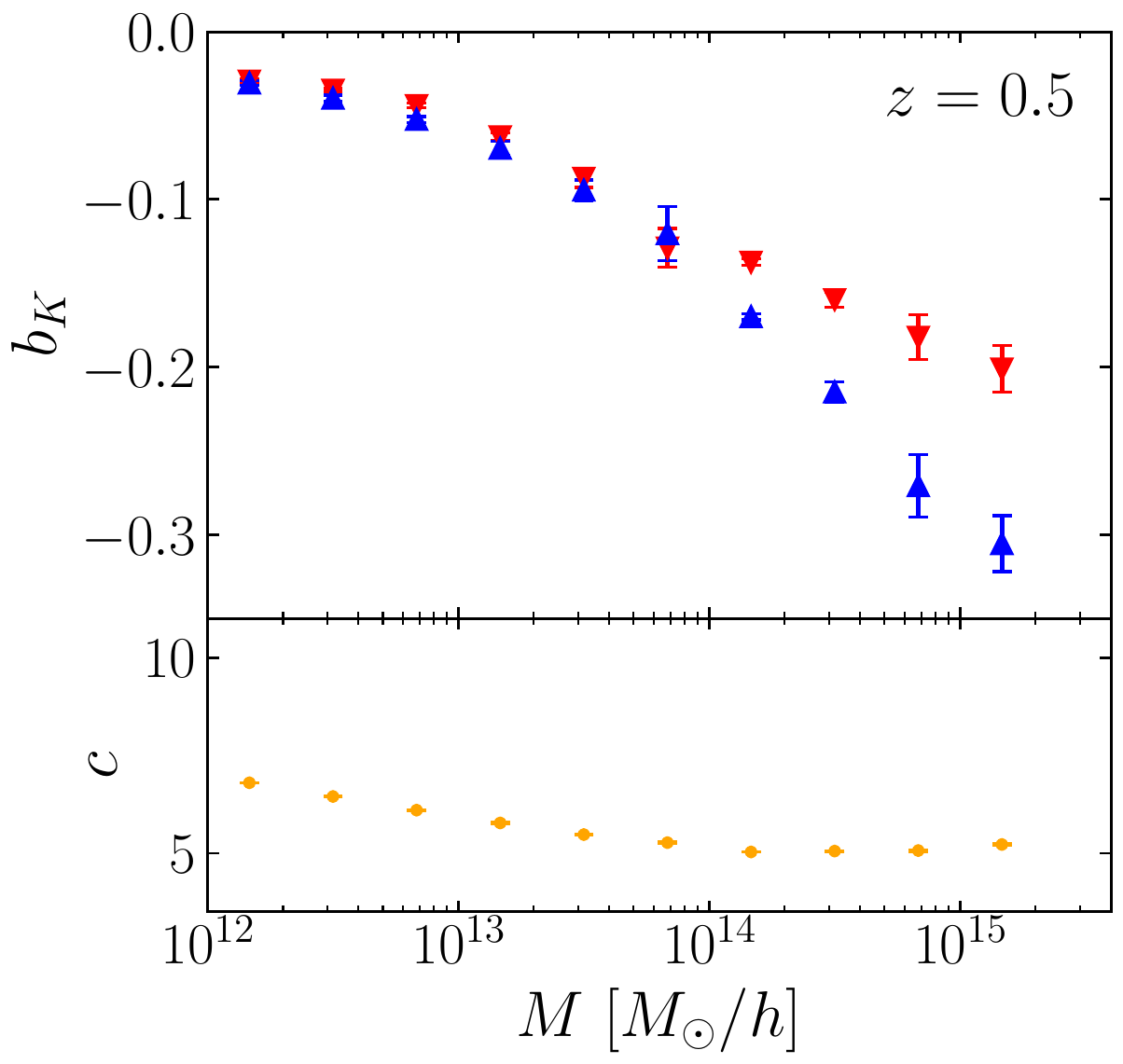}
\hspace{1em}
\includegraphics[width=0.52\textwidth]{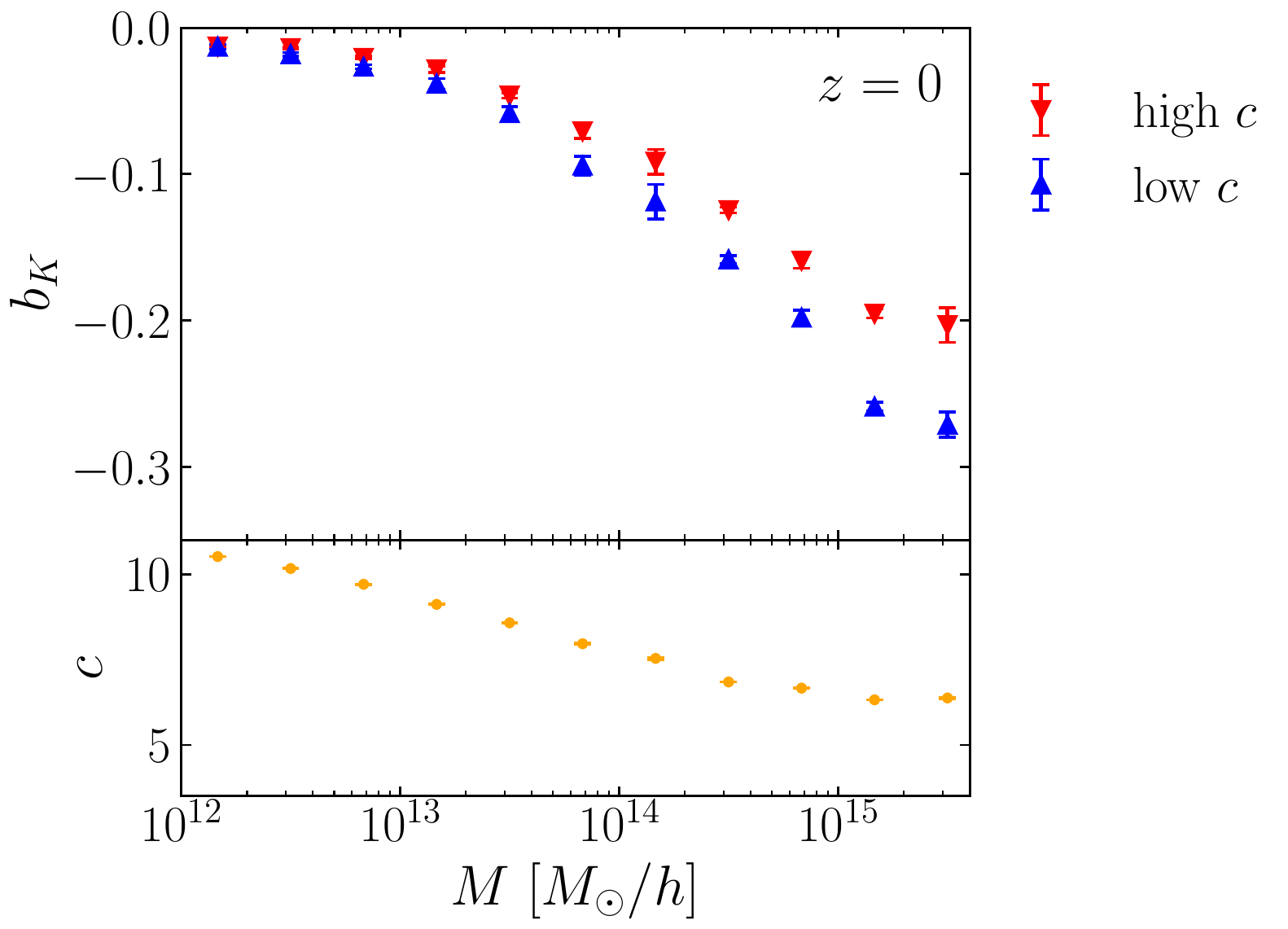}
\caption{
{\it Upper panels:} The halo concentration dependence of the linear alignment coefficient, $b_K$, at various redshifts and masses:
$b_K$ from high concentration (red) and low concentration (blue).
{\it Lower panels:} The median concentration on which we divided halo samples.
}
\label{fig:bK_nonreduced_c}
\end{figure}

\begin{figure}[tb]
\centering
\includegraphics[width=0.4\textwidth]{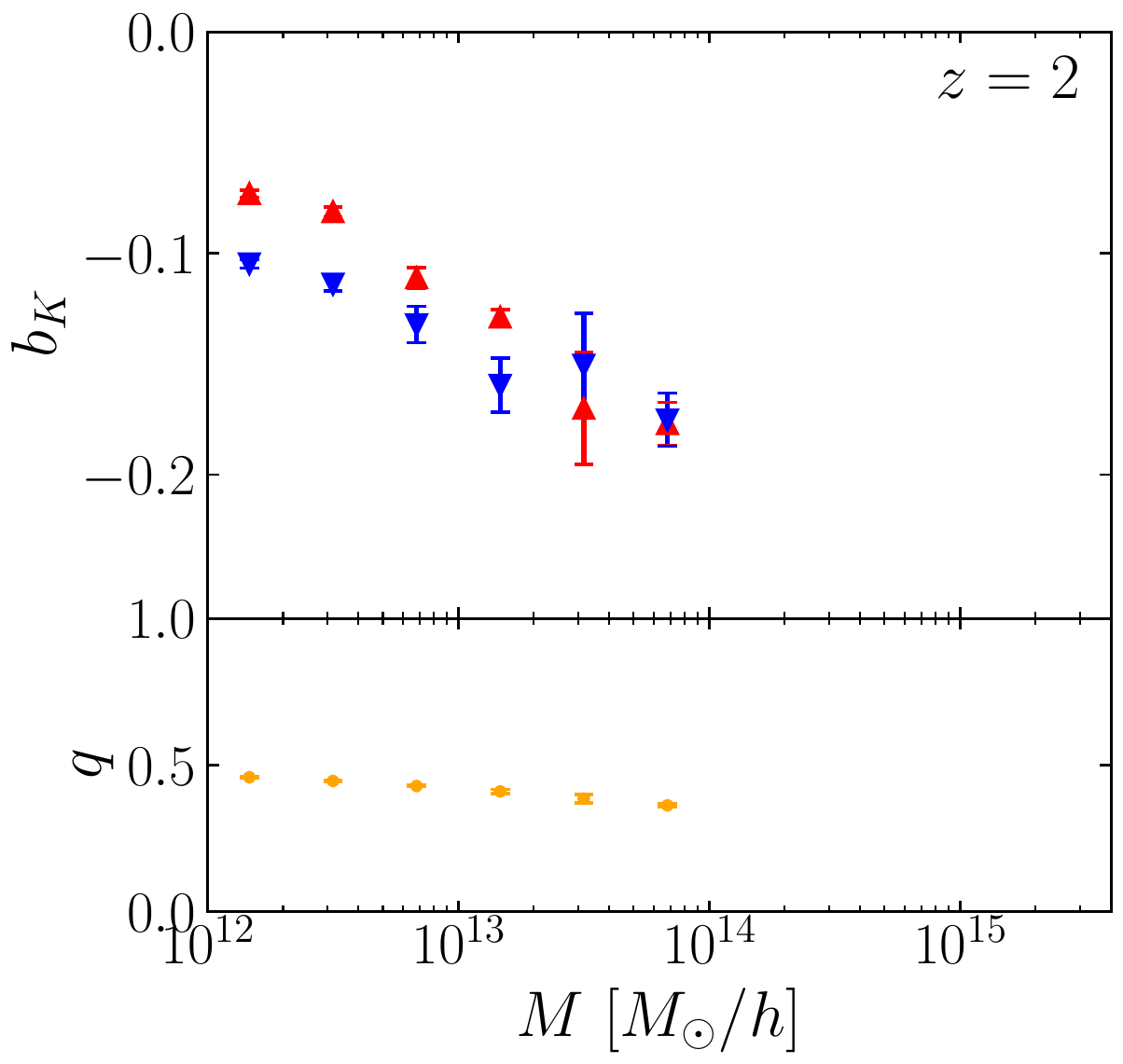}
\hspace{1em}
\includegraphics[width=0.52\textwidth]{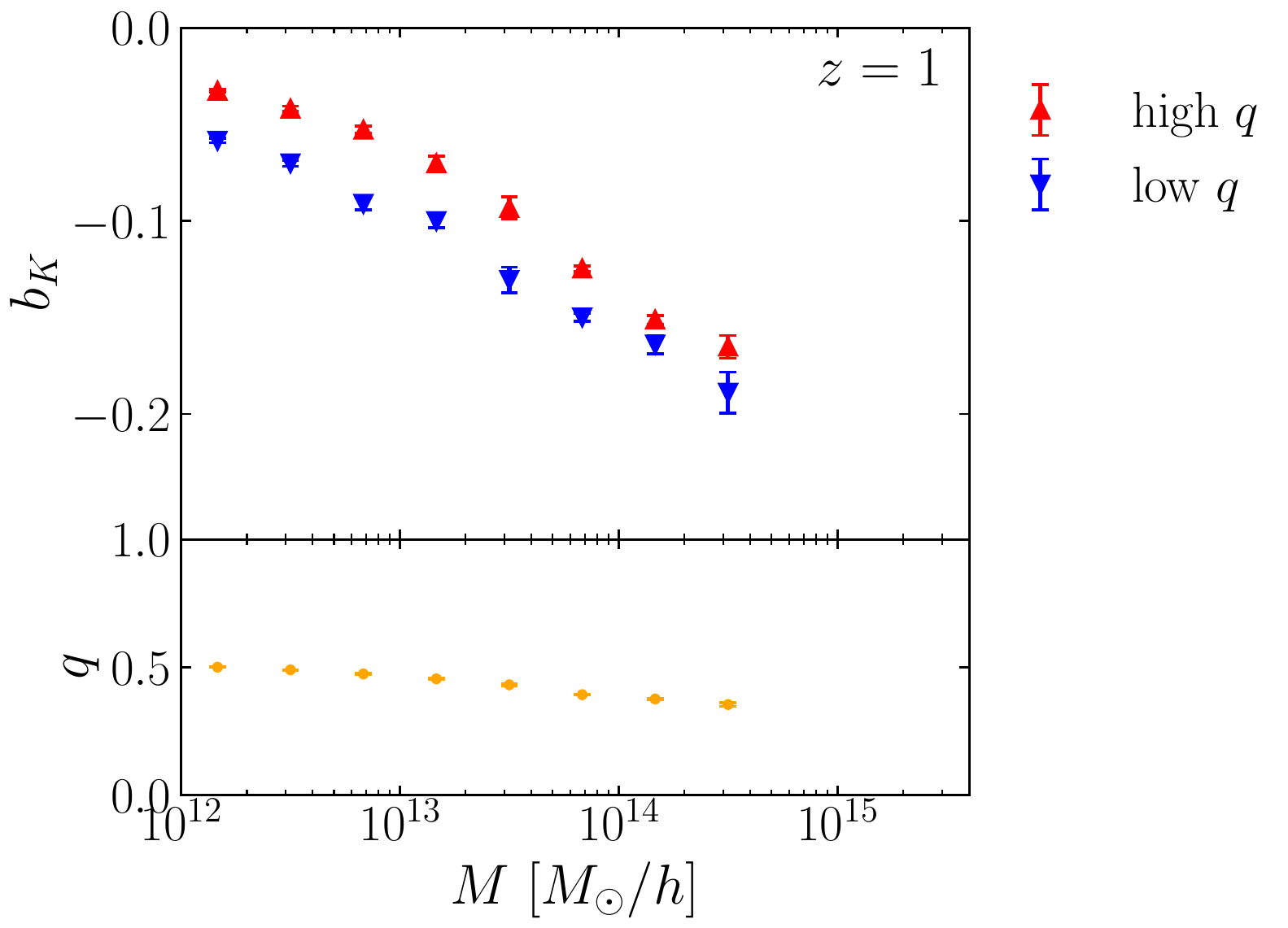}

\vspace{1em}
\includegraphics[width=0.4\textwidth]{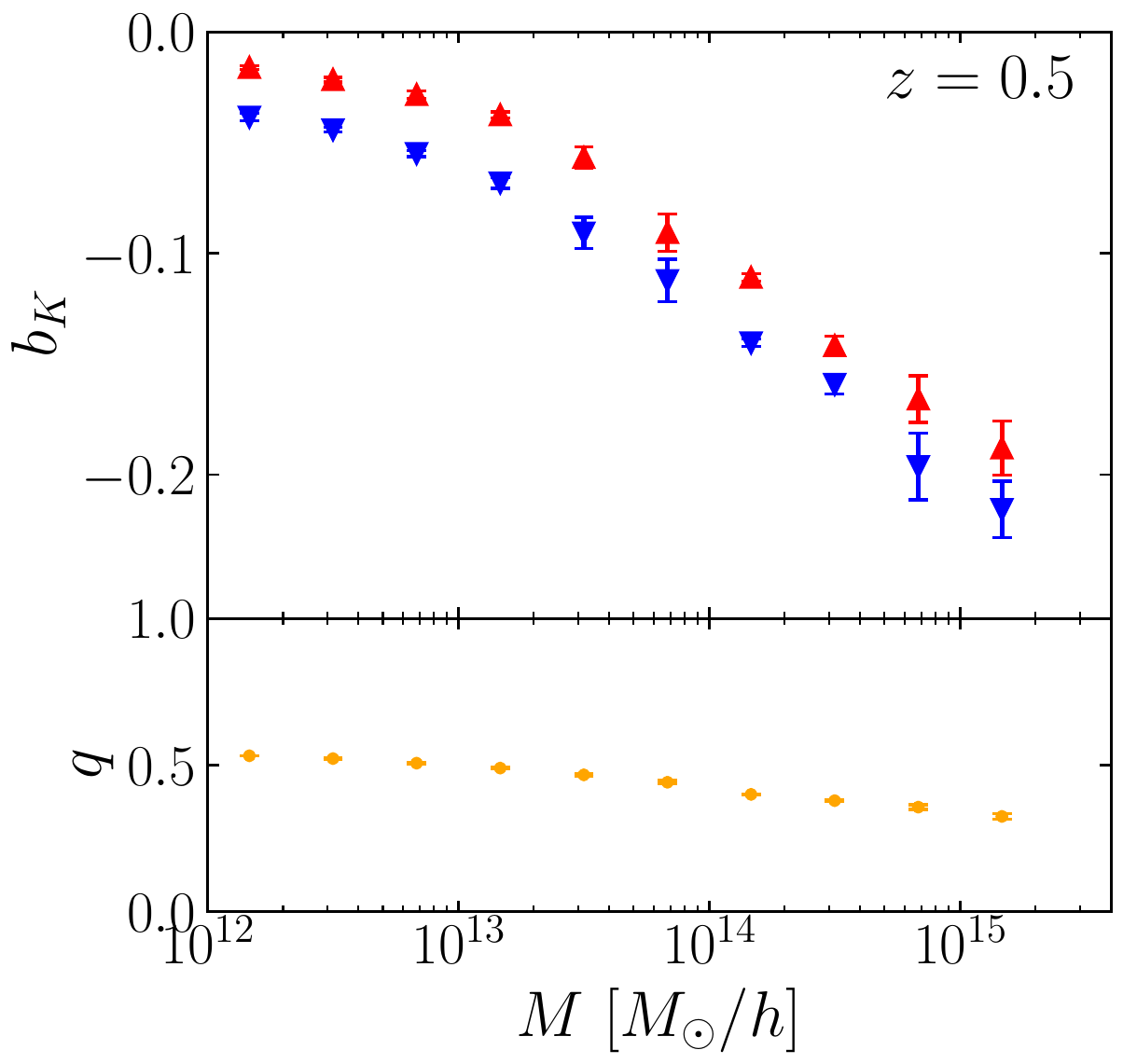}
\hspace{1em}
\includegraphics[width=0.52\textwidth]{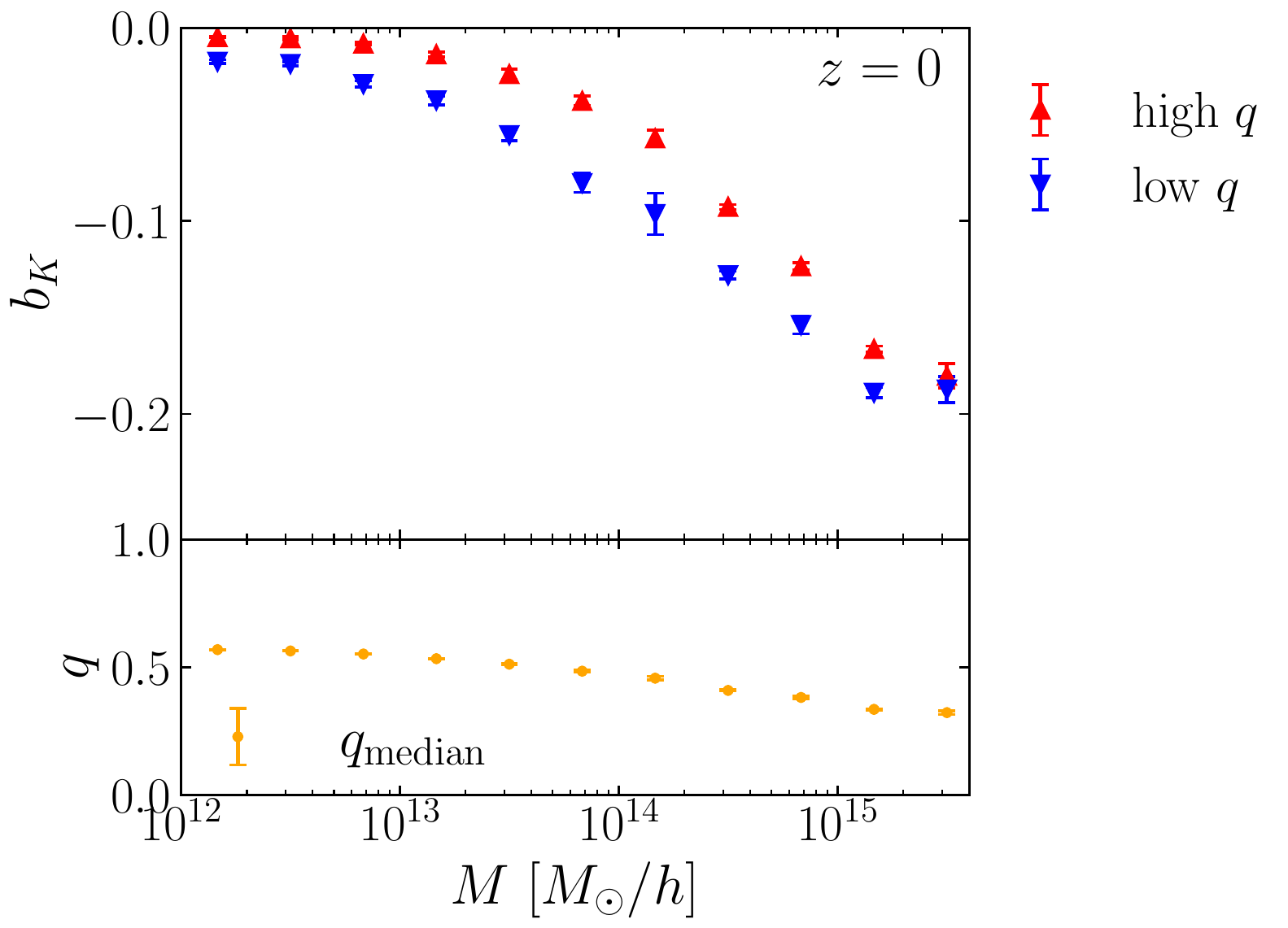}
\caption{
{\it Upper panels:} The axis-ratio ($q=I_3/I_1$ with $ I_1 \geq I_3$) dependence of the linear alignment coefficient, $b_K$, at various redshifts and masses: 
$b_K$ from high $q$ (low ellipticity, blue) and low $q$ (high ellipticity, red).
{\it Lower panels:} The median axis-ratio by which we divided halo samples.
}
\label{fig:bK_nonreduced_e}
\end{figure}

Finally, in Fig.~\ref{fig:bK_nonreduced_c} and Fig.~\ref{fig:bK_nonreduced_e},
we present that the secondary dependence of $b_K$ on halo concentration and the axis-ratio is also found when using $I_{ij}$
with the same trend.
This suggests that the secondary dependence of $b_K$ is genuine.

\clearpage

\bibliographystyle{JHEP}
\bibliography{main}
\end{document}